%% file: main.tex
\documentclass[12pt,a4paper]{article}
\pdfoutput=1
\usepackage{ifthen}
\newboolean{pdflatex}
\setboolean{pdflatex}{true}

\newboolean{articletitles}
\setboolean{articletitles}{true}

\newboolean{uprightparticles}
\setboolean{uprightparticles}{false}

\newboolean{inbibliography}
\setboolean{inbibliography}{false}

\input{preamble}
\usepackage{longtable}

\usepackage[separate-uncertainty=true]{siunitx}
\usepackage{lhcb-definitions}
\usepackage{charmproduction-definitions}
\usepackage{measurement-definitions}
\usepackage{subcaption}
\usepackage{booktabs}
\usepackage{lscape}
\usepackage{adjustbox}
\usepackage{hvfloat}

\begin{document}

\renewcommand{\thefootnote}{\fnsymbol{footnote}}
\setcounter{footnote}{1}

\input{title-LHCb-PAPER}
\input{./tables/syst_macros/macros_D0ToKpi.tex}
\input{./tables/syst_macros/macros_DpToKpipi.tex}
\input{./tables/syst_macros/macros_DsTophipi.tex}
\input{./tables/syst_macros/macros_DstToD0pi_D0ToKpi.tex}

\renewcommand{\thefootnote}{\arabic{footnote}}
\setcounter{footnote}{0}

\pagestyle{plain}
\setcounter{page}{1}
\pagenumbering{arabic}

\input{introduction}

\input{conditions}

\input{analysis}

\input{measurements}

\input{systematics}

\input{ratios}

\clearpage

\input{theory}

\input{summary}

\input{acknowledgements}

\clearpage

{\noindent\bf\Large Appendices}

\appendix

\input{appendix_xsec}

\input{appendix_ratios}

\clearpage

\input{appendix_mesonRatios}

\clearpage

\addcontentsline{toc}{section}{References}
\setboolean{inbibliography}{true}
\bibliographystyle{LHCb}
\bibliography{main,LHCb-PAPER,LHCb-CONF,LHCb-DP,LHCb-TDR,theory}

\newpage

\newpage
\input{LHCb_HD_authorlist_2015-07-21.tex}

\end{document}

%% file: preamble.tex
\usepackage{microtype}
\usepackage{lineno}
\usepackage{xspace}
\usepackage{caption}
\usepackage{multirow}

\usepackage{graphicx}
\usepackage{color}
\usepackage{colortbl}
\graphicspath{{./figs/}}

\usepackage{amsmath}
\usepackage{amssymb}
\usepackage{amsfonts}
\usepackage{upgreek}

\newcommand*\patchAmsMathEnvironmentForLineno[1]{%
\expandafter\let\csname old#1\expandafter\endcsname\csname #1\endcsname
\expandafter\let\csname oldend#1\expandafter\endcsname\csname
end#1\endcsname
 \renewenvironment{#1}%
   {\linenomath\csname old#1\endcsname}%
   {\csname oldend#1\endcsname\endlinenomath}%
}
\newcommand*\patchBothAmsMathEnvironmentsForLineno[1]{%
  \patchAmsMathEnvironmentForLineno{#1}%
  \patchAmsMathEnvironmentForLineno{#1*}%
}
\AtBeginDocument{%
\patchBothAmsMathEnvironmentsForLineno{equation}%
\patchBothAmsMathEnvironmentsForLineno{align}%
\patchBothAmsMathEnvironmentsForLineno{flalign}%
\patchBothAmsMathEnvironmentsForLineno{alignat}%
\patchBothAmsMathEnvironmentsForLineno{gather}%
\patchBothAmsMathEnvironmentsForLineno{multline}%
\patchBothAmsMathEnvironmentsForLineno{eqnarray}%
}

\usepackage{hyperref}
\usepackage[all]{hypcap}
\usepackage{cite}
\usepackage{mciteplus}

%% file: title-LHCb-PAPER.tex
% $Id: title-LHCb-PAPER.tex 97578 2016-09-02 13:33:07Z apearce $
% ===============================================================================
% Purpose: LHCb-PAPER journal paper title page template
% Author: 
% Created on: 2010-09-25
% ===============================================================================

\input{./tables/integrated/integrated_macros.tex}
%%%%%%%%%%%%%%%%%%%%%%%%%
%%%%%  TITLE PAGE  %%%%%%
%%%%%%%%%%%%%%%%%%%%%%%%%
\begin{titlepage}
\pagenumbering{roman}

% Header ---------------------------------------------------
\vspace*{-1.5cm}
\centerline{\large EUROPEAN ORGANIZATION FOR NUCLEAR RESEARCH (CERN)}
\vspace*{1.5cm}
\noindent
\begin{tabular*}{\linewidth}{lc@{\extracolsep{\fill}}r@{\extracolsep{0pt}}}
\ifthenelse{\boolean{pdflatex}}% Logo format choice
{\vspace*{-2.7cm}\mbox{\!\!\!\includegraphics[width=.14\textwidth]{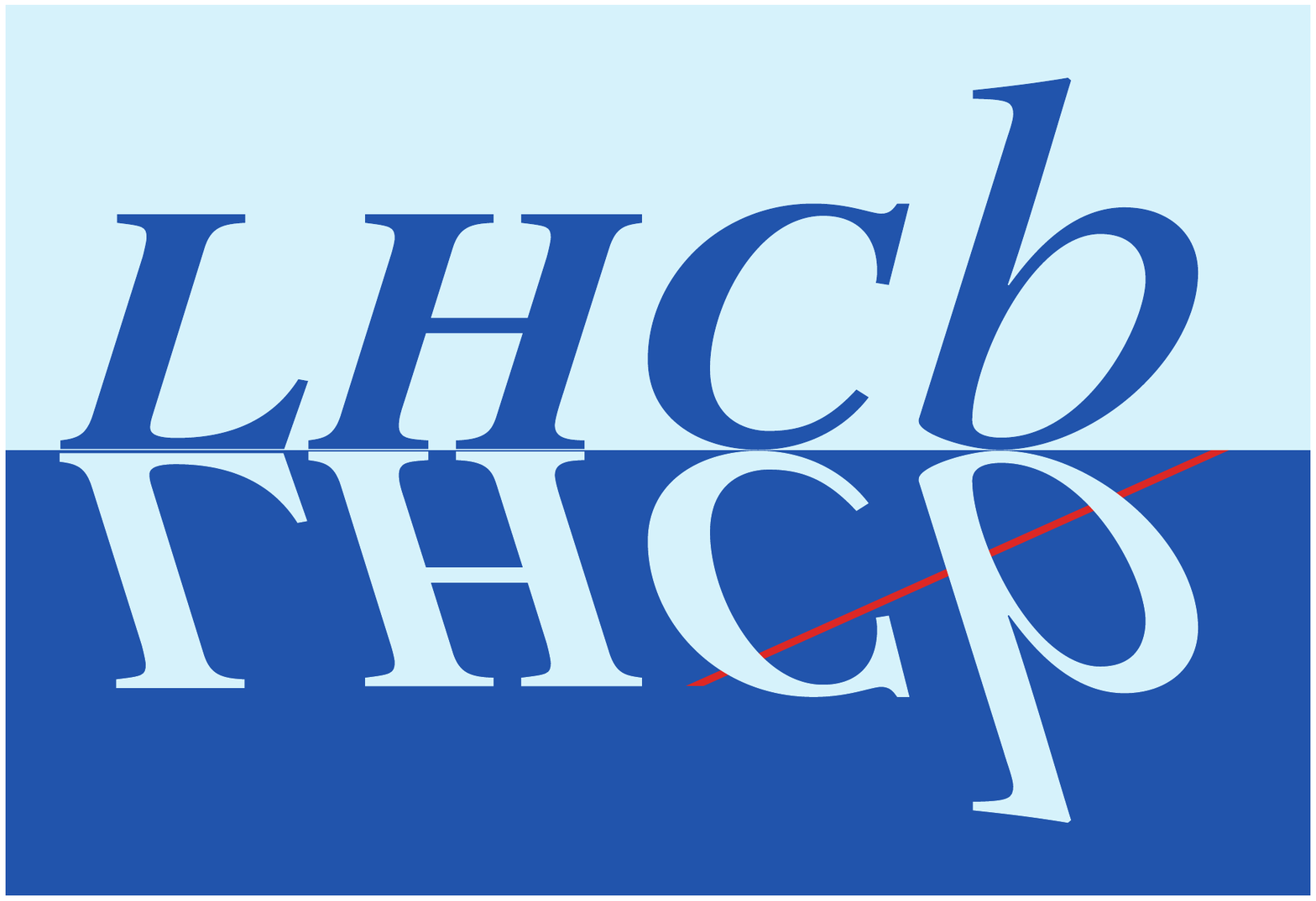}} & &}%
{\vspace*{-1.2cm}\mbox{\!\!\!\includegraphics[width=.12\textwidth]{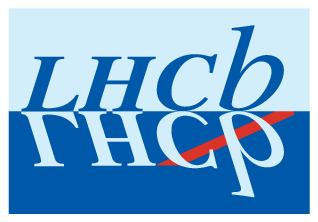}} & &}%
\\
 & & CERN-PH-EP-2015-272 \\  % ID 
 & & LHCb-PAPER-2015-041 \\  % ID 
 & & \today \\ % Date - Can also hardwire e.g.: 23 March 2010
% not in paper \hline
\end{tabular*}

\vspace*{1.5cm}

% Title --------------------------------------------------
{\bf\boldmath\huge
\begin{center}
  Measurements of prompt charm production cross-sections in \pp collisions at $\sqrt{s} = 13$\,TeV
\end{center}
}

\vspace*{0.5cm}

% Authors -------------------------------------------------
\begin{center}
%In the footnote, replace 'paper' by 'letter' in case of submission to PRL or PLB 
The LHCb collaboration\footnote{Authors are listed at the end of this paper.}
\end{center}

\vspace{0.5cm}

% Abstract -----------------------------------------------
\begin{abstract}
  \noindent
  Production cross-sections of prompt charm mesons are measured with the first data from \pp collisions at the LHC at a centre-of-mass energy of \SI{13}{\TeV}.
  The data sample corresponds to an integrated luminosity of
  \totlumi collected by the \lhcb experiment.
  The production cross-sections of \Dz, \Dp, \Dsp, and \Dstp mesons are measured in bins
  of charm meson transverse momentum, \pT, and rapidity, \rapidity, 
  and cover the range $0 < \pT < \SI{15}{\gevc}$ and \yrange.
  The inclusive cross-sections for the four mesons, including charge conjugation, within the range of $1 < \pT < \SI{8}{\gevc}$ are found to be
  \begin{align*}
    \begin{array}{lcr}
      \sigma(\decay{\pp}{\Dz X})  &=& \DzOnePtXsec, \\
      \sigma(\decay{\pp}{\Dp X})  &=& \DpOnePtXsec, \\
      \sigma(\decay{\pp}{\Dsp X}) &=& \DspOnePtXsec, \\
      \sigma(\decay{\pp}{\Dstp X})&=& \DstpOnePtXsec, \\
    \end{array}
  \end{align*}
where the uncertainties are due to statistical and systematic uncertainties, respectively.

\end{abstract}

\vspace*{\fill}

\begin{center}
  Published in JHEP 03 (2016) 159\\
  Errata published in JHEP 09 (2016) 013 and JHEP 05 (2017) 074\footnote{The contents of the errata are reflected in this manuscript.}
\end{center}

\vspace{\fill}

{\footnotesize 
\centerline{\copyright~CERN on behalf of the \lhcb collaboration, licence \href{http://creativecommons.org/licenses/by/4.0/}{CC-BY-4.0}.}}
\vspace*{2mm}

\end{titlepage}

%%%%%%%%%%%%%%%%%%%%%%%%%%%%%%%%
%%%%%  EOD OF TITLE PAGE  %%%%%%
%%%%%%%%%%%%%%%%%%%%%%%%%%%%%%%%

%  empty page follows the title page ----
\newpage
\setcounter{page}{2}
\mbox{~}
%\newpage
%
%% Author List ----------------------------

\cleardoublepage

%% file: tables/integrated/integrated_macros.tex
\newcommand{\DzXsec}{\SI[parse-numbers=false]{2709 \pm 2 \pm 165}{\micro\barn}}
\newcommand{\DpXsec}{\SI[parse-numbers=false]{1102 \pm 5 \pm 111}{\micro\barn}}
\newcommand{\DzOnePtXsec}{\SI[parse-numbers=false]{2072 \pm 2 \pm 124}{\micro\barn}}
\newcommand{\DpOnePtXsec}{\SI[parse-numbers=false]{834 \pm 2 \pm \phantom{1}78}{\micro\barn}}
\newcommand{\DspOnePtXsec}{\SI[parse-numbers=false]{353 \pm 9 \pm \phantom{1}76}{\micro\barn}}
\newcommand{\DstpOnePtXsec}{\SI[parse-numbers=false]{784 \pm 4 \pm \phantom{1}87}{\micro\barn}}
\newcommand{\ccbarXsec}{\SI[parse-numbers=false]{2369 \pm 3 \pm 152\pm 118}{\micro\barn}}
\newcommand{\ccbarOnePtXsec}{\SI[parse-numbers=false]{1802 \pm 2 \pm 115\pm 85}{\micro\barn}}
\newcommand{\DpDzint}{$\phantom{1}0.950 ^{+\phantom{1}0.003}_{-\phantom{1}0.003}$$^{+\phantom{1}0.040}_{-\phantom{1}0.040}$}
\newcommand{\DspDzint}{$0.0991 ^{+0.0026}_{-0.0026}$$^{+0.0053}_{-0.0079}$}
\newcommand{\DstpDzint}{$\phantom{1}0.256 ^{+\phantom{1}0.001}_{-\phantom{1}0.001}$$^{+\phantom{1}0.021}_{-\phantom{1}0.021}$}
\newcommand{\DspDpint}{$0.1043 ^{+0.0028}_{-0.0028}$$^{+0.0038}_{-0.0066}$}
\newcommand{\DstpDpint}{$\phantom{1}0.270 ^{+\phantom{1}0.002}_{-\phantom{1}0.002}$$^{+\phantom{1}0.023}_{-\phantom{1}0.022}$}
\newcommand{\DspDstpint}{$\phantom{1}0.390 ^{+\phantom{1}0.010}_{-\phantom{1}0.010}$$^{+\phantom{1}0.038}_{-\phantom{1}0.045}$}

%% file: tables/syst_macros/macros_D0ToKpi.tex
\newcommand{\suDzmcstat}{1--26} 
\newcommand{\suDztracking}{3--10}
\newcommand{\suDzmcagreement}{1}

%% file: tables/syst_macros/macros_DpToKpipi.tex
\newcommand{\suDpmcstat}{1--39}
\newcommand{\suDptracking}{4--14}
\newcommand{\suDpmcagreement}{1}

%% file: tables/syst_macros/macros_DsTophipi.tex
\newcommand{\suDspmcstat}{1--55}
\newcommand{\suDsptracking}{4--14}
\newcommand{\suDspmcagreement}{0.2}

%% file: tables/syst_macros/macros_DstToD0pi_D0ToKpi.tex
\newcommand{\suDstpmcstat}{1--23}
\newcommand{\suDstptracking}{5--11}
\newcommand{\suDstpmcagreement}{0.9}

%% file: introduction.tex
\section{Introduction}
\label{sec:Introduction}
Measurements of charm production cross-sections in proton-proton collisions are important tests of the predictions of perturbative quantum chromodynamics~\cite{Gauld:2015yia,Cacciari:2015fta,Kniehl:2012ti}. 
Predictions of charm meson cross-sections have been made 
at next-to-leading order using the generalized mass variable flavour number scheme~(GMVFNS)~\cite{Kniehl:2004fy,Kniehl:2005gz,Kniehl:2005ej,Kneesch:2007ey,Kniehl:2009ar,Kniehl:2012ti} and 
at fixed order with next-to-leading-log resummation~(FONLL)~\cite{Gauld:2015yia,Cacciari:1998it,Cacciari:2003zu,Cacciari:2005uk,Cacciari:2012ny, Cacciari:2015fta}.
These are based on a factorisation approach, where the cross-sections are calculated as a convolution of three terms: the parton distribution functions 
of the incoming protons; the partonic hard scattering rate, estimated as a perturbative series in the coupling constant of the strong interaction; and a fragmentation function 
that parametrises the hadronisation of the charm quark into a given type of charm hadron.
The range of \rapidity and \pT accessible to \lhcb enables quantum chromodynamics calculations to be tested in a region where the momentum fraction, $x$, of the initial state partons 
can reach values below $10^{-4}$. In this region the uncertainties on the gluon parton density functions are large, exceeding 30$\%$~\cite{Gauld:2015yia,Zenaiev:2015rfa}, and 
\lhcb measurements can be used to constrain them. 
For example, the predictions provided in Ref.~\cite{Gauld:2015yia} have made direct use of these constraints from \lhcb data, taking as input a set of parton density functions that is weighted to match the \lhcb measurements at $\sqrts = \SI{7}{\TeV}$.

The charm production cross-sections are also important in evaluating the rate of high-energy neutrinos created from the decay of charm hadrons produced in cosmic ray interactions with atmospheric nuclei~\cite{Bhattacharya:2015jpa,Gauld:2015yia}. 
Such neutrinos constitute an important background for experiments such as IceCube~\cite{IceCubeCollaboration} searching for neutrinos produced from astrophysical sources.
The previous measurements from \lhcb at $\sqrts = \SI{7}{\TeV}$~\cite{LHCb-PAPER-2012-041} permit the evaluation of this background for incoming cosmic rays with energy of \SI{26}{\PeV}.
In this paper measurements at $\sqrts = \SI{13}{\TeV}$ are presented, probing a new kinematic region that corresponds to a primary cosmic ray energy of \SI{90}{\PeV}.

Measurements of the charm production cross-sections have been performed in different kinematic regions and centre-of-mass energies.
Measurements by the CDF experiment cover the central rapidity region $|y| < 1$ and transverse momenta, \pT, 
between \SI{5.5}{\gevc} and \SI{20}{\gevc} at  $\sqrts = \SI{1.96}{\TeV} $ in \ppbar collisions~\cite{Acosta:2003ax}.
At the Large Hadron Collider~(\lhc), charm cross-sections in \pp collisions have been measured in 
the $|y| < 0.5$ region for $\pT>\SI{1}{\gevc}$ at $\sqrts =\SI{2.76}{\TeV}$ 
and $\sqrts = \SI{7}{\TeV}$  by the ALICE experiment~\cite{Abelev:2012sx,ALICE:2012ik,ALICE:2011aa}.
The \lhcb experiment has recorded the world's largest dataset of charm hadrons to date and this has led to numerous high-precision measurements of their production and decay properties. 
\lhcb measured the cross-sections in the forward region $2.0 < y < 4.5$ for  $0<\pT<\SI{8}{\gevc}$ at $\sqrts = \SI{7}{\TeV}$~\cite{LHCb-PAPER-2012-041}.

Charm mesons produced at the \pp collision point, either directly or as decay products of excited charm resonances, are referred to as promptly produced.
No attempt is made to distinguish between these two sources.
This paper presents measurements of the cross-sections for the prompt production of \Dz, \Dp, \Dsp, and $\PD^*(2010)^+$ (henceforth denoted as \Dstarp) mesons, 
based on data corresponding to an integrated luminosity of \totlumi.
Charm mesons produced through the decays of \bquark hadrons are referred to as secondary charm, and are considered as a background process.

Section~\ref{sec:Detector} describes the detector, data acquisition conditions, and the simulation; this is followed by a detailed account of the data analysis in Sec.~\ref{sec:analysis}.
The differential cross-section results are given in Sec.~\ref{sec:measurements}, followed by a discussion of systematic uncertainties in Sec.~\ref{sec:syst}.
Section~\ref{sec:ratios} presents the measurements of integrated cross-sections and of the ratios of the cross-sections measured at $\sqrts = \SI{13}{\TeV}$ to those at \SI{7}{\TeV}.
The theory predictions and their comparison with the results of this paper are discussed in Sec.~\ref{sec:theory}.
Sec.~\ref{sec:summary} provides a summary.
\vspace{-2mm}

%% file: conditions.tex
\section{Detector and simulation}
\label{sec:Detector}
The \lhcb detector~\cite{Alves:2008zz,LHCb-DP-2014-002} is a single-arm forward
spectrometer covering the \mbox{pseudorapidity} range $2<\eta <5$,
designed for the study of particles containing \bquark or \cquark
quarks. The detector includes a high-precision tracking system
consisting of a silicon-strip vertex detector surrounding the $pp$
interaction region, a large-area silicon-strip detector located
upstream of a dipole magnet with a bending power of about
$4{\rm\,Tm}$, and three stations of silicon-strip detectors and straw
drift tubes placed downstream of the magnet.
The tracking system provides a measurement of momentum of charged particles with
a relative uncertainty that varies from 0.5\% at low momentum to 1.0\% at \SI{200}{\gevc}.
The minimum distance of a track to a primary vertex, the impact parameter~(IP), is measured with a resolution of $(15+29/\pt)\,\si{\micro\meter}$,
where \pt is the component of the momentum transverse to the beam, in\,\si{\gevc}.
Different types of charged hadrons are distinguished by information
from two ring-imaging Cherenkov detectors. 
Photons, electrons and hadrons are identified by a calorimeter system consisting of
scintillating-pad and preshower detectors, an electromagnetic
calorimeter and a hadronic calorimeter. Muons are identified by a
system composed of alternating layers of iron and multiwire
proportional chambers.

The online event selection is performed by a trigger. This consists of a hardware stage, which for this analysis randomly selects 
a pre-defined fraction of all beam-beam crossings, followed by a software 
stage.  This analysis benefits from a new scheme for the \lhcb software trigger 
introduced for \lhc Run 2.
Alignment and calibration is performed in near real-time~\cite{Dujany:2017839} and updated constants are made available for the trigger.
The same alignment and calibration information is propagated to the offline reconstruction, ensuring consistent and high-quality
particle identification~(\pid) information between the trigger and offline software.
The larger timing budget available in the trigger compared to \lhcb Run 1 also results in the convergence of the online
and offline track reconstruction, such that offline performance is achieved in the trigger.
The identical performance of the online and offline 
reconstruction offers the opportunity to perform physics analyses directly using candidates
reconstructed in the trigger~\cite{LHCb-DP-2012-004}. The storage of only the triggered candidates enables a reduction
in the event size by an order of magnitude.

In the simulation, $pp$ collisions are generated with
\pythia~\cite{Sjostrand:2007gs} using a specific \lhcb
configuration~\cite{LHCb-PROC-2010-056}.  Decays of hadronic particles
are described by \evtgen~\cite{Lange:2001uf} in which final-state
radiation is generated with \photos~\cite{Golonka:2005pn}. The
implementation of the interaction of the generated particles with the detector, and its response,
uses the \geant
toolkit~\cite{Allison:2006ve, *Agostinelli:2002hh} as described in
Ref.~\cite{LHCb-PROC-2011-006}.

%% file: analysis.tex
\section{Analysis strategy}
\label{sec:analysis}

The analysis is based on fully reconstructed decays of charm mesons
in the following decay modes:
\DzToKpi, \DpToKpipi, \DstarpTopipDzToKmpip, \DsTophipi, and their charge conjugates.
The \DzToKpi sample contains the sum of the Cabibbo-favoured decays
\DzToKpi and the doubly Cabibbo-suppressed decays \DzbToKpi,
but for simplicity the combined sample is referred to by its dominant 
component.
The \DsTophipi sample comprises \DsToKKpi decays where the invariant mass 
of the $\Km\Kp$ pair is required to be within $\pm\SI{20}{\mevcc}$ of the nominal 
$\phi(1020)$ mass.
To allow cross-checks of the main results, the 
following decays are also reconstructed: \DpToKKpi, 
\DstarpTopipDzToKmpippimpip, and \DsToKKpi, where the \DsToKKpi sample here 
excludes candidates used in the \DsTophipi measurement. All decay modes are inclusive
with respect to final state radiation.

The cross-sections are measured in two-dimensional bins of \pT and \rapidity of 
the reconstructed mesons, where \pT and \rapidity are measured in the \pp 
centre-of-mass frame.  The bin widths are $0.5$ in \rapidity covering a range 
of \yrange,  \SI{1}{\gevc} in \pT for $0 < \pT < \SI{1}{\gevc}$, 
\SI{0.5}{\gevc} in \pT for $1 < \pT < \SI{3}{\gevc}$, and \SI{1}{\gevc} in \pT 
for $3 < \pT < \SI{15}{\gevc}$.

\subsection{Selection criteria}
\label{sec:ana_sel}

The selection of candidates is optimised independently for each decay mode.
For \DzToKpi decays the same criteria are used for 
both the \Dz and \Dstp cross-section measurements. All events are required to 
contain at least one reconstructed primary (\pp) interaction vertex~(PV). All 
final-state kaons and pions from the decays of \Dz, \Dp and \Dsp are required 
to be identified with high purity within the momentum and rapidity coverage of 
the \lhcb \pid system, \ie\ momentum between 3 and \SI{100}{\gevc} and pseudorapidity between 2 and 5.

The corresponding tracks must be of good quality and satisfy $\pT > 200$ or \SI{250}{\mevc}, depending on the decay mode.
At least one track must satisfy $\pT > \SI{800}{\mevc}$, while for three-body decays, one track has to satisfy $\pT>\SI{1000}{\mevc}$ and at least two tracks must have $\pT>\SI{400}{\mevc}$.
The lifetimes of the weakly decaying charm mesons are sufficiently long for the final-state particles 
to originate from a point away from the PV, and this characteristic is 
exploited by requiring that all final-state particles from these mesons are inconsistent with 
having originated from the PV\@.

When combining tracks to form \Dz, \Dp, and \Dsp meson candidates, requirements are made to 
ensure that the tracks are consistent with originating from a common decay 
vertex and that this vertex is significantly displaced from the PV\@.  
Additionally, the angle between the particle's momentum vector and the vector 
connecting the PV to the decay vertex of the \Dz (\Dp and \Dsp) candidate must 
not exceed $17(35)\,\si{\milli\radian}$.
Candidate \decay{\Dstp}{\Dz\pip} decays are formed by the combination of a \Dz candidate and 
a pion candidate, which are required to form a good quality vertex.
The \Dz candidates contained in the \Dstp sample are a subset of those used in the measurement of the \Dz cross-section.

\subsection{Selection efficiencies}
\label{sec:ana_eff}
The efficiencies for triggering, reconstructing and selecting signal decays 
are factorised into components that are measured in independent studies. 
These are the efficiency for decays to occur in the detector acceptance, for the final-state particles to be reconstructed, and for the decay to be selected.
To 
determine the efficiency of each of these components, the full event 
simulation is used, except for the \pid selection 
efficiencies, where a data-driven approach is adopted:
the efficiency with 
which pions and kaons are selected is measured using high-purity, independent calibration samples of 
pions and kaons from \DstarpTopipDzToKmpip decays identified without \pid 
requirements, but with otherwise tighter criteria. 
The efficiency in \pTy bins for each charm meson
decay mode is obtained with a weighting procedure 
to align the calibration and signal samples for the variables with respect to which the \pid selection efficiency varies.
These variables are the track momentum, track
pseudorapidity, and the number of hits in the scintillating-pad detector as a measure of the detector occupancy.
The 
signal distributions for this weighting are determined with the 
sPlot technique~\cite{Pivk:2004ty} with \lnipchisq as the discriminating variable, where \ipchisq is defined as the difference in $\chi^2$ 
of the PV reconstructed with and without the particle under consideration.

A correction factor is used to account for the difference between the tracking efficiencies measured in data and simulation
as described in Ref.~\cite{LHCb-DP-2013-002}.
This factor is computed in bins of track momentum and pseudorapidity and weighted to the kinematics of a given signal decay in the simulated sample to obtain a correction factor in each charm meson \pTy bin. 
This correction factor ranges from 0.98 to 1.16, depending on the decay mode. 

\vspace{-2mm}
\subsection{Determination of signal yields}
\label{sec:ana_yie}

The data contain a mixture of prompt signal decays, secondary charm mesons 
produced in decays of \bquark hadrons, and combinatorial background.
Secondary charm mesons will, in general, have a greater \IP with respect to the PV than prompt signal, and thus a greater 
value of \lnipchisq.
The number of prompt signal charm meson decays within each \pTy bin 
is determined with fits to the \lnipchisq distribution of the selected 
samples.
These fits are carried out in a signal window in the invariant mass of the candidates and background templates are obtained from regions outside the signal window.
Fits to the invariant mass distributions are used to constrain the level of combinatorial background in the subsequent fits to the \lnipchisq distributions.

In the case of the \Dz, \Dp, and \Dsp measurements, the signal window is 
defined as \SI{\pm 20}{\mevcc} around the known mass of the 
charm meson~\cite{PDG2014}, corresponding to approximately $2.5$ times the mass resolution.
Background samples are taken from two windows of width \SI{20}{\mevcc}, centred 
\SI{50}{\mevcc} below and \SI{50}{\mevcc} above the centre of the 
signal window.
For the \Dstp measurements, the signal window is defined in the 
distribution of the difference between the reconstructed \Dstp mass and the 
reconstructed \Dz mass, $\deltam = m(\Dstp) - m(\Dz)$, as \SI{\pm 3}{\mevcc} 
around the nominal \deltam value of \SI{145.43}{\mevcc}~\cite{PDG2014}.
The background sample is taken from the region \SI{4.5}{\mevcc} to 
\SI{9}{\mevcc} above the nominal \deltam value.

The number of combinatorial background candidates in the signal window of each decay mode is 
measured with binned extended maximum likelihood fits to either the 
mass or \deltam distribution, performed simultaneously across all \pTy 
bins for a given decay mode.
Prompt and secondary signals cannot be separated in mass or \deltam, so a single 
signal probability density function (PDF) is used to describe both components.
For the \Dz, \Dp, and \Dsp measurements the signal PDF is the sum 
of a Crystal Ball function~\cite{Skwarnicki:1986xj} and a Gaussian function, 
sharing a common mode but allowed to have different widths, whilst the 
combinatorial background is modelled as a first-order polynomial.
The signal PDF for the \Dstp measurement is the sum of three Gaussian functions with a common mean but different widths.
The combinatorial background component in \deltam is modelled as an 
empirically derived threshold function with an exponent $A$ and a turn-on parameter $\deltam_{0}$, 
fixed to be the nominal charged pion mass $\deltam_{0} = 
\SI{139.57}{\mevcc}$~\cite{PDG2014},
\begin{equation}
  g(\deltam; \deltam_{0}, A) = {(\deltam - \deltam_{0})}^{A}.
\end{equation}
Candidates entering the \deltam fit are required to be within the previously 
defined \Dz signal window.

Only candidates within the mass and \deltam signal windows are used in the \lnipchisq 
fits.
A Gaussian constraint is applied to the background yield in each \pTy bin, 
requiring it to be consistent with the integral of the background 
PDF in the signal window of the mass or \deltam fit.

Extended likelihood functions are constructed from one-dimensional 
PDFs in the \lnipchisq observable, with one set of signal and background PDFs for each 
\pTy bin.
The set of these PDFs is fitted simultaneously to the data in each 
\pTy bin, where all shape parameters other than the peak value of the prompt 
signal PDF are shared between bins.

The signal PDF in \lnipchisq is a bifurcated Gaussian with exponential tails, 
defined as
\begin{equation}
  f_{\text{S}}(\lnipchisq; \mu, \sigma, \epsilon, \rho_{L}, \rho_{R}) =
  \begin{cases}
    \exp\left(\frac{\rho_{L}^{2}}{2} + \rho_{L}\frac{\lnipchisq - \mu}{(1 - 
    \epsilon)\sigma}\right) & \lnipchisq < \mu - (\rho_{L}\sigma(1 - 
        \epsilon)), \\
    \exp\left(-\left(\frac{\lnipchisq - \mu}{\sqrt{2}\sigma(1 - 
    \epsilon)}\right)^{2}\right) & \mu - (\rho_{L}\sigma(1 - \epsilon)) 
          \leq \lnipchisq < \mu, \\
    \exp\left(-\left(\frac{\lnipchisq - \mu}{\sqrt{2}\sigma(1 + 
    \epsilon)}\right)^{2}\right) & \mu \leq \lnipchisq < \mu + (\rho_{R}\sigma(1 + 
          \epsilon)), \\
    \exp\left(\frac{\rho_{R}^{2}}{2} - \rho_{R}\frac{\lnipchisq - \mu}{(1 + 
    \epsilon)\sigma}\right) & \lnipchisq \geq \mu + (\rho_{R}\sigma(1 + 
        \epsilon)),
  \end{cases}
\end{equation}
where $\mu$ is the mode of the distribution, $\sigma$ is the average of the 
left and right Gaussian widths, $\epsilon$ is the asymmetry of the left and 
right Gaussian widths, and $\rho_{L(R)}$ is the exponent for the left (right) 
tail.
The PDF for secondary charm decays is a Gaussian function.

\begin{figure}
  \begin{subfigure}[b]{0.5\textwidth}
    \includegraphics[width=\textwidth]{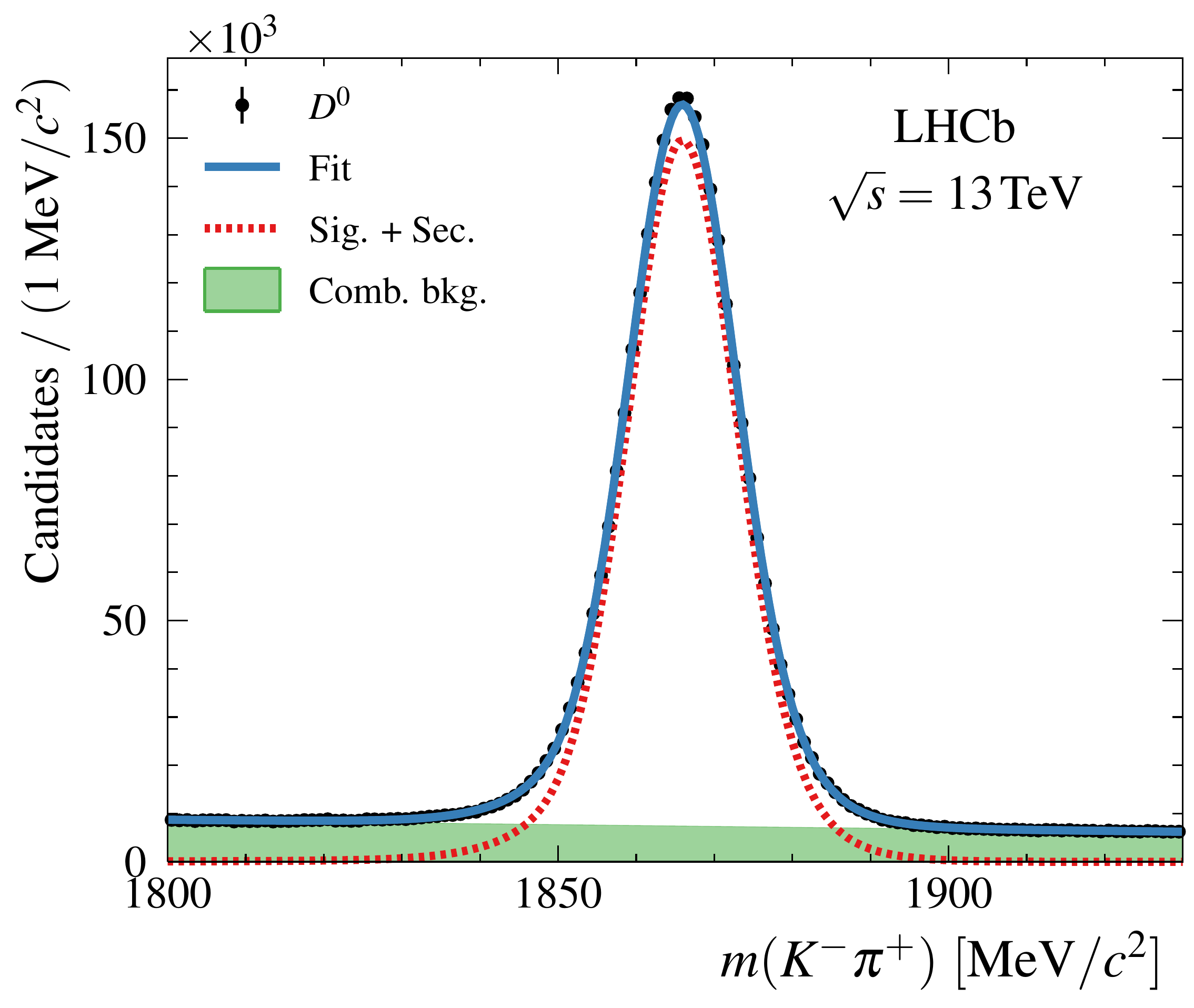}
  \end{subfigure}
  \begin{subfigure}[b]{0.5\textwidth}
    \includegraphics[width=\textwidth]{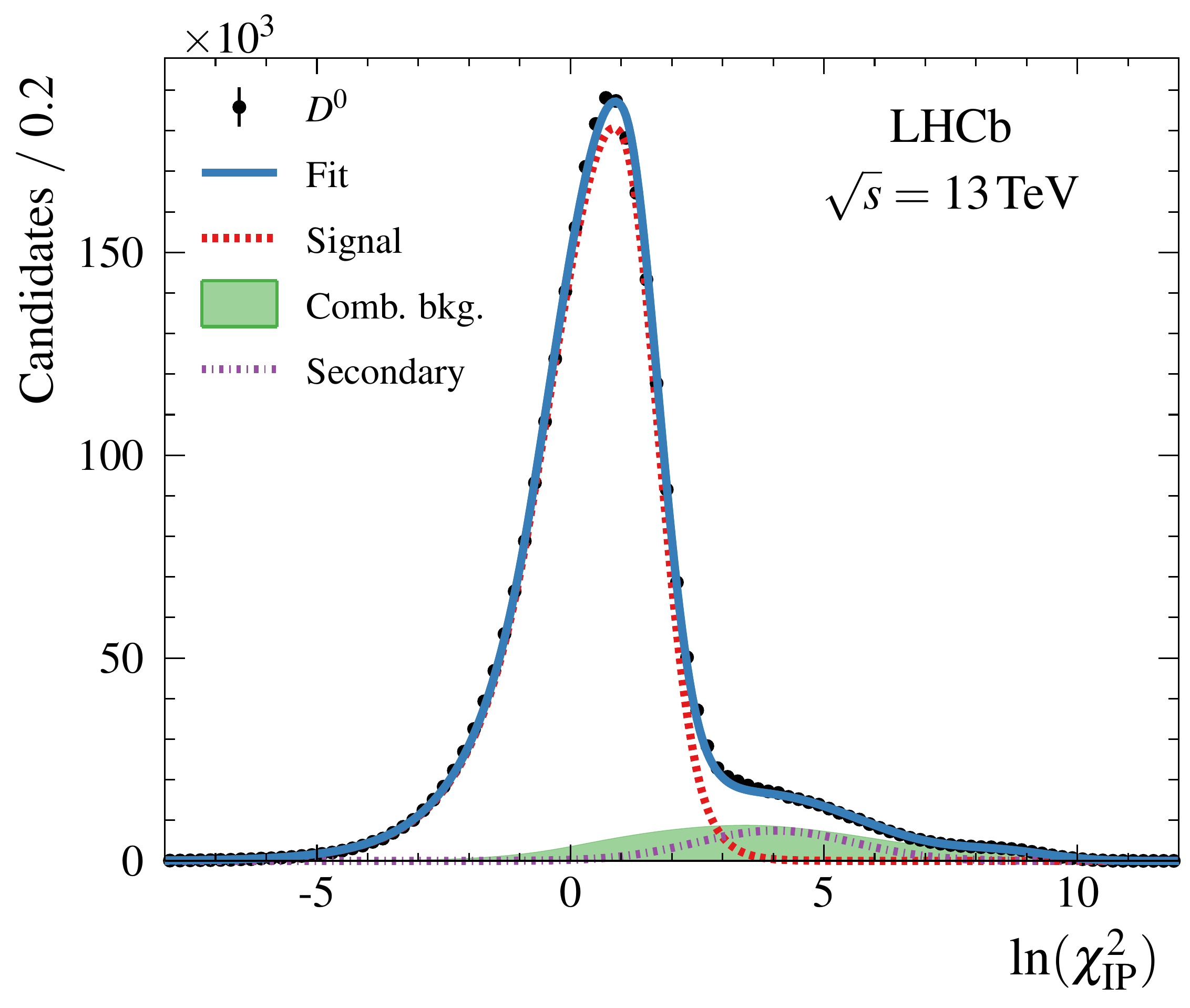}
  \end{subfigure}
  \caption{%
    Distributions for selected \DzToKpi candidates: (left) $\Km\pip$ invariant mass and (right) \lnipchisq for a mass window of $\pm\SI{20}{\mevcc}$ around the nominal \Dz mass.
    The sum of the simultaneous likelihood fits in each 
    \pTy bin is shown, with components as indicated in the legends.
  }
  \label{fig:analysis:fits:D0ToKpi}
\end{figure}

\begin{figure}
  \begin{subfigure}[b]{0.5\textwidth}
    \includegraphics[width=\textwidth]{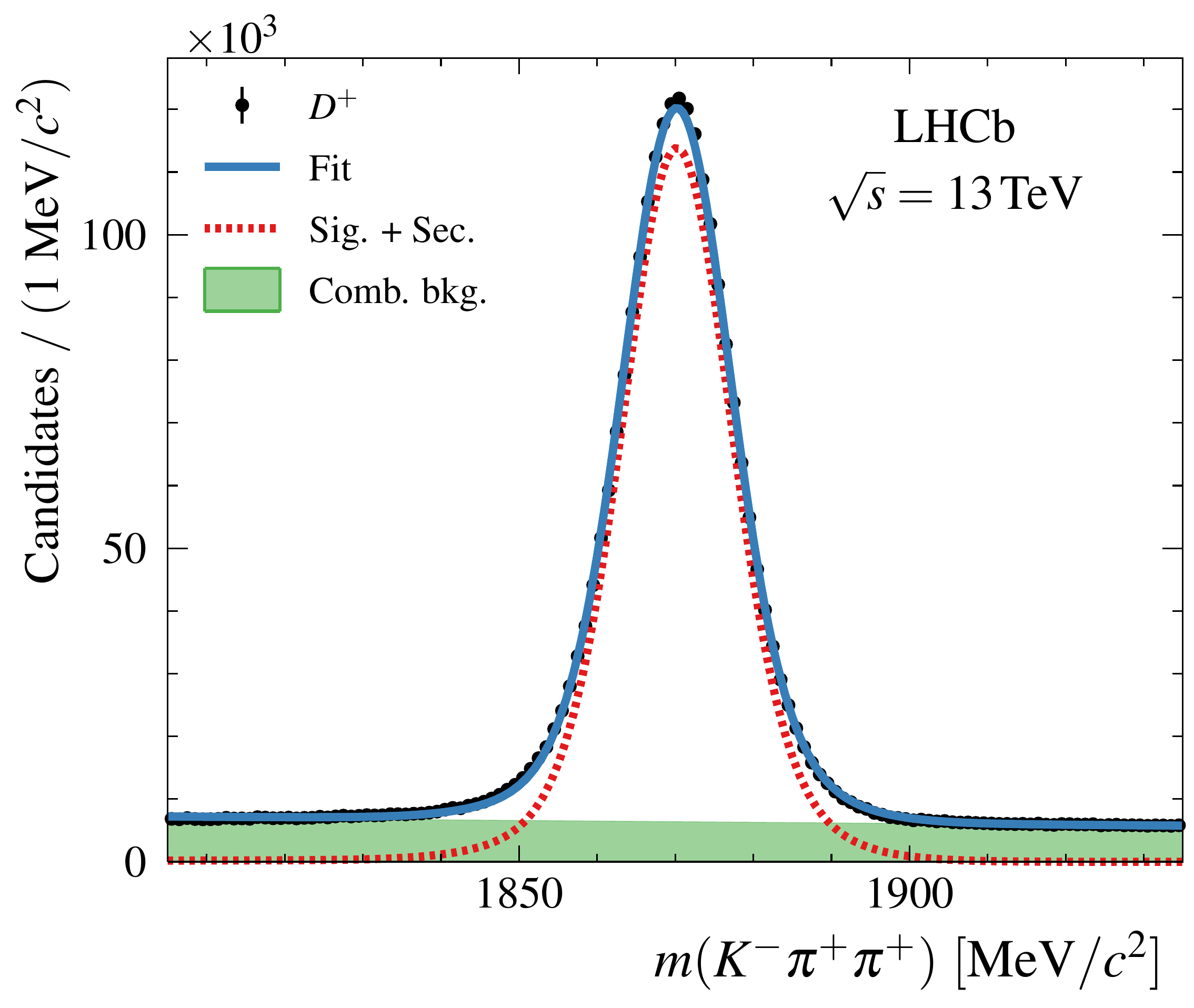}
  \end{subfigure}
  \begin{subfigure}[b]{0.5\textwidth}
    \includegraphics[width=\textwidth]{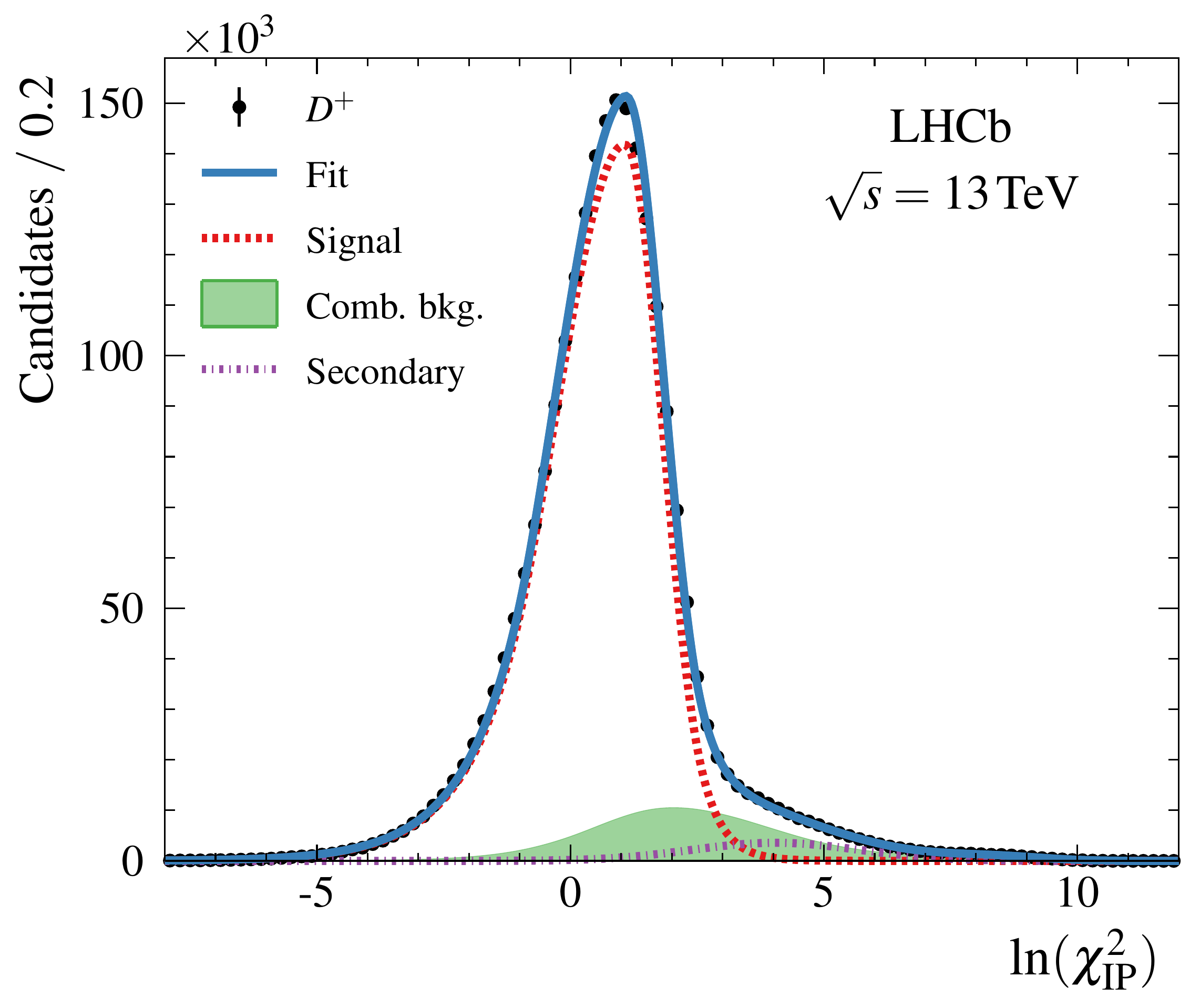}
  \end{subfigure}
  \caption{%
    Distributions for selected \DpToKpipi candidates: (left) $\Km\pip\pip$ invariant mass and (right) \lnipchisq for a mass window of $\pm\SI{20}{\mevcc}$ around the nominal \Dp mass.
    The sum of the simultaneous likelihood fits in each 
    \pTy bin is shown, with components as indicated in the legends.
  }
  \label{fig:analysis:fits:DpToKpipi}
\end{figure}

\begin{figure}
  \begin{subfigure}[b]{0.5\textwidth}
    \includegraphics[width=\textwidth]{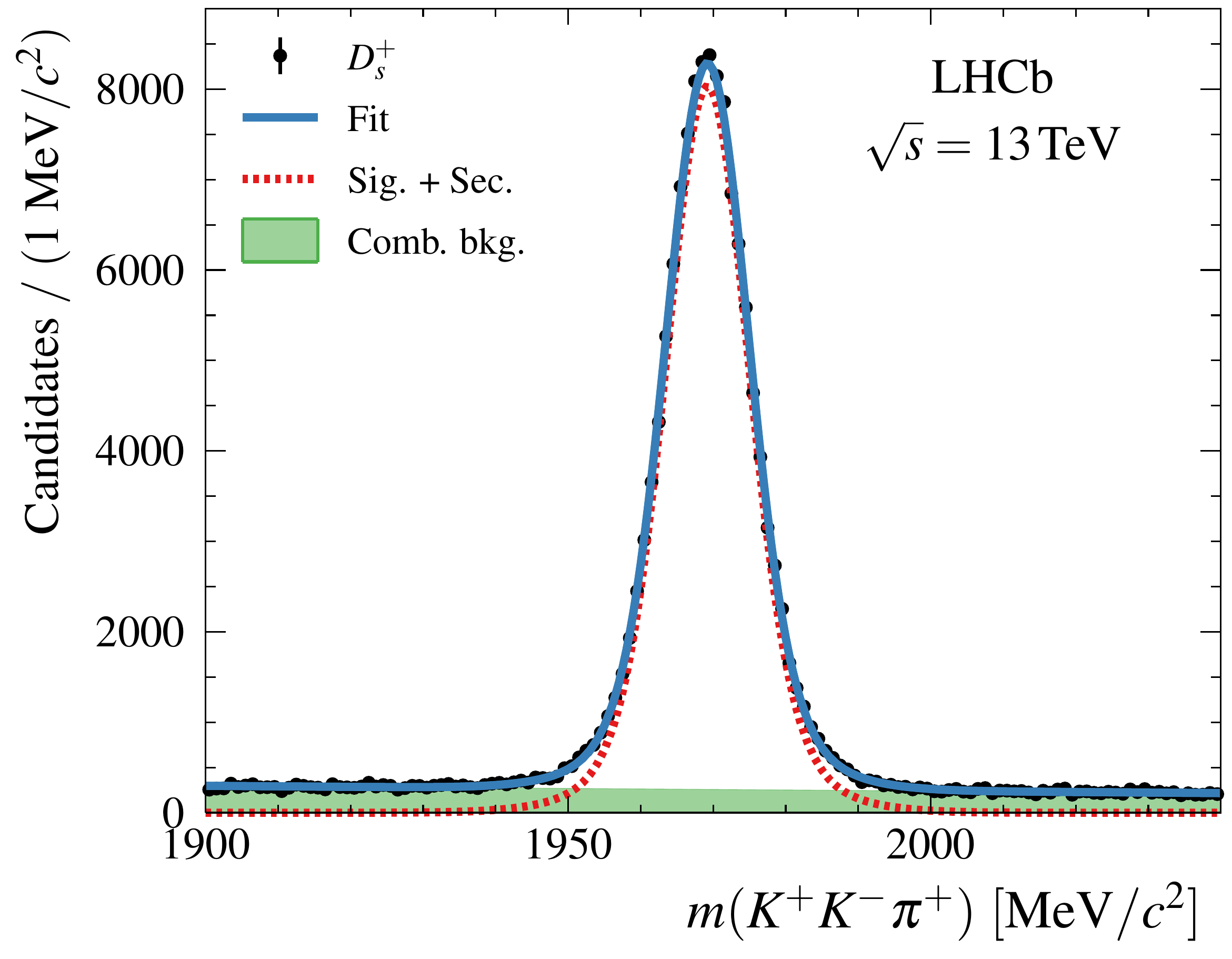}
  \end{subfigure}
  \begin{subfigure}[b]{0.5\textwidth}
    \includegraphics[width=\textwidth]{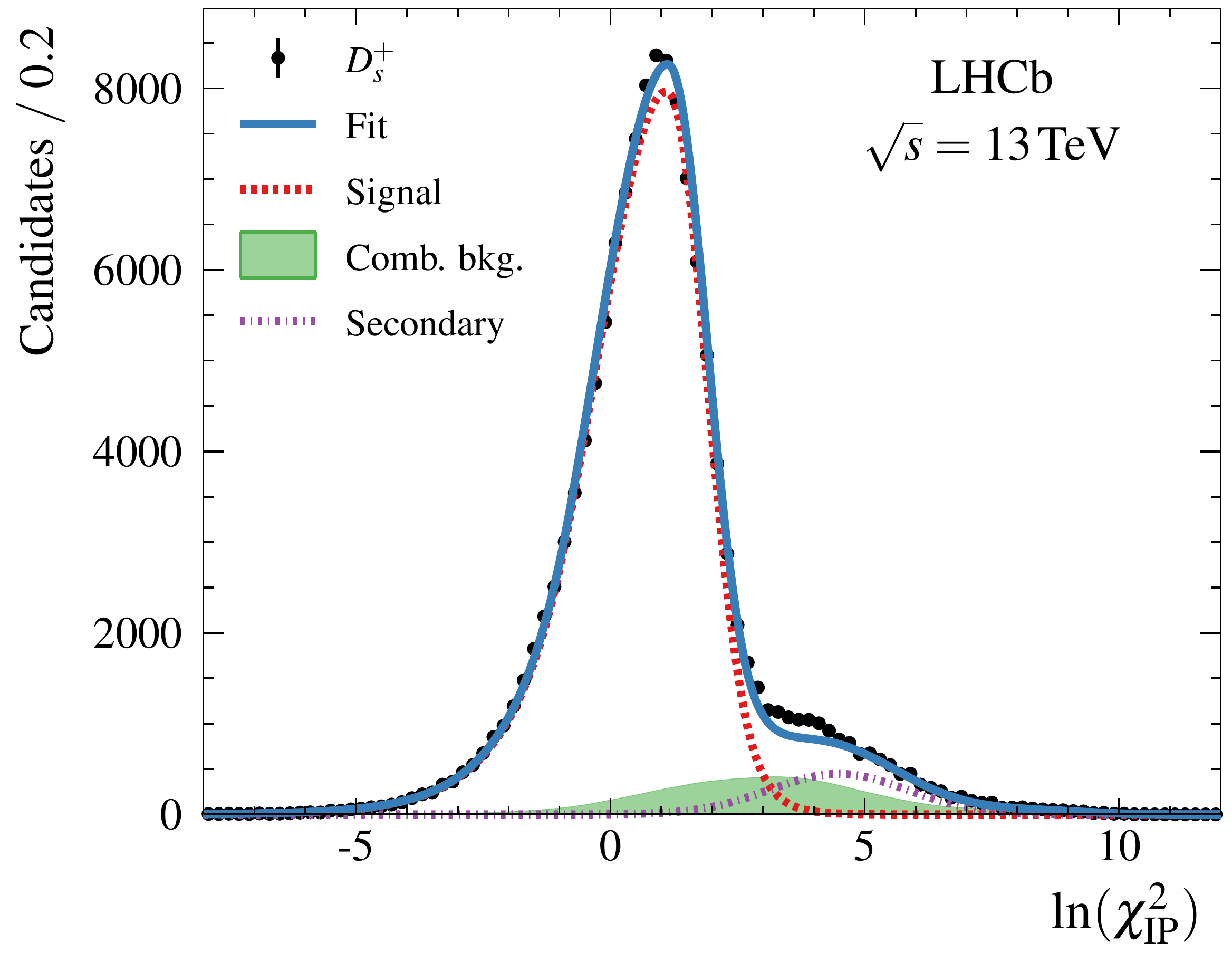}
  \end{subfigure}
  \caption{%
    Distributions for selected \DsTophipi candidates: (left) $\Kp\Km\pip$ invariant mass and (right) \lnipchisq for a mass window of $\pm\SI{20}{\mevcc}$ around the nominal \Dsp mass.
    The sum of the simultaneous likelihood fits in each 
    \pTy bin is shown, with components as indicated in the legends.
  }
  \label{fig:analysis:fits:DsTophipi}
\end{figure}

\begin{figure}
  \begin{subfigure}[b]{0.5\textwidth}
    \includegraphics[width=\textwidth]{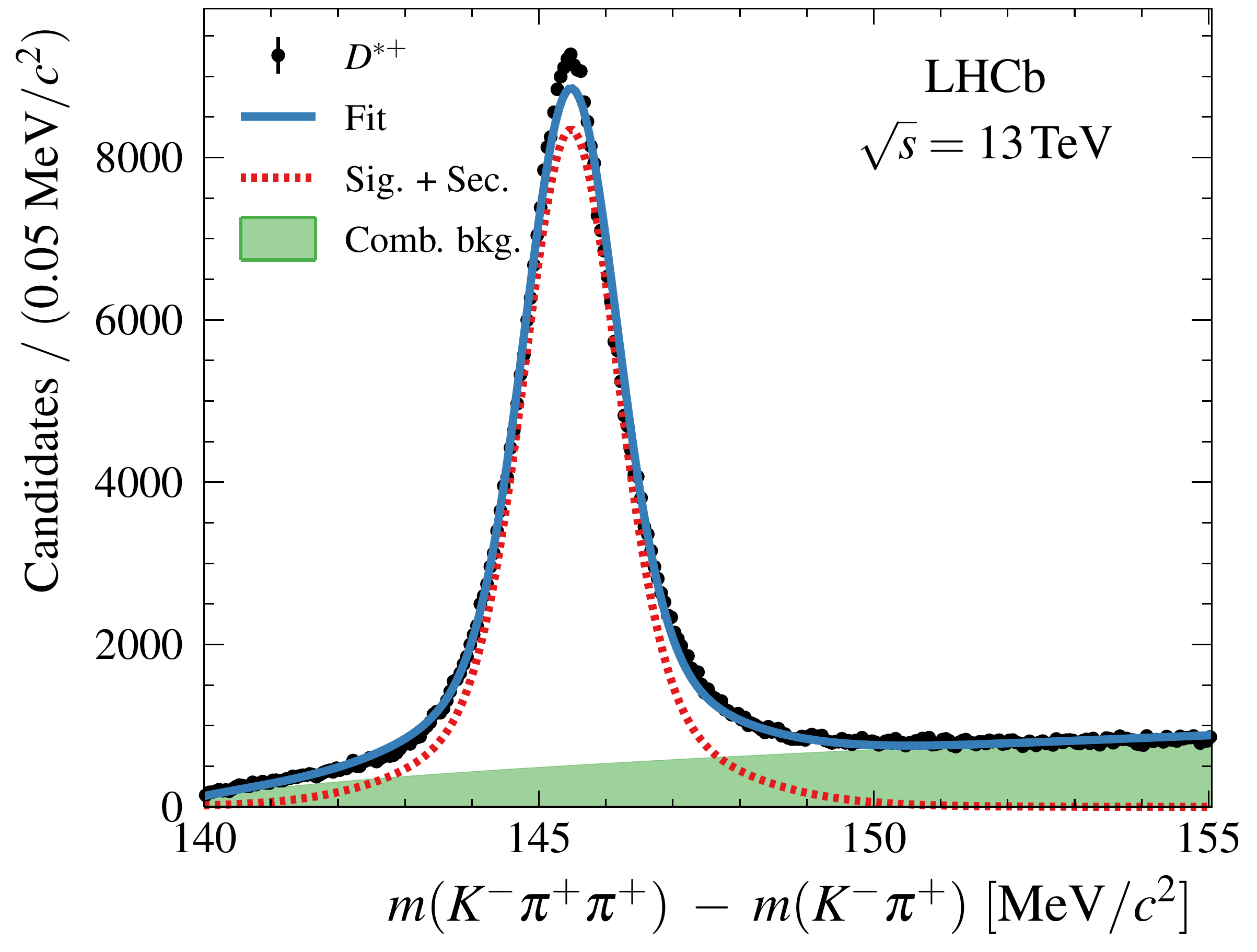}
  \end{subfigure}
  \begin{subfigure}[b]{0.5\textwidth}
    \includegraphics[width=\textwidth]{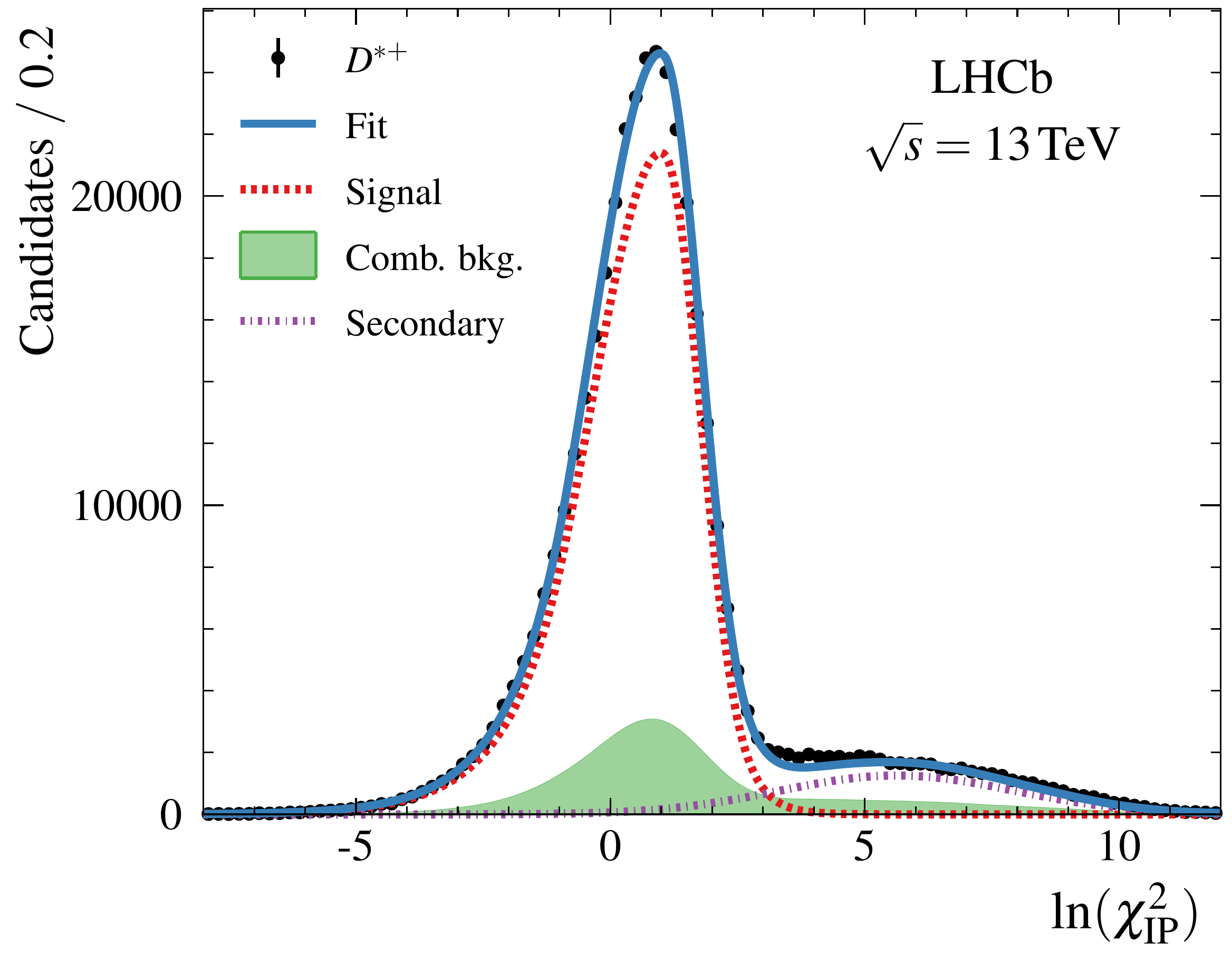}
  \end{subfigure}
  \caption{%
    Distributions for selected \DstToDzpi candidates, with \DzToKpi: (left) 
    $\deltam = m(\Dstp) - m(\Dz)$ for a mass window of $\pm\SI{20}{\mevcc}$ 
    around the nominal \Dz mass and (right) \lnipchisq with an additional mass 
    window of $\pm\SI{3}{\mevcc}$ around the nominal \Dstp{\kern 0.2em}-\,\Dz mass difference.
    The sum of the simultaneous likelihood fits in each \pTy bin is shown, with 
    components as indicated in the legends.
  }
  \label{fig:analysis:fits:DstToD0pi_D0ToKpi}
\end{figure}

The tail parameters $\rho_{L}$ and $\rho_{R}$ and the asymmetry parameter 
$\epsilon$ of the \lnipchisq prompt signal PDFs are fixed to values obtained 
from unbinned maximum likelihood fits to simulated signal samples.
All other parameters are determined in the fit.
The sums of the simultaneous likelihood fits in each \pTy bin are given in 
Figures~\ref{fig:analysis:fits:D0ToKpi}--\ref{fig:analysis:fits:DstToD0pi_D0ToKpi}.
The fits generally describe the data well.
The systematic uncertainty due to fit inaccuracies is determined as described in Sec.~\ref{sec:syst}.
The sums of the prompt signal yields, as determined by the fits, are given in 
Table~\ref{table:analysis:yields}.

\clearpage
\input{tables/prompt_signal_yields}

%% file: tables/prompt_signal_yields.tex
\begin{table}
  \caption{%
    Prompt signal yields in the fully selected dataset, summed over all 
    (\pT,\rapidity) bins in which a measurement is made.
  }
  \label{table:analysis:yields}
  \centering{%
    \begin{tabular}{lr}
      Hadron & Prompt signal yield              \\
      \midrule
      \Dz    & $(25.77 \pm 0.02) \times 10^{5}$ \\
      \Dp    & $(19.74 \pm 0.02) \times 10^{5}$ \\
      \Dsp   & $(11.32 \pm 0.04) \times 10^{4}$    \\
      \Dstp  & $(30.12 \pm 0.06) \times 10^{4}$    \\
    \end{tabular}
  }
\end{table}

%% file: measurements.tex
\section{Cross-section measurements}
\label{sec:measurements}

The signal yields are used to measure differential cross-sections in bins of \pT and $y$ in the range \ptrange and \yrange.
The differential cross-section for producing the charm meson species \Hc in bin 
$i$ is calculated from the relation
\begin{equation}
  \frac{\text{d}^2\sigma_i(\Hc)}{\text{d}\pT\,\text{d}y} =
    \frac{1}{\Delta\pT\Delta\rapidity} \cdot
    \frac{%
      N_i(\decay{\Hc}{f} + \text{c.c.})
    }{%
      \varepsilon_{i,\text{tot}}(\decay{\Hc}{f}) \bfrac(\decay{\Hc}{f}) \kappa\lum_{\text{int}}
    },
  \label{eq:xsec}
\end{equation}
where $\Delta\pT$ and $\Delta\rapidity$ are the widths in \pT and \rapidity of bin $i$, 
$N_{i}(\decay{\Hc}{f} + \text{c.c.})$ is the measured yield of prompt \Hc decays 
to the final state $f$ in bin $i$ plus the charge-conjugate decay, $\bfrac(\decay{\Hc}{f})$ is the known 
branching fraction of the decay, and $\varepsilon_{i,\text{tot}}(\decay{\Hc}{f})$ 
is the total efficiency for observing the signal decay in bin $i$.
The total integrated luminosity collected $\lum_{\text{int}}$ is \totlumi and $\kappa=10.7\%$ is the average fraction of events passed by the prescaled hardware trigger.
The integrated luminosity of the dataset is evaluated with a precision of $3.8\%$ from the number of visible $\proton\proton$ collisions and a constant of proportionality that is measured in a dedicated calibration dataset.
The absolute luminosity for the calibration dataset is determined from the beam currents, which are measured by LHC instruments, and the beam profiles and overlap integral, which are measured with a beam-gas imaging method~\cite{LHCb-PAPER-2014-047}.

The following branching fractions taken from Ref.~\cite{PDG2014} are used:
\mbox{$\bfrac(\DpToKmpippip) = (9.13 \pm 0.19)\%$}, \mbox{$\bfrac(\DstarpTopipDzToKmpip) = (2.63 \pm 0.04)\%$},~and \mbox{$\bfrac(\decay{\Dz}{\Kmp\pipm}) = (3.89 \pm 0.05)\%$}.
The last is the sum of Cabibbo-favoured and doubly Cabibbo-suppressed branching 
fractions, which agrees to better than 1\% with the HFAG result that accounts 
for the effects of final-state radiation~\cite{hfag}.
For the \Dsp measurement the fraction of \DspToKmKppip decays with a $\Km\Kp$ invariant mass in the range $1000<m_{\Km\Kp}<\SI{1040}{\mevcc}$ is taken as $(2.24 \pm 0.13)\%$~\cite{Alexander:2008aa}.

\begin{figure}[htb]
  \centering
  \vspace{1cm}
  \includegraphics[width=\textwidth]{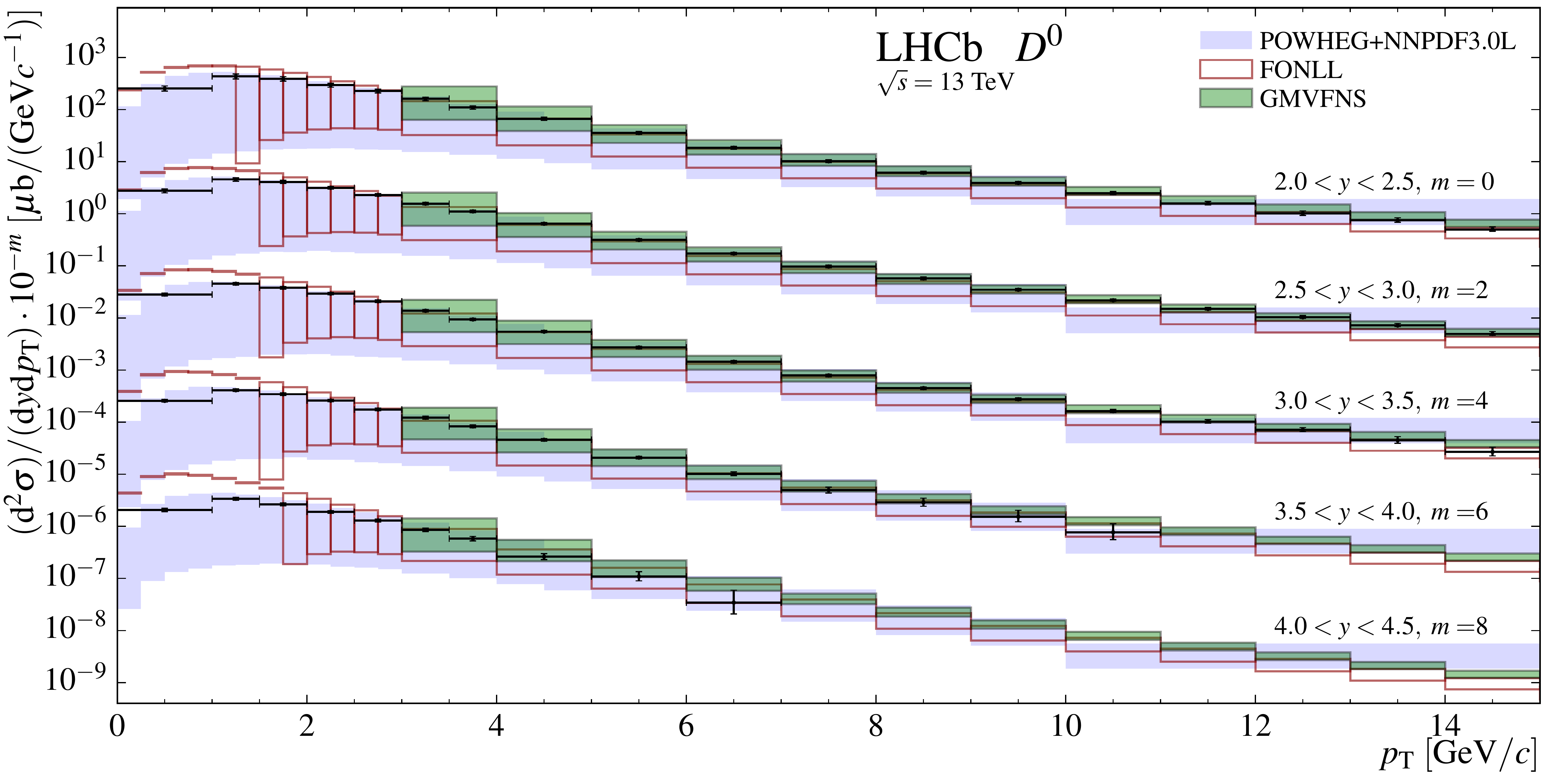}
  \vspace{1cm}
  \includegraphics[width=\textwidth]{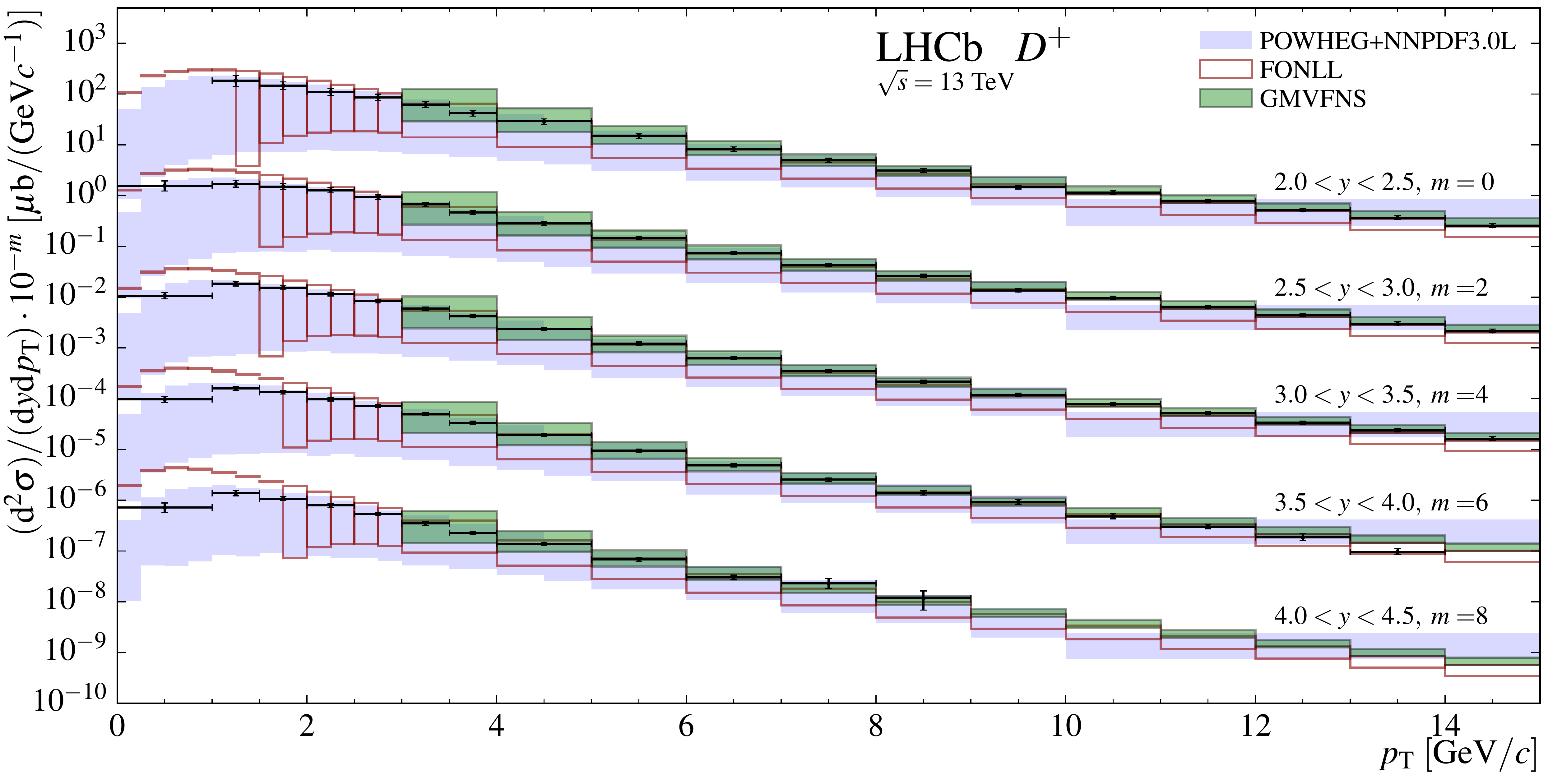}
  \caption{Measurements and predictions for the absolute prompt (top) \Dz, and (bottom) \Dp cross-sections at \comenergy.
           Each set of measurements and predictions in a given rapidity bin is 
           offset by a multiplicative factor $10^{-m}$, where the factor $m$ is 
           shown on the plots.
           The boxes indicate the $\pm1\sigma$ uncertainty band on the theory 
           predictions.
           In cases where this band spans more than two orders of magnitude only its upper edge is indicated.
  \label{fig:theory_absolute_D0_Dp}}
\end{figure}

\begin{figure}[htb]
  \centering
  \vspace{1cm}
  \includegraphics[width=\textwidth]{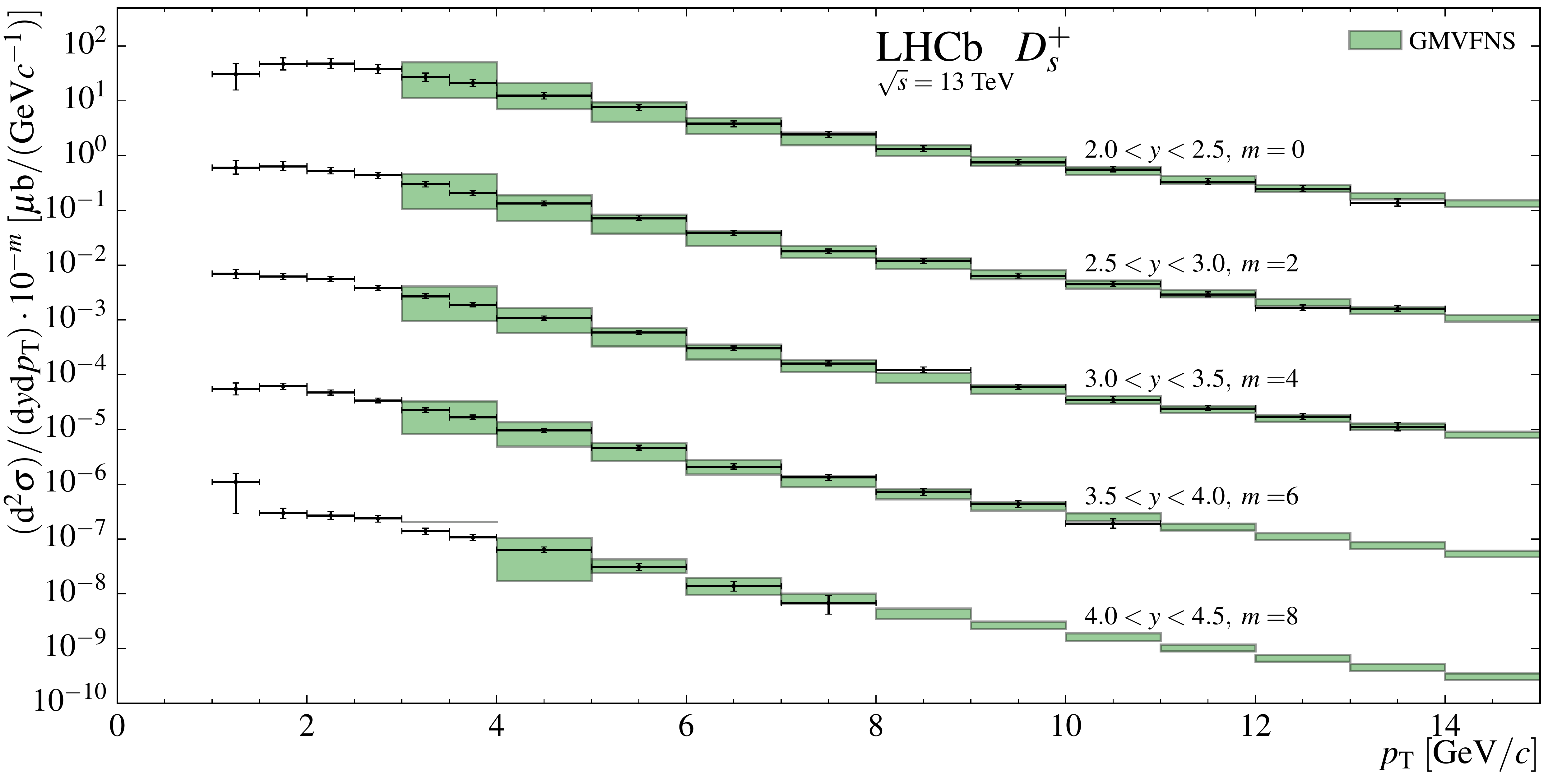}
  \vspace{1cm}
  \includegraphics[width=\textwidth]{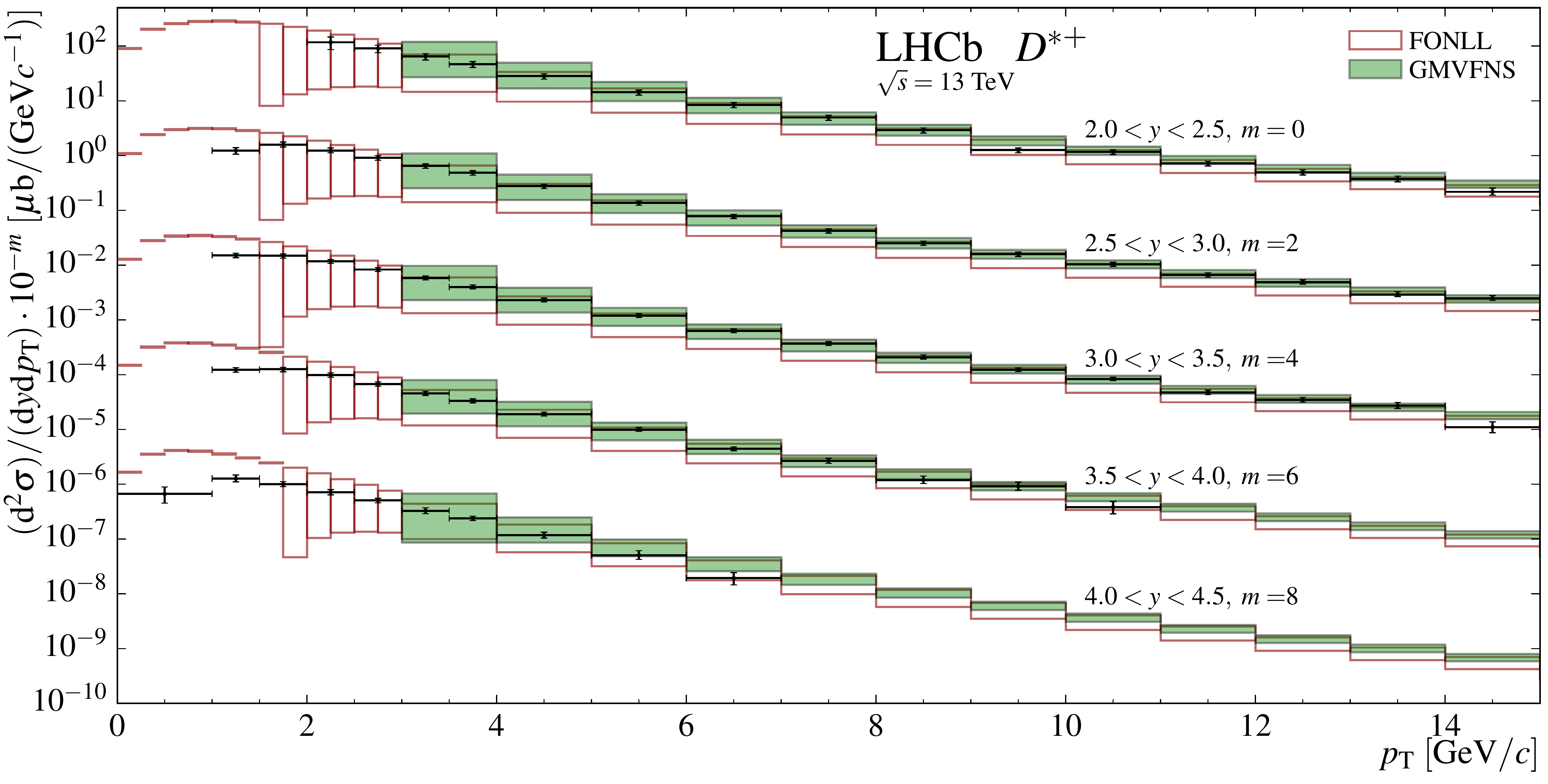}
  \caption{Measurements and predictions for the absolute prompt (top) \Dsp, and (bottom) \Dstarp cross-sections at \comenergy.
           Each set of measurements and predictions in a given rapidity bin is 
           offset by a multiplicative factor $10^{-m}$, where the factor $m$ is 
           shown on the plots.
           The boxes indicate the $\pm1\sigma$ uncertainty band on the theory 
           predictions.
           In cases where this band spans more than two orders of magnitude only its upper edge is indicated.
  \label{fig:theory_absolute_Dsp_Dstar}}
\end{figure}

The measured differential cross-sections are tabulated in Appendix~\ref{app:xsec}.
These results agree with the absolute 
cross-sections measured using the cross-check modes that are listed in 
Section~\ref{sec:analysis}.
Figures~\ref{fig:theory_absolute_D0_Dp} and~\ref{fig:theory_absolute_Dsp_Dstar} show the \Dz, \Dp, \Dsp, and \Dstarp cross-section 
measurements and predictions.
The systematic uncertainties are discussed in Sec.~\ref{sec:syst} and the theory contributions are provided in Refs.~\cite{Gauld:2015yia,Cacciari:2015fta,Kniehl:2012ti} and described in Sec.~\ref{sec:theory}.
\clearpage

\clearpage

%% file: systematics.tex
\section{Systematic uncertainties}
\label{sec:syst}

Several sources of systematic uncertainty are identified and evaluated separately for each decay mode and \pTy bin. 
In all cases, the dominant systematic uncertainties originate from the luminosity and the estimation of the tracking efficiencies, amounting to 3.9\% and 5--10\%, respectively. 
Uncertainties in the branching fractions give rise to systematic uncertainties between 1\% and 5\%, depending on the decay mode.
Systematic uncertainties are also evaluated to account for the modelling in the simulation, the \pid calibration procedure, and the PDF shapes used in the determination of the signal yields.
These sum in quadrature to around 5\%. 
Table~\ref{tab:sys:summary} lists the fractional systematic uncertainties for the different decay modes. 
Also given are the correlations of each uncertainty between different \pTy bins and between different decay modes.
The systematic uncertainties can be grouped into three categories:
those highly correlated between different decay modes and \pTy bins, those
that are only correlated between different bins but independent between
decay modes, and those that are independent between different decay 
modes and bins.

The systematic uncertainty on the luminosity is identical for all \pTy bins and decay modes.
The uncertainty on the tracking efficiency correction is a strongly correlated contribution. 
It includes a per-track uncertainty on the correction factor that originates from the finite size of the calibration sample, a 0.4\% uncertainty stemming from the weighting in different event multiplicity variables, and an additional 1.1\% (1.4\%) uncertainty for kaon (pion) tracks, due to uncertainties on the amount of material in the detector. The per-track uncertainties are propagated to obtain uncertainties on the correction factor in each \pTy bin of the charm meson, and are included as systematic uncertainties, resulting in a 5--10\% uncertainty on the measured cross-sections, depending on the decay mode.

The finite sizes of the simulated samples limit the statistical precision of the estimated efficiencies, leading to a systematic uncertainty on the measured cross-sections. As different simulated samples are used for each decay mode, the resulting uncertainty is independent between different decay modes and \pTy bins. 

Imperfect modelling of variables used in the selection can lead to differences between data and simulation, giving rise to a biased estimate of selection efficiencies. 
The effect is estimated by comparing the efficiencies when using modified selection criteria. 
The simulated sample is used to define a tighter requirement for each variable used in the selection, such that 50\% of the simulated events are accepted. 
The same requirement is then applied to the collision data sample, and the signal yield in this subset of the data is compared to the 50\% reduction expected from simulation.
The procedure is performed separately for each variable used in the selection. 
The sum of the individual differences, taking the correlations between the variables into account, is assigned as an uncertainty on the signal yield.
The corresponding uncertainty on the measured cross-sections is evaluated with Eq.~\ref{eq:xsec}.

The systematic uncertainties associated with the \pid calibration procedure result from the finite size of the calibration sample and binning effects of the weighting procedure.
The \pid efficiency in this calibration sample is determined in bins of track momentum, track pseudorapidity, and detector occupancy.
The statistical uncertainties of these efficiencies are propagated to obtain systematic uncertainties on the cross-sections.
In the weighting procedure, it is assumed that the \pid efficiencies for all candidates in a given bin are identical.
A systematic uncertainty is assigned to account for deviations from this approximation by sampling from kernel density estimates~\cite{Poluektov:2014rxa} created from the calibration samples, and recomputing the total \pid efficiency with the sampled data using a progressively finer binning.
The efficiency converges to a value that is offset from the value measured with the nominal binning.
This deviation is assigned as a systematic uncertainty on the \pid efficiency.
As all decay modes and \pTy bins use the same calibration data, this systematic uncertainty is highly correlated between different modes and bins.

Lastly, the systematic uncertainty on the signal yield extracted from the fits is dominated by the uncertainties on the choice of fit model.
This is evaluated by refitting the data with different sets of PDFs that are also compatible with the data, and assigning a systematic uncertainty based on the largest deviation in the prompt signal yield.

\input{./tables/syst_summary.tex}

%% file: tables/syst_summary.tex
% Systematic uncertainty overview table
\begin{table}
  \caption[Systematic uncertainties summary]{%
    Systematic uncertainties expressed as fractions of the cross-section 
    measurements, in percent. Uncertainties that are computed bin-by-bin are 
    expressed as ranges giving the minimum to maximum values.
    Ranges for the correlations between \pT-\rapidity bins and between modes 
    are also given, expressed in percent.
  }
  \label{tab:sys:summary}
  \centering
  \begin{tabular}{lcccccc}
                        & \multicolumn{4}{c}{Uncertainties (\%)} & \multicolumn{2}{c}{Correlations (\%)} \\
                        & \Dz                     & \Dp   & \Dsp   & \Dstp & Bins    & Modes   \\
    \midrule
    Luminosity          & \multicolumn{4}{c}{3.9} & 100   & 100   \\
    Tracking                  & \suDztracking            & \suDptracking      & \suDsptracking     & \suDstptracking     & 90--100   & 90--100 \\
    Branching fractions & 1.2                     & 2.1   & 5.8    & 1.5   & 100     & 0--95   \\
    Simulation sample size    & \suDzmcstat              & \suDpmcstat        & \suDspmcstat       & \suDstpmcstat       & 0      & 0    \\
    Simulation modelling      & \suDzmcagreement         & \suDpmcagreement   & \suDspmcagreement  & \suDstpmcagreement  & 0      & 0    \\
    \pid sample size    & 0--2                    & 0--1  & 0--2   & 0--1  & 0--100  & 0--100  \\
    \pid binning        & 0--44                   & 0--10 & 0--20  & 0-15  & 100     & 100     \\
    PDF shapes          & 1--6                    & 1--5  & 1--2   & 1--2  & -       & -       \\
  \end{tabular}
\end{table}

%% file: ratios.tex
\section{Production ratios and integrated cross-sections}
\label{sec:ratios}
\subsection{Production ratios}

\begin{figure}[p!]
  \centering
  \vspace{0.5cm}
  \includegraphics[width=0.49\textwidth]{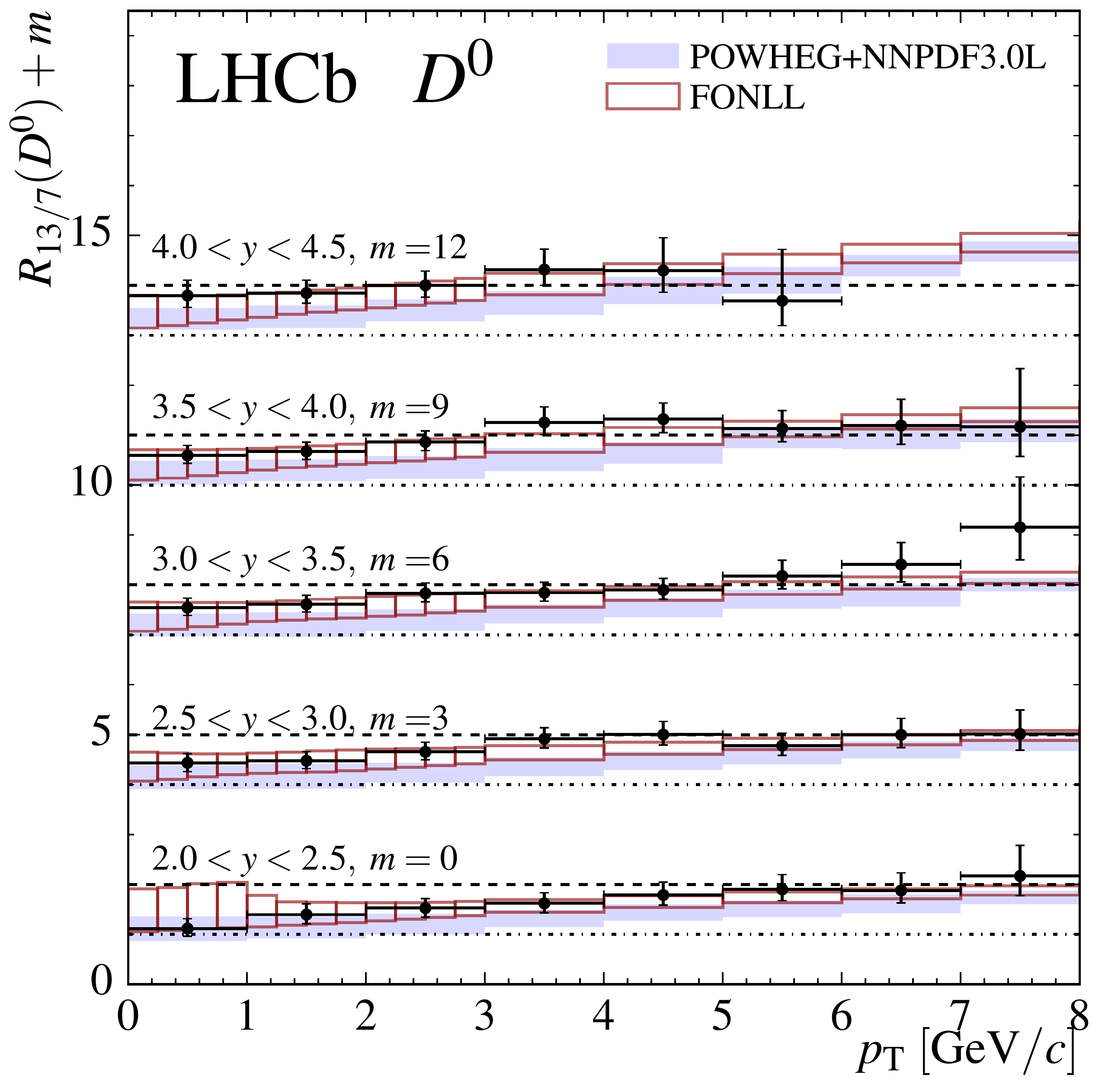}
  \includegraphics[width=0.49\textwidth]{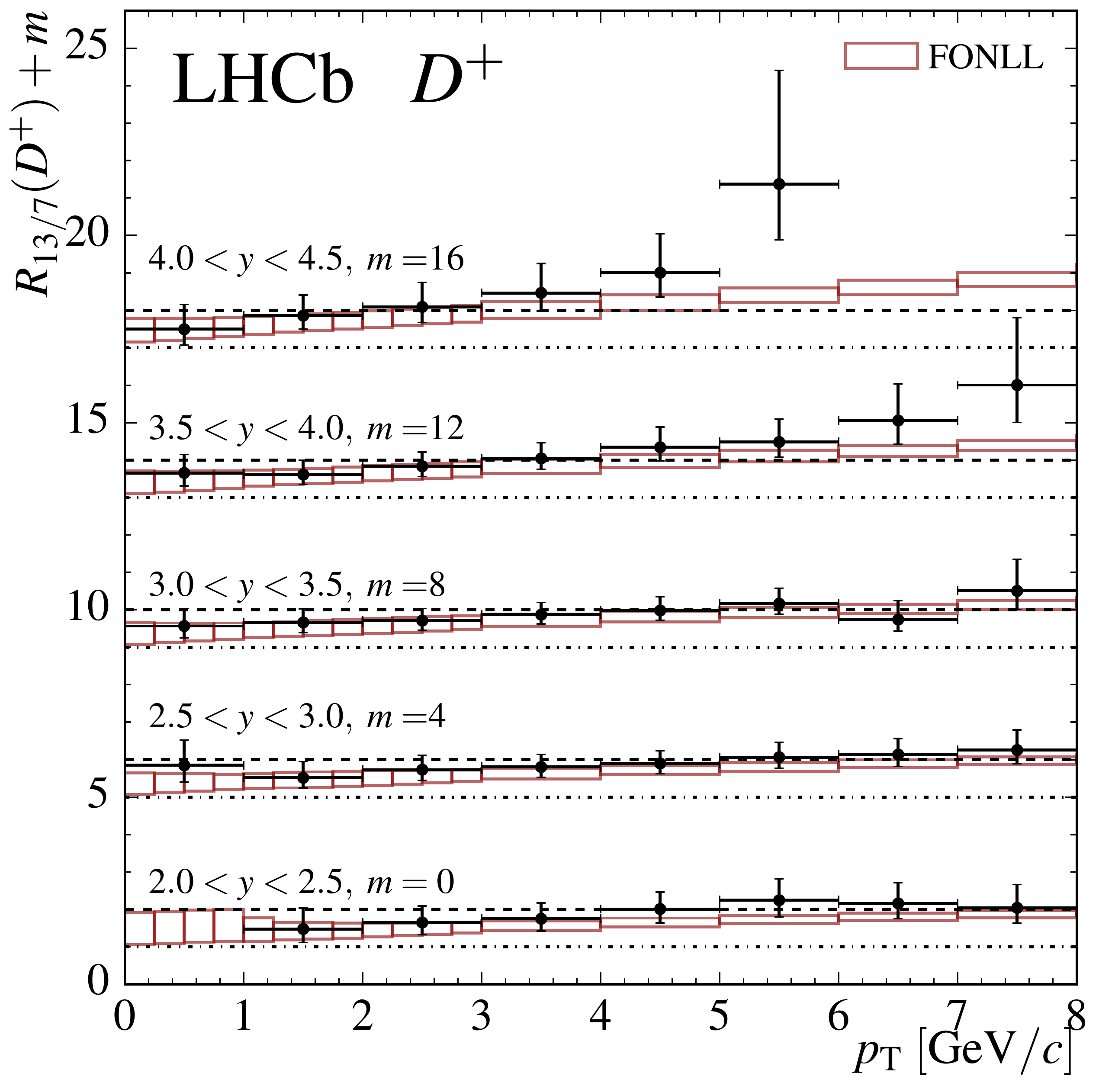}\\
  \vspace{0.5cm}
  \includegraphics[width=0.49\textwidth]{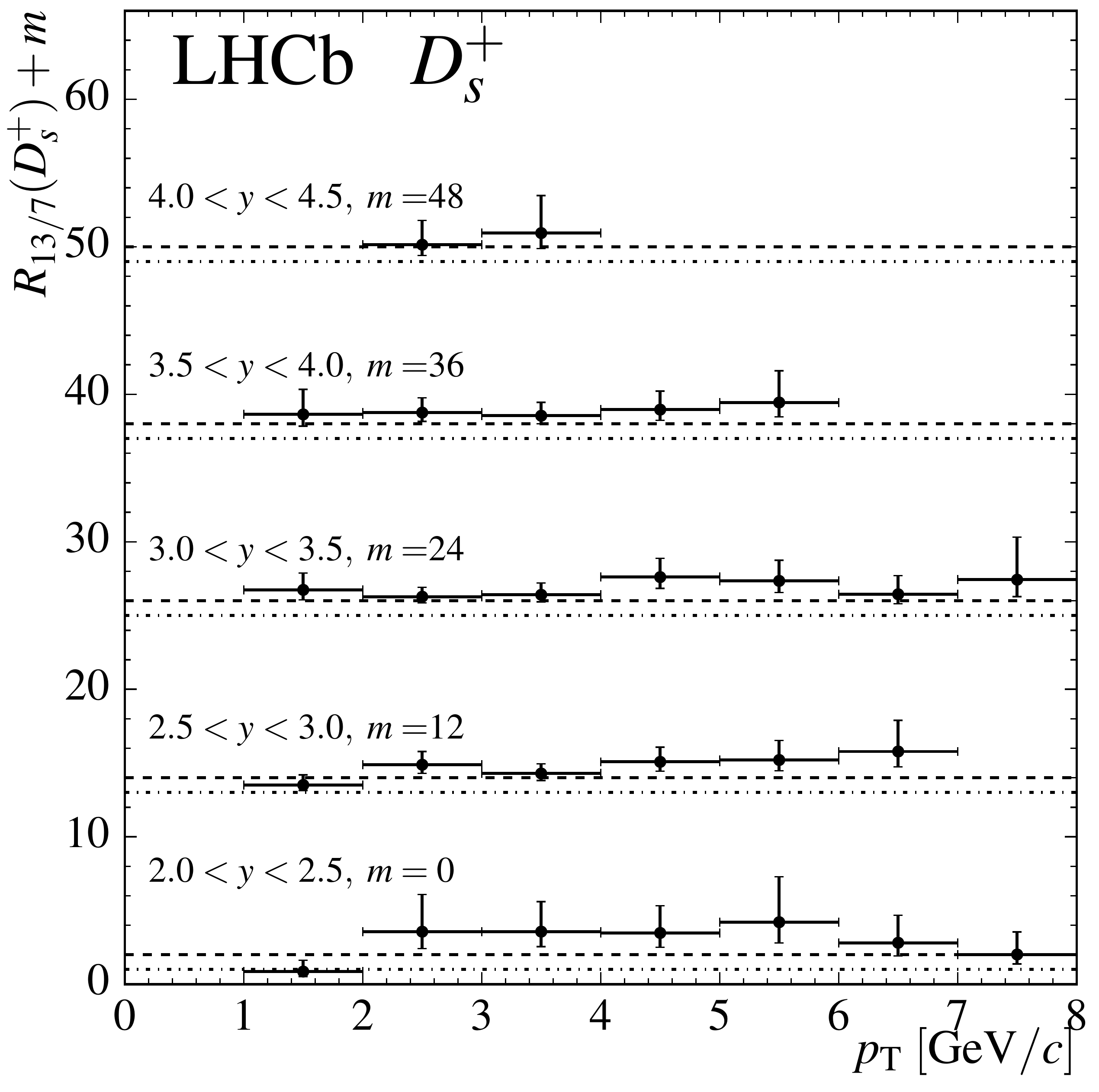}
  \includegraphics[width=0.49\textwidth]{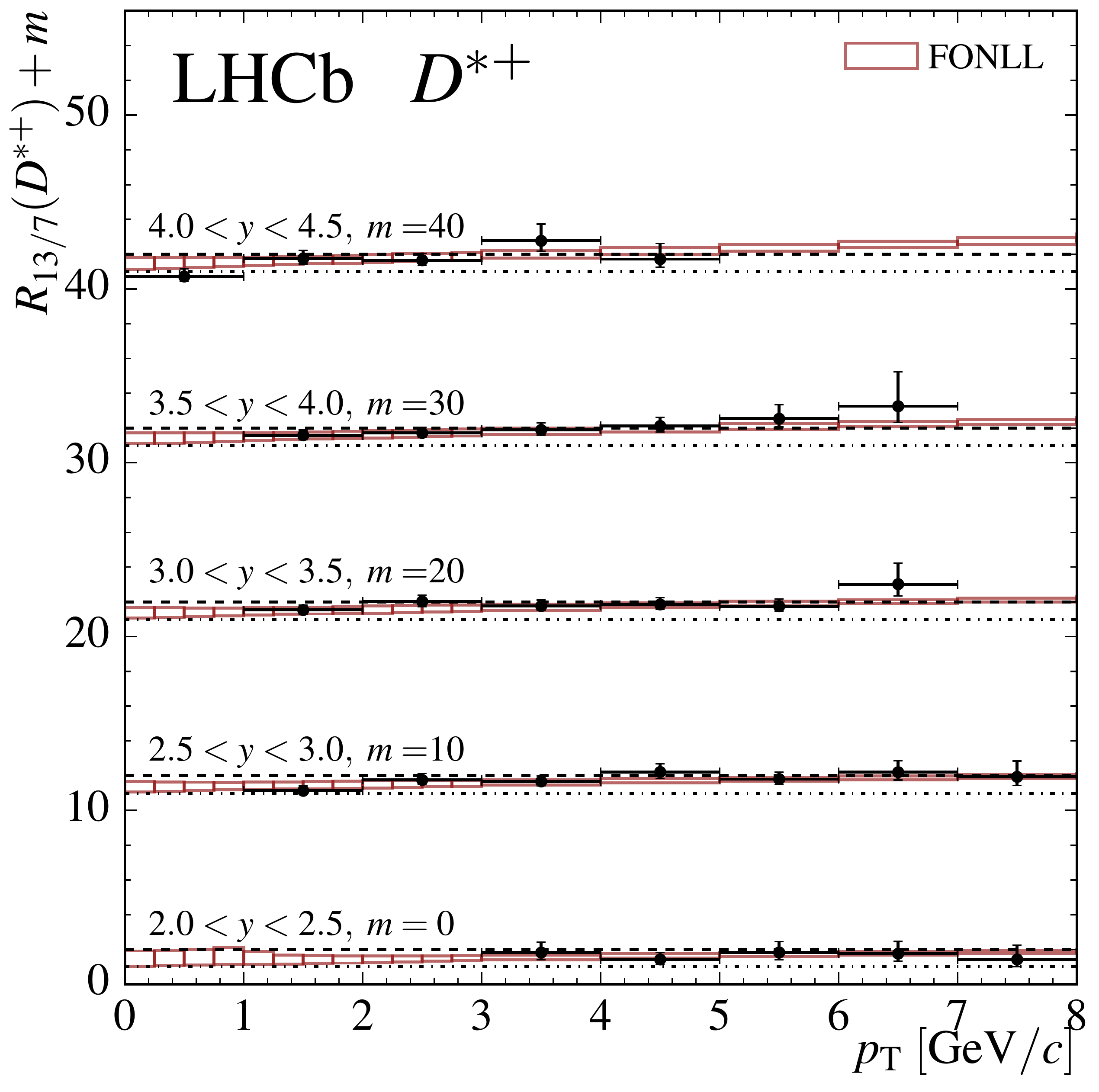}
    \caption{%
      Measurements and predictions of the prompt \Dz, \Dp, \Dsp, and \Dstarp 
      cross-section ratios.
      The dash-dotted lines indicate the unit ratio for each of the rapidity intervals and the dashed lines indicate a ratio of two.
      Each set of measurements and predictions in a given rapidity bin is 
      offset by an additive constant $m$, which is shown on the plot.
      No prediction is available for the \Dsp ratio.
      \label{fig:TheoryPredRatio}}
\end{figure}

The predicted ratios of prompt charm production cross-sections between different centre-of-mass energies are devoid of several theoretical  
uncertainties~\cite{Gauld:2015yia,Cacciari:2015fta,Kniehl:2012ti} that are inherent in the corresponding absolute cross-sections. Using the
present results obtained at $\sqrts=\SI{13}{\TeV}$ and the corresponding results from \lhcb data at $\sqrts=\SI{7}{\TeV}$~\cite{LHCb-PAPER-2012-041},  
these ratios, $R_{13/7}$, are measured for  \Dz, \Dp, \Dsp, and \Dstp mesons. 
The $\sqrts=\SI{13}{\TeV}$ measurements are rebinned to match the binning used in the $\sqrts=\SI{7}{\TeV}$ results and the 
production ratios are presented for  $0<\pT<\SI{8}{\gevc}$ and $2.0 < y < 4.5$ in Appendix~\ref{app:ratios}.
In the calculation of the uncertainties the branching fraction uncertainties cancel, and correlations of $30\%$ and $50\%$ are assumed for the uncertainties of luminosity and tracking, respectively.
All other uncertainties are assumed to be uncorrelated.
Figure~\ref{fig:TheoryPredRatio} shows
the measured ratios compared with predictions from theory calculations~\cite{Gauld:2015yia,Cacciari:2015fta,Kniehl:2012ti}.

%\vspace{2mm}
\subsection{Integrated cross-sections}
%\vspace{1mm}
Integrated production cross-sections, $\sigma(D)$, for each charm meson are computed as the 
sum of the per-bin measurements, where the uncertainty on the sum takes into 
account the correlations between bins discussed in Section~\ref{sec:syst}.
For \Dsp and \Dstp mesons, the kinematic region considered is $1 < \pT 
< \SI{8}{\gevc}$ and $2.0 < \rapidity < 4.5$ due to insufficient data below $\pT=\SI{1}{\gevc}$, while for \Dz and \Dp the same kinematic region as for the ratio measurements is used.
The upper limit is chosen to coincide with that of the measurements at $\sqrt{s}=\SI{7}{\TeV}$.

The \Dz and \Dstp cross-section results contain bins in which a measurement was not possible and which require a correction that is based on theory calculations.
A multiplicative correction factor is computed as the ratio between the predicted integrated cross-section within the considered kinematic region and the sum of all per-bin cross-section predictions for bins for which a measurement exists. 
This method is based on the \Rhorry predictions~\cite{Gauld:2015yia} for \Dz and the \Matteo predictions~\cite{Cacciari:2015fta} for \Dstp.
The uncertainty on the extrapolation factor is taken as the difference between 
factors computed using the upper and lower bounds of the theory predictions 
and is propagated to the integrated cross-sections as a 
systematic uncertainty.
Table~\ref{table:ratios:integrated} gives the integrated cross-sections for 
\Dz, \Dp, \Dsp, and \Dstp mesons.

The ratios of the cross-sections, with the \Dz and \Dp results re-evaluated in 
the kinematic region of the \Dsp and \Dstarp measurements, can be compared with 
the ratios of the cross-sections measured at $\ep\en$\/ colliders operating at 
a centre-of-mass energy close to the $\Upsilon(4S)$ 
resonance~\cite{Artuso:2004pj,Seuster:2005tr,Aubert:2002ue}.
A more precise comparison is made here by computing the ratios of 
cross-section-times-branching-fractions, \xsectimesbfrac, where the final 
states $f$ are the same between the \lhcb measurements and those made at 
$\ep\en$ experiments.
Differential ratios are shown in Fig.~\ref{fig:MesonRatio} and tabulated 
results and remaining figures are presented in Appendix~\ref{app:MesonRatio}.
They exhibit a \pT dependence that is consistent with heavier particles 
having a harder \pT spectrum.
% TODO compare our results with those from the B-factories.

The integrated charm cross-section, $\sigma(\ppToccbarX)$, is calculated as 
$\sigma(D)/(2f(\decay{\cquark}{D}))$ for each decay mode.
The term $f(\decay{\cquark}{D})$ is the quark to hadron transition 
probability, and the factor 2 accounts for the inclusion of charge conjugate 
states in the measurement.
The transition probabilities have been computed using measurements at $\ep\en$ 
colliders operating at a centre-of-mass energy close to the $\Upsilon(4S)$ 
resonance~\cite{Amsler:2008zzb:Frag} to be
\mbox{$f(\decay{\cquark}{\Dz}) = 0.565 \pm 0.032$},
\mbox{$f(\decay{\cquark}{\Dp}) = 0.246 \pm 0.020$},
\mbox{$f(\decay{\cquark}{\Dsp}) = 0.080 \pm 0.017$}, and
\mbox{$f(\decay{\cquark}{\Dstarp}) = 0.224 \pm 0.028$}.
The fragmentation fraction $f(\decay{\cquark}{\Dz})$ has an overlapping 
contribution
from $f(\decay{\cquark}{\Dstarp})$.

The combination of the \Dz and \Dp measurements, based on the {\sc Blue} 
method~\cite{Lyons:1988rp}, gives
\begin{equation*}
  \sigma(\ppToccbarX)_{\pT\,<\,8\,\si{\gevc},\,2.0\,<\,y\,<\,4.5} = \ccbarXsec,
\end{equation*}
where the uncertainties are due to statistical, systematic and fragmentation 
fraction uncertainties, respectively.
A comparison with predictions is given in~Fig.~\ref{fig:TotalCrossSec}.
The same figure also shows a comparison of $\sigma(\ppToccbarX)$ for $1 < \pT < \SI{8}{\gevc}$ based on the measurements of all four mesons.
Ratios of the integrated cross-section-time-branching-fraction measurements are 
given in Table~\ref{table:ratios:integrated_ratios}.

\input{tables/integrated_cross_sections.tex}

\input{tables/integrated_cross_section_ratios.tex}

\begin{figure}[p!]
  \centering
  \includegraphics[width=0.70\textwidth]{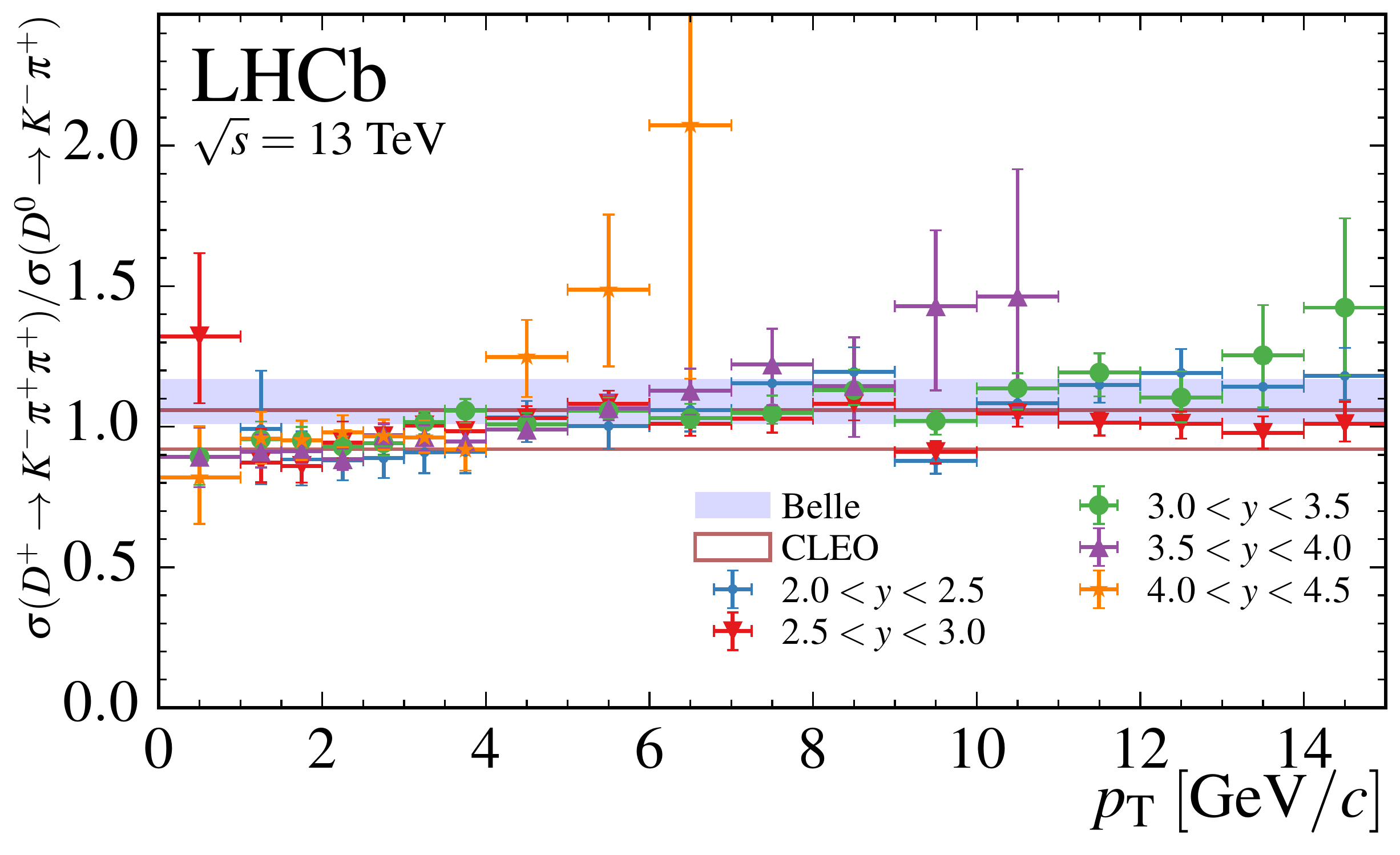}
  \includegraphics[width=0.70\textwidth]{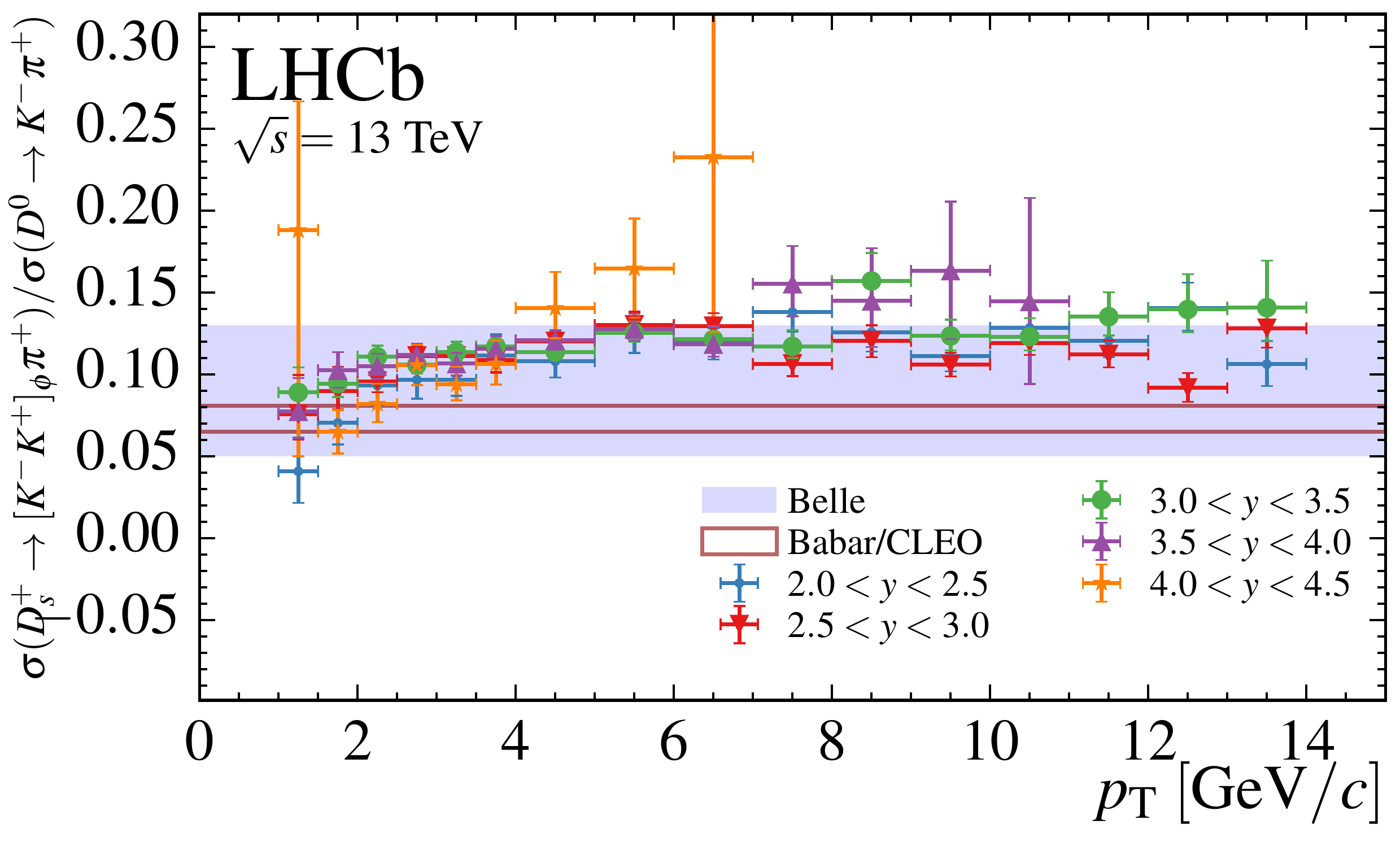}
  \includegraphics[width=0.70\textwidth]{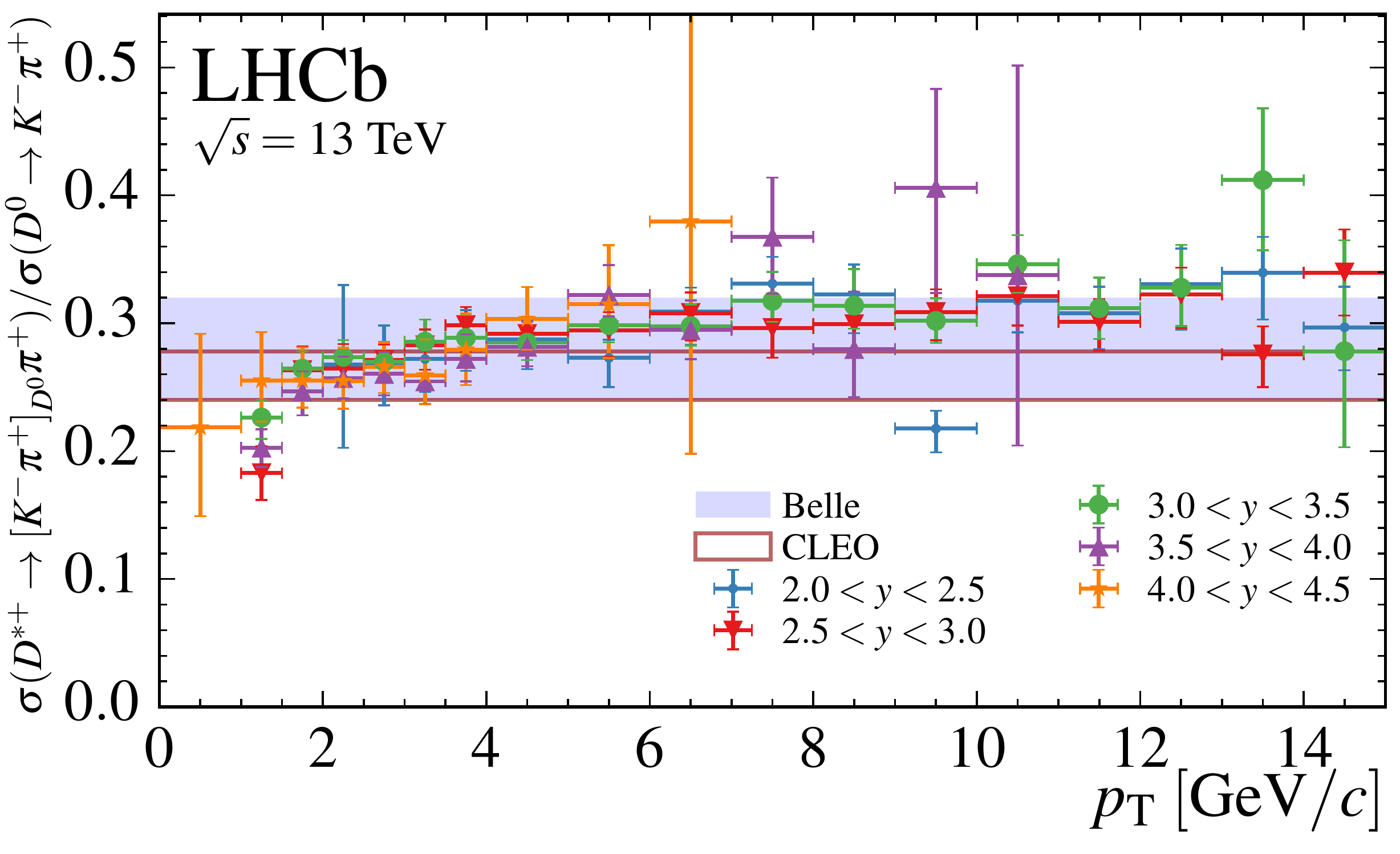}
    \caption{%
      Ratios of cross-section-times-branching-fraction measurements of (top) 
      \Dp, (middle) \Dsp, and (bottom) \Dstarp mesons with respect to the \Dz 
      measurements.
      The bands indicate the corresponding ratios computed using measurements 
      from $\ep\en$ collider 
      experiments~\cite{Artuso:2004pj,Seuster:2005tr,Aubert:2002ue}.
      The ratios are given as a function of \pT and different symbols indicate different ranges in \rapidity.
      The notation $\sigma(\decay{D}{f})$ is shorthand for \xsectimesbfrac.
      \label{fig:MesonRatio}}
\end{figure}

%% file: tables/integrated_cross_sections.tex
\begin{table}[tb]
  \caption[Integrated charm hadron cross-sections]{%
    Prompt charm production cross-sections in the kinematic ranges given.
    The computation of the extrapolation factors is described in the text.
    The first uncertainty on the cross-section is statistical, and the second 
    is systematic and includes the contribution from the extrapolation factor.
    No extrapolation factor is given for $\Dp_{(\squark)}$ as a measurement is available
    in every bin of the integrated phase space.
  }
  \label{table:ratios:integrated}
  \input{./tables/integrated/integrated_crosssections.tex}
\end{table}

%% file: tables/integrated/integrated_crosssections.tex
\begin{adjustbox}{center}
\begin{tabular}{ccccr}
&&& Extrapolation factor & Cross-section~(\si{\micro\barn}) \\
\midrule
\Dz   & $0 < \pT < \SI{8}{\GeVc}$ & $2 < \rapidity < 4.5$ & $1.0014 \pm 0.0024$  & $2709 \pm \phantom{1}2  \pm 165$\\
\Dp   & $0 < \pT < \SI{8}{\GeVc}$ & $2 < \rapidity < 4.5$ & $1.049 \pm 0.031$  & $1102 \pm \phantom{1}5  \pm 111$\\
\midrule
\Dz   & $1 < \pT < \SI{8}{\GeVc}$ & $2 < \rapidity < 4.5$ & $1.0018 \pm 0.0025$  & $2072 \pm \phantom{1}2  \pm 124$\\
\Dp   & $1 < \pT < \SI{8}{\GeVc}$ & $2 < \rapidity < 4.5$ & ---  & $834 \pm \phantom{1}2  \pm \phantom{1}78$\\
\Dsp   & $1 < \pT < \SI{8}{\GeVc}$ & $2 < \rapidity < 4.5$ & ---  & $353 \pm \phantom{1}9  \pm \phantom{1}76$\\
\Dstp   & $1 < \pT < \SI{8}{\GeVc}$ & $2 < \rapidity < 4.5$ & $1.102 \pm 0.081$  & $784 \pm \phantom{1}4  \pm \phantom{1}87$\\
\midrule
\Dz   & $0 < \pT < \SI{8}{\GeVc}$ & $2.5 < \rapidity < 4$ & ---  & $1720 \pm \phantom{1}1  \pm \phantom{1}98$\\
\Dp   & $0 < \pT < \SI{8}{\GeVc}$ & $2.5 < \rapidity < 4$ & ---  & $706 \pm \phantom{1}4  \pm \phantom{1}66$\\
\midrule
\Dz   & $1 < \pT < \SI{8}{\GeVc}$ & $2.5 < \rapidity < 4$ & ---  & $1313 \pm \phantom{1}1  \pm \phantom{1}73$\\
\Dp   & $1 < \pT < \SI{8}{\GeVc}$ & $2.5 < \rapidity < 4$ & ---  & $527 \pm \phantom{1}1  \pm\phantom{1} 45$\\
\Dsp   & $1 < \pT < \SI{8}{\GeVc}$ & $2.5 < \rapidity < 4$ & ---  & $227 \pm \phantom{1}2  \pm \phantom{1}24$\\
\Dstp   & $1 < \pT < \SI{8}{\GeVc}$ & $2.5 < \rapidity < 4$ & ---  & $493 \pm \phantom{1}2  \pm \phantom{1}41$\\
\end{tabular}
\end{adjustbox}

%% file: tables/integrated_cross_section_ratios.tex
\begin{table}[tb]
  \renewcommand{\arraystretch}{1.3}
  \caption[Ratios of integrated charm hadron cross-section-times-branching-fractions]{%
    Ratios of integrated cross-section-times-branching-fraction measurements in 
    the kinematic range $1 < \pT < \SI{8}{\gevc}$ and $2 < \rapidity < 4.5$. 
    The first uncertainty on the ratio is statistical and the second 
    is systematic. The notation $\sigma(\decay{D}{f})$ is shorthand for \xsectimesbfrac.
  }
  \label{table:ratios:integrated_ratios}
  \begin{adjustbox}{center}
  \begin{tabular}{lc}
    Quantity                                & Measurement \\
    \midrule
    $\sigma(\DpToKpipi)/\sigma(\DzToKpi)$                                                       & \DpDzint    \\
    $\sigma(\Dsp \to [K^{-}K^{+}]_{\phi}\pi^{+})/\sigma(\DzToKpi)$                              & \DspDzint    \\
    $\sigma(\Dstp \to [K^{-}\pi^{+}]_{\Dz}\pi^{+})/\sigma(\DzToKpi)$                            & \DstpDzint   \\
    \midrule                                                                                      
    $\sigma(\Dsp \to [K^{-}K^{+}]_{\phi}\pi^{+})/\sigma(\DpToKpipi)$                            & \DspDpint    \\
    $\sigma(\Dstp \to [K^{-}\pi^{+}]_{\Dz}\pi^{+})/\sigma(\DpToKpipi)$                          & \DstpDpint   \\
    \midrule                                                                                      
    $\sigma(\Dsp \to [K^{-}K^{+}]_{\phi}\pi^{+})/\sigma(\Dstp \to [K^{-}\pi^{+}]_{\Dz}\pi^{+})$ & \DspDstpint   \\
  \end{tabular}
  \end{adjustbox}
\end{table}

%% file: theory.tex
\section{Comparison to theory}
\label{sec:theory}

Theoretical calculations for charm meson production cross-sections in
\pp collisions at \comenergy have been provided in Refs.~\cite{Gauld:2015yia} (\Rhorry), \cite{Cacciari:2015fta} (\Matteo), and \cite{Kniehl:2012ti} (\Hubert).
All three sets of calculations are performed at NLO precision, and each
includes estimates of theoretical uncertainties due to the renormalisation
and factorisation scales.
The theoretical uncertainties provided with the \Matteo and \Rhorry predictions
also include contributions due to uncertainties in the effective charm quark mass and the 
parton distribution functions.

The \Matteo predictions are provided in the form of \Dz, \Dp,
and \Dstp production cross-sections for \pp collisions at \comenergy for each
bin in a subdivision of the phase space, \mbox{$\pT < \SI{30}{\gevc}$} and
\mbox{$2.0 < y < 4.5$}.
Ratios of these cross-sections to those computed for \pp collisions at
\SI{7}{\TeV}\ are also supplied.
The calculations use the NNPDF3.0 NLO~\cite{Ball:2014uwa} parton densities.
These \Matteo calculations of the meson differential production cross-sections 
assume $f(\decay{\cquark}{\PD}) = 1$ and are multiplied by the
transition probabilities measured at $\ep\en$\/ colliders for comparison to the current measurements.
No dedicated \Matteo cross-section calculation for \Dsp production is available.

%\clearpage
%
\begin{figure}[h!]
  \centering
  \includegraphics[width=0.49\textwidth]{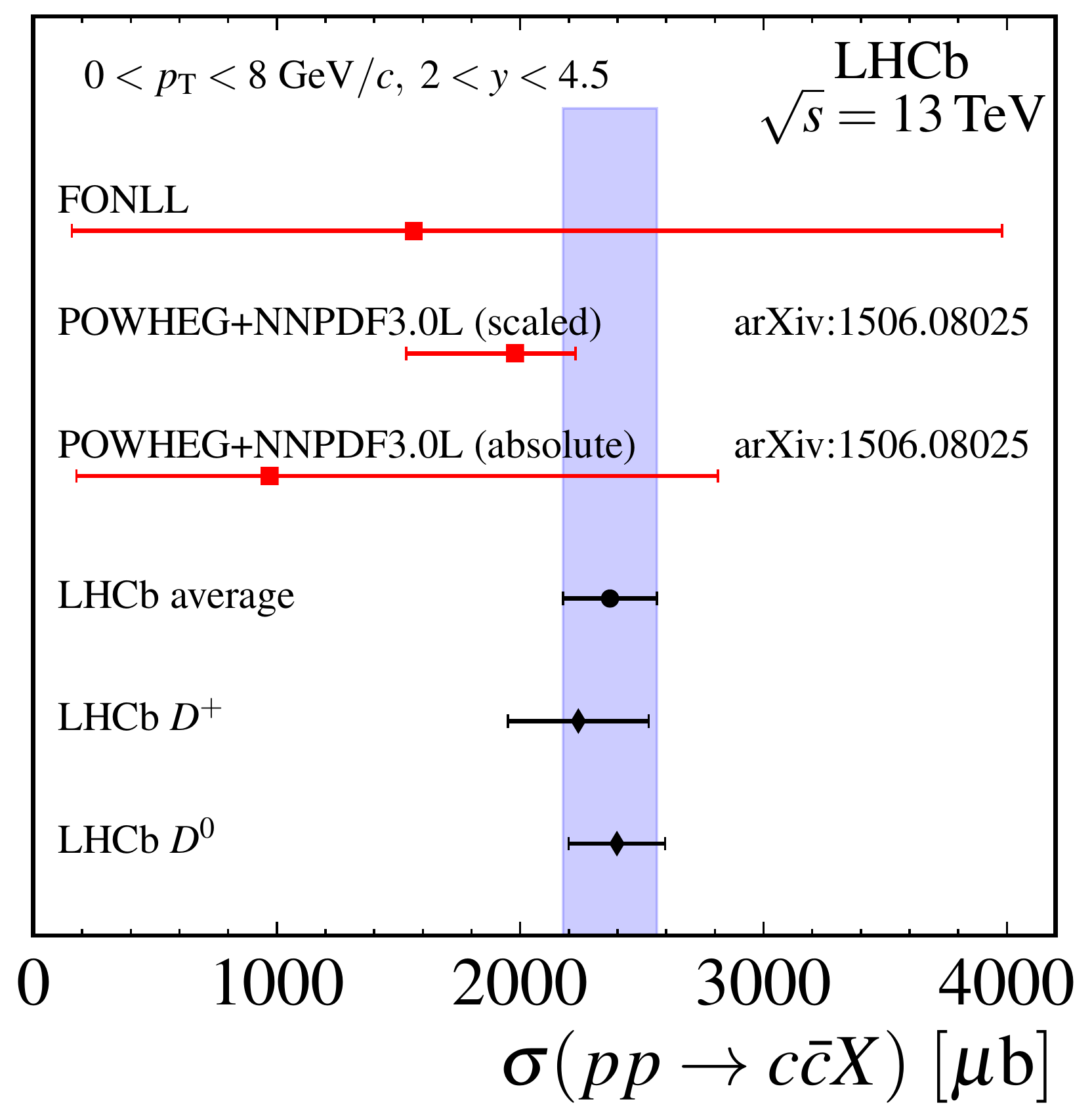}
  \includegraphics[width=0.49\textwidth]{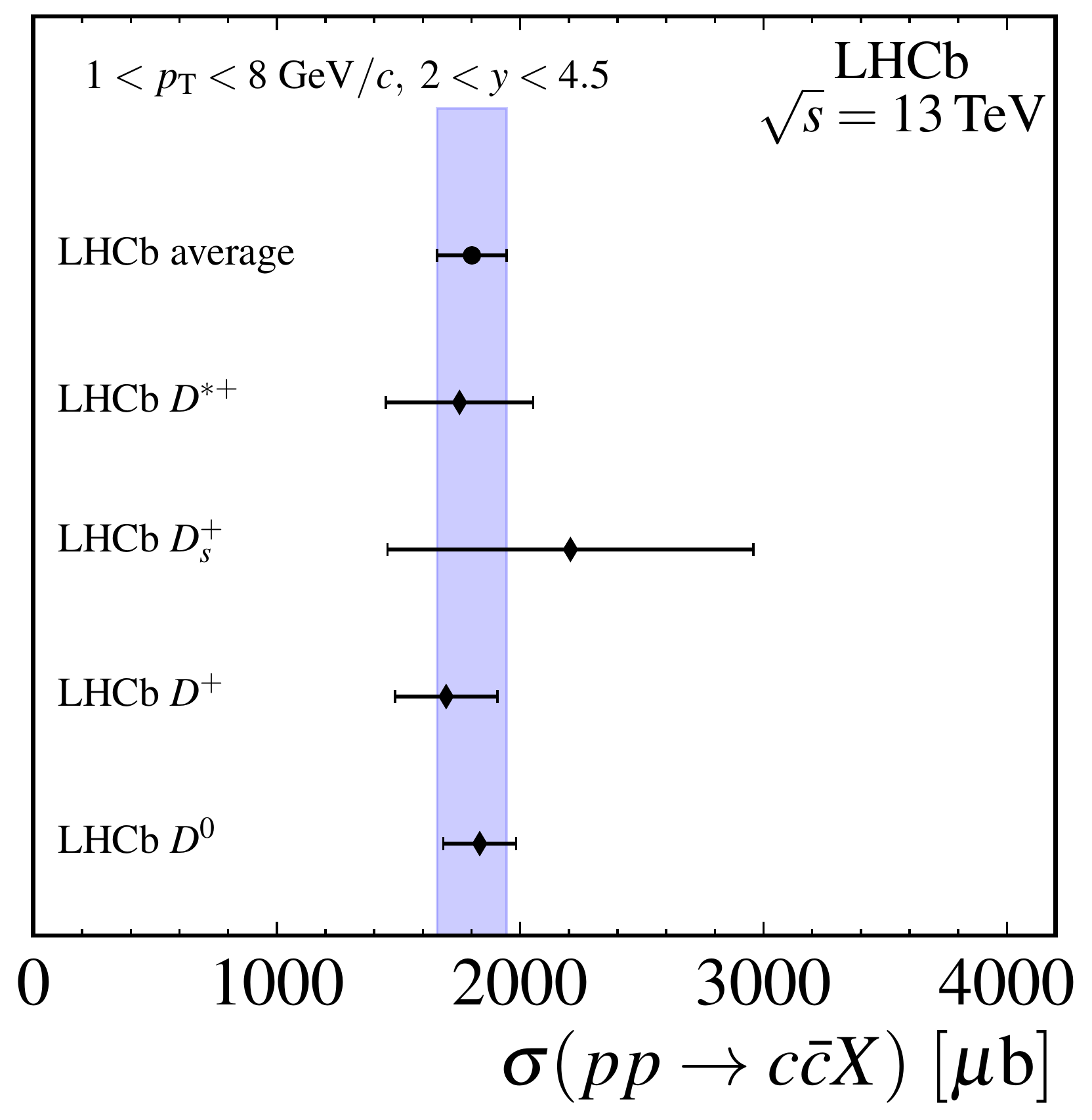}
    \caption{Integrated cross-sections (black diamonds), their average (black circle and blue band) and theory predictions (red squares)~\cite{Gauld:2015yia,Cacciari:2015fta} are shown (left) based on the \Dz and \Dp for $0 < \pT < \SI{8}{\gevc}$ and (right) for measurements based on all four mesons for $1 < \pT < \SI{8}{\gevc}$. 
    The ``absolute'' predictions are based on calculations of the \SI{13}{\TeV} cross-section, while the ``scaled'' predictions are based on calculations of the $13$ to \SI{7}{\TeV} ratio multiplied with the \lhcb measurement at \SI{7}{\TeV}~\cite{LHCb-PAPER-2012-041}.
  \label{fig:TotalCrossSec}}
\end{figure}

The \Rhorry predictions are also provided in the form of \Dz and \Dp
production cross-sections for \pp collisions at \comenergy for each
bin in a subdivision of the phase space, \mbox{$\pT < \SI{30}{\gevc}$} and
\mbox{$2.0 < y < 4.5$}.
Ratios of $13$ to \SI{7}{\TeV} cross-sections are given as well.
They are obtained with \powheg~\cite{Alioli:2010xd} matched to
\pythiaeight~\cite{Sjostrand:2014zea} parton showers and an improved
 version of the NNPDF3.0 NLO parton distribution function set 
 designated NNPDF3.0+LHCb~\cite{Gauld:2015yia}. To produce this improved set, 
the authors of Ref.~\cite{Gauld:2015yia}
weight the NNPDF3.0 NLO set in order to match FONLL calculations to \lhcb's
charm cross-section measurements at \SI{7}{\TeV}~\cite{LHCb-PAPER-2012-041}.
This results in a significant improvement in the uncertainties for the gluon
distribution function at small momentum fraction $x$.
Two predictions for the integrated cross-section are provided, one an absolute 
calculation, identical to that for the differential cross-sections, and the other scaled from the \SI{7}{\TeV} measurement.

The \Hubert calculations include theoretical predictions for all mesons
studied in this analysis.
Results are provided for \mbox{$3 < \pT < \SI{30}{\gevc}$}.
Here the CT10~\cite{Lai:2010vv} set of parton distributions is used.
The \Hubert theoretical framework includes the convolution with fragmentation
functions describing the transition \decay{\cquark}{\ensuremath{\PH_c}}
that are normalised to the respective total transition
probabilities~\cite{Kneesch:2007ey,Kniehl:2006mw}.
The fragmentation functions are taken from a fit to production
measurements at $\ep\en$\/ colliders, where no attempt is made to
separate direct production and feed-down from higher resonances.

In general, the data shown in Figures~\ref{fig:theory_absolute_D0_Dp} and~\ref{fig:theory_absolute_Dsp_Dstar} agree with the predicted shapes of the cross-sections at $\sqrt{s}=\SI{13}{\TeV}$ for all three sets of calculations.
The central values of the measurements generally lie above those of the \Rhorry 
and \Matteo predictions, albeit within the uncertainties provided.
The \Hubert predictions provide the best description of the data, although the 
cross-section measurements decrease with \pT at a higher rate than the predictions.
Similar behaviour is observed for the $\sqrt{s}=\SI{7}{\TeV}$ measurement~\cite{LHCb-PAPER-2012-041}, where only central values are shown for the \Matteo prediction which give lower cross-sections than the data.

The predictions are in agreement with the data for the ratios of cross-sections at $\sqrt{s}=\SI{13}{\TeV}$ and \SI{7}{\TeV}, shown in Figure~\ref{fig:TheoryPredRatio}, though consistently at the upper limit of the predicted values.

The absolute predictions for the integrated cross-sections show agreement with data within their large uncertainties, with central values below the measurements.
The scaled \Rhorry prediction, which has smaller uncertainties than the absolute prediction, agrees with the data.
The measurements are consistent with a linear scaling of the cross-section with the collision energy.

%% file: summary.tex
\section{Summary}
\label{sec:summary}

A measurement of charm production in \pp collisions at a
centre-of-mass energy of \mbox{\comenergy} has been performed with data 
collected with the \lhcb detector.
The shapes of the differential cross-sections for
\Dz, \Dp, \Dstarp, and \Dsp mesons are found to be in
agreement with NLO predictions while the predicted central values generally 
lie below the data, albeit mostly within uncertainties.
The ratios of the production cross-sections for centre-of-mass energies of \SI{13}{TeV} and \SI{7}{TeV} 
have been measured and also show consistency with theoretical predictions.
The \ccbar cross-section for production of a charm meson in \pp collisions at $\sqrt{s}=\SI{13}{TeV}$ and
in the range $0 < \pT < \SI{8}{\gevc}$ and $2 < \rapidity < 4.5$ is
\mbox{\ccbarXsec}.

%% file: acknowledgements.tex
\section*{Acknowledgements}

\noindent The authors would like to thank R. Gauld, J. Rojo, L. Rottoli, and J. Talbert for the provision of 
the \Rhorry numbers; M. Cacciari, M.L. Mangano, and P. Nason for the \Matteo 
predictions; and H. Spiesberger, B.A. Kniehl, G. Kramer, and I. Schienbein 
for the \Hubert calculations.
We express our gratitude to our colleagues in the CERN
accelerator departments for the excellent performance of the LHC. We
thank the technical and administrative staff at the LHCb
institutes. We acknowledge support from CERN and from the national
agencies: CAPES, CNPq, FAPERJ and FINEP (Brazil); NSFC (China);
CNRS/IN2P3 (France); BMBF, DFG and MPG (Germany); INFN (Italy); 
FOM and NWO (The Netherlands); MNiSW and NCN (Poland); MEN/IFA (Romania); 
MinES and FANO (Russia); MinECo (Spain); SNSF and SER (Switzerland); 
NASU (Ukraine); STFC (United Kingdom); NSF (USA).
We acknowledge the computing resources that are provided by CERN, IN2P3 (France), KIT and DESY (Germany), INFN (Italy), SURF (The Netherlands), PIC (Spain), GridPP (United Kingdom), RRCKI (Russia), CSCS (Switzerland), IFIN-HH (Romania), CBPF (Brazil), PL-GRID (Poland) and OSC (USA). We are indebted to the communities behind the multiple open 
source software packages on which we depend. We are also thankful for the 
computing resources and the access to software R\&D tools provided by Yandex LLC (Russia).
Individual groups or members have received support from AvH Foundation (Germany),
EPLANET, Marie Sk\l{}odowska-Curie Actions and ERC (European Union), 
Conseil G\'{e}n\'{e}ral de Haute-Savoie, Labex ENIGMASS and OCEVU, 
R\'{e}gion Auvergne (France), RFBR (Russia), XuntaGal and GENCAT (Spain), The Royal Society 
and Royal Commission for the Exhibition of 1851 (United Kingdom).

%% file: appendix_xsec.tex
\section{Absolute cross-sections}
\label{app:xsec}

Tables~\ref{table:differential:D0}--\ref{table:differential:Dstp} give the numerical results for the differential cross-sections.

\input{tables/differential/D0.tex}

\input{tables/differential/Dp.tex}

\input{tables/differential/Dsp.tex}

\input{tables/differential/Dstp.tex}

%% file: tables/differential/D0.tex
\hvFloat[
  nonFloat=true,
  objectAngle=90,
  capPos=l,
  capAngle=90,
  capWidth=h]{table}{
    \centering
    \input{./tables/differential/D0ToKpi.tex}
}{Differential production cross-sections, $\text{d}^2\sigma/(\text{d}\pT\,\text{d}y)$, in 
$\si{\micro\barn}/(\si{\gevc})$ for prompt $\Dz + \Dzbar$ mesons in bins 
of \pTy.
The first uncertainty is statistical, and the second is the total 
systematic.}{table:differential:D0}

%% file: tables/differential/D0ToKpi.tex
\renewcommand{\arraystretch}{1.3}
\begin{tabular}{l|r@{\hskip+0.2em}c@{\hskip+0.2em}r@{\hskip+0.2em}c@{\hskip+0.2em}rr@{\hskip+0.2em}c@{\hskip+0.2em}r@{\hskip+0.2em}c@{\hskip+0.2em}rr@{\hskip+0.2em}c@{\hskip+0.2em}r@{\hskip+0.2em}c@{\hskip+0.2em}rr@{\hskip+0.2em}c@{\hskip+0.2em}r@{\hskip+0.2em}c@{\hskip+0.2em}rr@{\hskip+0.2em}c@{\hskip+0.2em}r@{\hskip+0.2em}c@{\hskip+0.2em}r}
\toprule&\multicolumn{25}{c}{$\text{$y$}$}\\
$\text{$p_{\text{T}}$ [\text{MeV}/c]}$ & \multicolumn{5}{c}{$[2,2.5]$} & \multicolumn{5}{c}{$[2.5,3]$} & \multicolumn{5}{c}{$[3,3.5]$} & \multicolumn{5}{c}{$[3.5,4]$} & \multicolumn{5}{c}{$[4,4.5]$} \\
\midrule$[0,1000]$ & $254$ & $^+_-$ & $^{2}_{2}$&$^+_-$ & $^{31}_{27}$ & $277.3$ & $^+_-$ & $^{1.0}_{1.0}$&$^+_-$ & $^{21.4}_{24.8}$ & $281.1$ & $^+_-$ & $^{0.9}_{0.9}$&$^+_-$ & $^{19.4}_{17.4}$ & $256$ & $^+_-$ & $^{1}_{1}$&$^+_-$ & $^{16}_{14}$ & $206$ & $^+_-$ & $^{2}_{2}$&$^+_-$ & $^{16}_{14}$ \\
$[1000,1500]$ & $433$ & $^+_-$ & $^{3}_{3}$&$^+_-$ & $^{50}_{45}$ & $457$ & $^+_-$ & $^{2}_{2}$&$^+_-$ & $^{32}_{37}$ & $454$ & $^+_-$ & $^{1}_{1}$&$^+_-$ & $^{28}_{27}$ & $409$ & $^+_-$ & $^{2}_{2}$&$^+_-$ & $^{25}_{23}$ & $337$ & $^+_-$ & $^{3}_{3}$&$^+_-$ & $^{25}_{22}$ \\
$[1500,2000]$ & $388$ & $^+_-$ & $^{2}_{2}$&$^+_-$ & $^{40}_{37}$ & $409$ & $^+_-$ & $^{1}_{1}$&$^+_-$ & $^{30}_{26}$ & $380$ & $^+_-$ & $^{1}_{1}$&$^+_-$ & $^{23}_{22}$ & $344$ & $^+_-$ & $^{1}_{1}$&$^+_-$ & $^{21}_{19}$ & $265$ & $^+_-$ & $^{2}_{2}$&$^+_-$ & $^{18}_{17}$ \\
$[2000,2500]$ & $296$ & $^+_-$ & $^{2}_{2}$&$^+_-$ & $^{24}_{28}$ & $314.4$ & $^+_-$ & $^{1.0}_{1.0}$&$^+_-$ & $^{21.1}_{18.7}$ & $292.5$ & $^+_-$ & $^{0.9}_{0.9}$&$^+_-$ & $^{17.4}_{15.8}$ & $259$ & $^+_-$ & $^{1}_{1}$&$^+_-$ & $^{16}_{14}$ & $190$ & $^+_-$ & $^{2}_{2}$&$^+_-$ & $^{11}_{12}$ \\
$[2500,3000]$ & $228$ & $^+_-$ & $^{1}_{1}$&$^+_-$ & $^{17}_{20}$ & $228.9$ & $^+_-$ & $^{0.7}_{0.7}$&$^+_-$ & $^{14.7}_{13.2}$ & $208.8$ & $^+_-$ & $^{0.7}_{0.7}$&$^+_-$ & $^{12.6}_{10.4}$ & $175.0$ & $^+_-$ & $^{0.8}_{0.8}$&$^+_-$ & $^{11.3}_{\phantom{0}9.5}$ & $129.6$ & $^+_-$ & $^{1.2}_{1.2}$&$^+_-$ & $^{8.3}_{8.9}$ \\
$[3000,3500]$ & $161.3$ & $^+_-$ & $^{0.9}_{0.9}$&$^+_-$ & $^{11.4}_{13.0}$ & $156.1$ & $^+_-$ & $^{0.5}_{0.5}$&$^+_-$ & $^{9.6}_{8.7}$ & $137.3$ & $^+_-$ & $^{0.5}_{0.5}$&$^+_-$ & $^{8.1}_{6.9}$ & $121.4$ & $^+_-$ & $^{0.6}_{0.6}$&$^+_-$ & $^{8.5}_{6.9}$ & $85.6$ & $^+_-$ & $^{0.9}_{0.9}$&$^+_-$ & $^{6.3}_{6.4}$ \\
$[3500,4000]$ & $109.7$ & $^+_-$ & $^{0.6}_{0.6}$&$^+_-$ & $^{7.3}_{8.2}$ & $111.0$ & $^+_-$ & $^{0.4}_{0.4}$&$^+_-$ & $^{6.6}_{6.0}$ & $93.5$ & $^+_-$ & $^{0.4}_{0.4}$&$^+_-$ & $^{5.5}_{4.7}$ & $82.7$ & $^+_-$ & $^{0.4}_{0.4}$&$^+_-$ & $^{5.5}_{4.9}$ & $58.0$ & $^+_-$ & $^{0.8}_{0.8}$&$^+_-$ & $^{5.9}_{5.5}$ \\
$[4000,5000]$ & $66.5$ & $^+_-$ & $^{0.3}_{0.3}$&$^+_-$ & $^{4.7}_{4.2}$ & $64.5$ & $^+_-$ & $^{0.2}_{0.2}$&$^+_-$ & $^{3.9}_{3.3}$ & $54.5$ & $^+_-$ & $^{0.2}_{0.2}$&$^+_-$ & $^{3.2}_{2.7}$ & $45.9$ & $^+_-$ & $^{0.2}_{0.2}$&$^+_-$ & $^{2.5}_{2.7}$ & $26.1$ & $^+_-$ & $^{0.4}_{0.4}$&$^+_-$ & $^{3.6}_{3.0}$ \\
$[5000,6000]$ & $35.5$ & $^+_-$ & $^{0.2}_{0.2}$&$^+_-$ & $^{2.6}_{2.1}$ & $31.5$ & $^+_-$ & $^{0.1}_{0.1}$&$^+_-$ & $^{1.8}_{1.6}$ & $27.1$ & $^+_-$ & $^{0.1}_{0.1}$&$^+_-$ & $^{1.5}_{1.6}$ & $20.8$ & $^+_-$ & $^{0.1}_{0.1}$&$^+_-$ & $^{1.3}_{1.2}$ & $10.9$ & $^+_-$ & $^{0.4}_{0.4}$&$^+_-$ & $^{2.5}_{1.8}$ \\
$[6000,7000]$ & $18.5$ & $^+_-$ & $^{0.1}_{0.1}$&$^+_-$ & $^{1.3}_{1.2}$ & $17.20$ & $^+_-$ & $^{0.09}_{0.09}$&$^+_-$ & $^{0.96}_{0.92}$ & $14.40$ & $^+_-$ & $^{0.09}_{0.09}$&$^+_-$ & $^{0.82}_{0.85}$ & $10.20$ & $^+_-$ & $^{0.12}_{0.12}$&$^+_-$ & $^{0.87}_{0.81}$ & $3.4$ & $^+_-$ & $^{0.5}_{0.5}$&$^+_-$ & $^{2.4}_{1.3}$ \\
$[7000,8000]$ & $10.15$ & $^+_-$ & $^{0.09}_{0.09}$&$^+_-$ & $^{0.72}_{0.63}$ & $9.71$ & $^+_-$ & $^{0.07}_{0.07}$&$^+_-$ & $^{0.60}_{0.52}$ & $7.88$ & $^+_-$ & $^{0.07}_{0.07}$&$^+_-$ & $^{0.42}_{0.48}$ & $4.94$ & $^+_-$ & $^{0.10}_{0.10}$&$^+_-$ & $^{0.71}_{0.58}$ & \multicolumn{5}{c}{ } \\
$[8000,9000]$ & $6.11$ & $^+_-$ & $^{0.07}_{0.07}$&$^+_-$ & $^{0.46}_{0.39}$ & $5.72$ & $^+_-$ & $^{0.05}_{0.05}$&$^+_-$ & $^{0.38}_{0.32}$ & $4.49$ & $^+_-$ & $^{0.05}_{0.05}$&$^+_-$ & $^{0.27}_{0.29}$ & $2.88$ & $^+_-$ & $^{0.11}_{0.11}$&$^+_-$ & $^{0.57}_{0.44}$ & \multicolumn{5}{c}{ } \\
$[9000,10000]$ & $3.92$ & $^+_-$ & $^{0.06}_{0.06}$&$^+_-$ & $^{0.26}_{0.22}$ & $3.48$ & $^+_-$ & $^{0.04}_{0.04}$&$^+_-$ & $^{0.21}_{0.15}$ & $2.75$ & $^+_-$ & $^{0.04}_{0.04}$&$^+_-$ & $^{0.16}_{0.13}$ & $1.53$ & $^+_-$ & $^{0.11}_{0.11}$&$^+_-$ & $^{0.49}_{0.28}$ & \multicolumn{5}{c}{ } \\
$[10000,11000]$ & $2.48$ & $^+_-$ & $^{0.05}_{0.05}$&$^+_-$ & $^{0.18}_{0.16}$ & $2.18$ & $^+_-$ & $^{0.03}_{0.03}$&$^+_-$ & $^{0.14}_{0.09}$ & $1.62$ & $^+_-$ & $^{0.04}_{0.04}$&$^+_-$ & $^{0.12}_{0.09}$ & $0.77$ & $^+_-$ & $^{0.13}_{0.13}$&$^+_-$ & $^{0.33}_{0.16}$ & \multicolumn{5}{c}{ } \\
$[11000,12000]$ & $1.59$ & $^+_-$ & $^{0.04}_{0.04}$&$^+_-$ & $^{0.14}_{0.10}$ & $1.49$ & $^+_-$ & $^{0.03}_{0.03}$&$^+_-$ & $^{0.10}_{0.06}$ & $1.025$ & $^+_-$ & $^{0.034}_{0.034}$&$^+_-$ & $^{0.083}_{0.056}$ & \multicolumn{5}{c}{ } & \multicolumn{5}{c}{ } \\
$[12000,13000]$ & $1.01$ & $^+_-$ & $^{0.03}_{0.03}$&$^+_-$ & $^{0.12}_{0.08}$ & $1.031$ & $^+_-$ & $^{0.026}_{0.026}$&$^+_-$ & $^{0.078}_{0.037}$ & $0.716$ & $^+_-$ & $^{0.034}_{0.034}$&$^+_-$ & $^{0.060}_{0.048}$ & \multicolumn{5}{c}{ } & \multicolumn{5}{c}{ } \\
$[13000,14000]$ & $0.746$ & $^+_-$ & $^{0.026}_{0.026}$&$^+_-$ & $^{0.089}_{0.055}$ & $0.718$ & $^+_-$ & $^{0.023}_{0.023}$&$^+_-$ & $^{0.060}_{0.025}$ & $0.447$ & $^+_-$ & $^{0.031}_{0.031}$&$^+_-$ & $^{0.075}_{0.045}$ & \multicolumn{5}{c}{ } & \multicolumn{5}{c}{ } \\
$[14000,15000]$ & $0.498$ & $^+_-$ & $^{0.022}_{0.022}$&$^+_-$ & $^{0.058}_{0.030}$ & $0.495$ & $^+_-$ & $^{0.020}_{0.020}$&$^+_-$ & $^{0.048}_{0.022}$ & $0.268$ & $^+_-$ & $^{0.031}_{0.031}$&$^+_-$ & $^{0.053}_{0.031}$ & \multicolumn{5}{c}{ } & \multicolumn{5}{c}{ } \\
\bottomrule\end{tabular}

%% file: tables/differential/Dp.tex
\hvFloat[
  nonFloat=true,
  objectAngle=90,
  capPos=l,
  capAngle=90,
  capWidth=h]{table}{
    \centering
    \input{./tables/differential/DpToKpipi.tex}
}{Differential production cross-sections, $\text{d}^2\sigma/(\text{d}\pT\,\text{d}y)$, in 
$\si{\micro\barn}/(\si{\gevc})$ for prompt $\Dp + \Dm$ mesons in bins of 
\pTy. 
The first uncertainty is statistical, and the second is the total 
systematic.}{table:differential:Dp}

%% file: tables/differential/DpToKpipi.tex
\renewcommand{\arraystretch}{1.3}
\begin{tabular}{l|r@{\hskip+0.2em}c@{\hskip+0.2em}r@{\hskip+0.2em}c@{\hskip+0.2em}rr@{\hskip+0.2em}c@{\hskip+0.2em}r@{\hskip+0.2em}c@{\hskip+0.2em}rr@{\hskip+0.2em}c@{\hskip+0.2em}r@{\hskip+0.2em}c@{\hskip+0.2em}rr@{\hskip+0.2em}c@{\hskip+0.2em}r@{\hskip+0.2em}c@{\hskip+0.2em}rr@{\hskip+0.2em}c@{\hskip+0.2em}r@{\hskip+0.2em}c@{\hskip+0.2em}r}
\toprule&\multicolumn{25}{c}{$\text{$y$}$}\\
$\text{$p_{\text{T}}$ [\text{MeV}/c]}$ & \multicolumn{5}{c}{$[2,2.5]$} & \multicolumn{5}{c}{$[2.5,3]$} & \multicolumn{5}{c}{$[3,3.5]$} & \multicolumn{5}{c}{$[3.5,4]$} & \multicolumn{5}{c}{$[4,4.5]$} \\
\midrule$[0,1000]$ & \multicolumn{5}{c}{ } & $156$ & $^+_-$ & $^{6}_{6}$&$^+_-$ & $^{38}_{32}$ & $107$ & $^+_-$ & $^{3}_{3}$&$^+_-$ & $^{15}_{14}$ & $97$ & $^+_-$ & $^{3}_{3}$&$^+_-$ & $^{13}_{13}$ & $72$ & $^+_-$ & $^{5}_{5}$&$^+_-$ & $^{16}_{14}$ \\
$[1000,1500]$ & $182$ & $^+_-$ & $^{7}_{7}$&$^+_-$ & $^{47}_{42}$ & $170$ & $^+_-$ & $^{2}_{2}$&$^+_-$ & $^{31}_{22}$ & $184$ & $^+_-$ & $^{1}_{1}$&$^+_-$ & $^{19}_{18}$ & $158$ & $^+_-$ & $^{2}_{2}$&$^+_-$ & $^{15}_{14}$ & $137$ & $^+_-$ & $^{3}_{3}$&$^+_-$ & $^{16}_{15}$ \\
$[1500,2000]$ & $146$ & $^+_-$ & $^{2}_{2}$&$^+_-$ & $^{27}_{23}$ & $149.5$ & $^+_-$ & $^{0.8}_{0.8}$&$^+_-$ & $^{23.0}_{16.6}$ & $153.5$ & $^+_-$ & $^{0.7}_{0.7}$&$^+_-$ & $^{13.7}_{14.2}$ & $133.6$ & $^+_-$ & $^{0.7}_{0.7}$&$^+_-$ & $^{11.4}_{10.4}$ & $107.1$ & $^+_-$ & $^{1.2}_{1.2}$&$^+_-$ & $^{9.6}_{9.1}$ \\
$[2000,2500]$ & $111$ & $^+_-$ & $^{1}_{1}$&$^+_-$ & $^{18}_{16}$ & $126.1$ & $^+_-$ & $^{0.5}_{0.5}$&$^+_-$ & $^{16.4}_{12.9}$ & $115.4$ & $^+_-$ & $^{0.4}_{0.4}$&$^+_-$ & $^{9.5}_{9.7}$ & $97.4$ & $^+_-$ & $^{0.4}_{0.4}$&$^+_-$ & $^{7.8}_{7.2}$ & $78.9$ & $^+_-$ & $^{0.7}_{0.7}$&$^+_-$ & $^{6.3}_{6.1}$ \\
$[2500,3000]$ & $86.1$ & $^+_-$ & $^{0.7}_{0.7}$&$^+_-$ & $^{13.1}_{12.3}$ & $94.0$ & $^+_-$ & $^{0.3}_{0.3}$&$^+_-$ & $^{\phantom{0}8.8}_{10.0}$ & $83.5$ & $^+_-$ & $^{0.3}_{0.3}$&$^+_-$ & $^{6.5}_{6.5}$ & $72.1$ & $^+_-$ & $^{0.3}_{0.3}$&$^+_-$ & $^{5.4}_{5.1}$ & $53.2$ & $^+_-$ & $^{0.4}_{0.4}$&$^+_-$ & $^{4.4}_{4.0}$ \\
$[3000,3500]$ & $62.3$ & $^+_-$ & $^{0.5}_{0.5}$&$^+_-$ & $^{8.8}_{8.6}$ & $66.7$ & $^+_-$ & $^{0.2}_{0.2}$&$^+_-$ & $^{6.0}_{6.6}$ & $59.3$ & $^+_-$ & $^{0.2}_{0.2}$&$^+_-$ & $^{4.6}_{4.3}$ & $49.7$ & $^+_-$ & $^{0.2}_{0.2}$&$^+_-$ & $^{3.8}_{3.4}$ & $35.0$ & $^+_-$ & $^{0.3}_{0.3}$&$^+_-$ & $^{2.7}_{2.7}$ \\
$[3500,4000]$ & $42.4$ & $^+_-$ & $^{0.3}_{0.3}$&$^+_-$ & $^{5.5}_{5.6}$ & $46.4$ & $^+_-$ & $^{0.2}_{0.2}$&$^+_-$ & $^{4.2}_{4.2}$ & $42.0$ & $^+_-$ & $^{0.2}_{0.2}$&$^+_-$ & $^{3.3}_{2.9}$ & $33.3$ & $^+_-$ & $^{0.2}_{0.2}$&$^+_-$ & $^{2.7}_{2.2}$ & $22.6$ & $^+_-$ & $^{0.2}_{0.2}$&$^+_-$ & $^{1.7}_{1.8}$ \\
$[4000,5000]$ & $29.2$ & $^+_-$ & $^{0.2}_{0.2}$&$^+_-$ & $^{3.2}_{3.7}$ & $28.22$ & $^+_-$ & $^{0.09}_{0.09}$&$^+_-$ & $^{2.35}_{2.34}$ & $23.36$ & $^+_-$ & $^{0.08}_{0.08}$&$^+_-$ & $^{1.83}_{1.48}$ & $19.29$ & $^+_-$ & $^{0.08}_{0.08}$&$^+_-$ & $^{1.51}_{1.26}$ & $13.9$ & $^+_-$ & $^{0.1}_{0.1}$&$^+_-$ & $^{0.9}_{1.1}$ \\
$[5000,6000]$ & $15.1$ & $^+_-$ & $^{0.1}_{0.1}$&$^+_-$ & $^{1.5}_{1.7}$ & $14.49$ & $^+_-$ & $^{0.06}_{0.06}$&$^+_-$ & $^{1.18}_{1.10}$ & $12.18$ & $^+_-$ & $^{0.05}_{0.05}$&$^+_-$ & $^{0.94}_{0.78}$ & $9.44$ & $^+_-$ & $^{0.05}_{0.05}$&$^+_-$ & $^{0.71}_{0.63}$ & $6.87$ & $^+_-$ & $^{0.11}_{0.11}$&$^+_-$ & $^{0.60}_{0.64}$ \\
$[6000,7000]$ & $8.34$ & $^+_-$ & $^{0.07}_{0.07}$&$^+_-$ & $^{0.76}_{0.84}$ & $7.39$ & $^+_-$ & $^{0.04}_{0.04}$&$^+_-$ & $^{0.59}_{0.54}$ & $6.30$ & $^+_-$ & $^{0.04}_{0.04}$&$^+_-$ & $^{0.49}_{0.40}$ & $4.89$ & $^+_-$ & $^{0.04}_{0.04}$&$^+_-$ & $^{0.32}_{0.37}$ & $3.03$ & $^+_-$ & $^{0.09}_{0.09}$&$^+_-$ & $^{0.32}_{0.30}$ \\
$[7000,8000]$ & $4.98$ & $^+_-$ & $^{0.05}_{0.05}$&$^+_-$ & $^{0.47}_{0.46}$ & $4.24$ & $^+_-$ & $^{0.03}_{0.03}$&$^+_-$ & $^{0.33}_{0.31}$ & $3.51$ & $^+_-$ & $^{0.03}_{0.03}$&$^+_-$ & $^{0.28}_{0.23}$ & $2.56$ & $^+_-$ & $^{0.03}_{0.03}$&$^+_-$ & $^{0.18}_{0.21}$ & $2.31$ & $^+_-$ & $^{0.13}_{0.13}$&$^+_-$ & $^{0.53}_{0.47}$ \\
$[8000,9000]$ & $3.11$ & $^+_-$ & $^{0.04}_{0.04}$&$^+_-$ & $^{0.31}_{0.25}$ & $2.63$ & $^+_-$ & $^{0.02}_{0.02}$&$^+_-$ & $^{0.21}_{0.19}$ & $2.16$ & $^+_-$ & $^{0.02}_{0.02}$&$^+_-$ & $^{0.17}_{0.14}$ & $1.40$ & $^+_-$ & $^{0.02}_{0.02}$&$^+_-$ & $^{0.12}_{0.13}$ & $1.17$ & $^+_-$ & $^{0.09}_{0.09}$&$^+_-$ & $^{0.46}_{0.47}$ \\
$[9000,10000]$ & $1.46$ & $^+_-$ & $^{0.02}_{0.02}$&$^+_-$ & $^{0.13}_{0.11}$ & $1.35$ & $^+_-$ & $^{0.01}_{0.01}$&$^+_-$ & $^{0.11}_{0.09}$ & $1.192$ & $^+_-$ & $^{0.015}_{0.015}$&$^+_-$ & $^{0.083}_{0.083}$ & $0.928$ & $^+_-$ & $^{0.025}_{0.025}$&$^+_-$ & $^{0.096}_{0.087}$ & \multicolumn{5}{c}{ } \\
$[10000,11000]$ & $1.144$ & $^+_-$ & $^{0.020}_{0.020}$&$^+_-$ & $^{0.098}_{0.081}$ & $0.970$ & $^+_-$ & $^{0.013}_{0.013}$&$^+_-$ & $^{0.072}_{0.061}$ & $0.784$ & $^+_-$ & $^{0.013}_{0.013}$&$^+_-$ & $^{0.050}_{0.052}$ & $0.477$ & $^+_-$ & $^{0.020}_{0.020}$&$^+_-$ & $^{0.057}_{0.044}$ & \multicolumn{5}{c}{ } \\
$[11000,12000]$ & $0.774$ & $^+_-$ & $^{0.016}_{0.016}$&$^+_-$ & $^{0.070}_{0.054}$ & $0.641$ & $^+_-$ & $^{0.011}_{0.011}$&$^+_-$ & $^{0.050}_{0.037}$ & $0.520$ & $^+_-$ & $^{0.011}_{0.011}$&$^+_-$ & $^{0.035}_{0.034}$ & $0.301$ & $^+_-$ & $^{0.017}_{0.017}$&$^+_-$ & $^{0.031}_{0.021}$ & \multicolumn{5}{c}{ } \\
$[12000,13000]$ & $0.513$ & $^+_-$ & $^{0.012}_{0.012}$&$^+_-$ & $^{0.049}_{0.033}$ & $0.443$ & $^+_-$ & $^{0.009}_{0.009}$&$^+_-$ & $^{0.036}_{0.025}$ & $0.336$ & $^+_-$ & $^{0.009}_{0.009}$&$^+_-$ & $^{0.024}_{0.021}$ & $0.185$ & $^+_-$ & $^{0.011}_{0.011}$&$^+_-$ & $^{0.034}_{0.021}$ & \multicolumn{5}{c}{ } \\
$[13000,14000]$ & $0.362$ & $^+_-$ & $^{0.010}_{0.010}$&$^+_-$ & $^{0.037}_{0.023}$ & $0.298$ & $^+_-$ & $^{0.007}_{0.007}$&$^+_-$ & $^{0.028}_{0.015}$ & $0.238$ & $^+_-$ & $^{0.008}_{0.008}$&$^+_-$ & $^{0.019}_{0.014}$ & $0.096$ & $^+_-$ & $^{0.009}_{0.009}$&$^+_-$ & $^{0.015}_{0.006}$ & \multicolumn{5}{c}{ } \\
$[14000,15000]$ & $0.250$ & $^+_-$ & $^{0.008}_{0.008}$&$^+_-$ & $^{0.028}_{0.014}$ & $0.212$ & $^+_-$ & $^{0.006}_{0.006}$&$^+_-$ & $^{0.022}_{0.009}$ & $0.162$ & $^+_-$ & $^{0.007}_{0.007}$&$^+_-$ & $^{0.017}_{0.007}$ & \multicolumn{5}{c}{ } & \multicolumn{5}{c}{ } \\
\bottomrule\end{tabular}

%% file: tables/differential/Dsp.tex
\hvFloat[
  nonFloat=true,
  objectAngle=90,
  capPos=l,
  capAngle=90,
  capWidth=h]{table}{
  \centering
  \input{./tables/differential/DsTophipi.tex}
}{Differential production cross-sections, $\text{d}^2\sigma/(\text{d}\pT\,\text{d}y)$, in 
$\si{\micro\barn}/(\si{\gevc})$ for prompt $\Dsp + \Dsm$ mesons in bins of 
\pTy.
The first uncertainty is statistical, and the second is the total 
systematic.}{table:differential:Dsp}

%% file: tables/differential/DsTophipi.tex
\renewcommand{\arraystretch}{1.3}
\begin{tabular}{l|r@{\hskip+0.2em}c@{\hskip+0.2em}r@{\hskip+0.2em}c@{\hskip+0.2em}rr@{\hskip+0.2em}c@{\hskip+0.2em}r@{\hskip+0.2em}c@{\hskip+0.2em}rr@{\hskip+0.2em}c@{\hskip+0.2em}r@{\hskip+0.2em}c@{\hskip+0.2em}rr@{\hskip+0.2em}c@{\hskip+0.2em}r@{\hskip+0.2em}c@{\hskip+0.2em}rr@{\hskip+0.2em}c@{\hskip+0.2em}r@{\hskip+0.2em}c@{\hskip+0.2em}r}
\toprule&\multicolumn{25}{c}{$\text{$y$}$}\\
$\text{$p_{\text{T}}$ [\text{MeV}/c]}$ & \multicolumn{5}{c}{$[2,2.5]$} & \multicolumn{5}{c}{$[2.5,3]$} & \multicolumn{5}{c}{$[3,3.5]$} & \multicolumn{5}{c}{$[3.5,4]$} & \multicolumn{5}{c}{$[4,4.5]$} \\
\midrule$[1000,1500]$ & $31$ & $^+_-$ & $^{8}_{8}$&$^+_-$ & $^{15}_{13}$ & $60$ & $^+_-$ & $^{4}_{4}$&$^+_-$ & $^{21}_{13}$ & $70$ & $^+_-$ & $^{4}_{4}$&$^+_-$ & $^{13}_{13}$ & $55$ & $^+_-$ & $^{5}_{5}$&$^+_-$ & $^{14}_{11}$ & $110$ & $^+_-$ & $^{36}_{36}$&$^+_-$ & $^{34}_{72}$ \\
$[1500,2000]$ & $47$ & $^+_-$ & $^{3}_{3}$&$^+_-$ & $^{13}_{10}$ & $63$ & $^+_-$ & $^{2}_{2}$&$^+_-$ & $^{14}_{10}$ & $62.1$ & $^+_-$ & $^{1.6}_{1.6}$&$^+_-$ & $^{7.5}_{7.2}$ & $61.2$ & $^+_-$ & $^{2.2}_{2.2}$&$^+_-$ & $^{8.5}_{7.2}$ & $29.8$ & $^+_-$ & $^{3.1}_{3.1}$&$^+_-$ & $^{5.9}_{5.5}$ \\
$[2000,2500]$ & $48$ & $^+_-$ & $^{2}_{2}$&$^+_-$ & $^{11}_{\phantom{0}9}$ & $52.2$ & $^+_-$ & $^{0.9}_{0.9}$&$^+_-$ & $^{8.4}_{5.8}$ & $56.1$ & $^+_-$ & $^{0.9}_{0.9}$&$^+_-$ & $^{6.2}_{5.6}$ & $47.1$ & $^+_-$ & $^{1.1}_{1.1}$&$^+_-$ & $^{5.3}_{4.7}$ & $26.9$ & $^+_-$ & $^{1.8}_{1.8}$&$^+_-$ & $^{4.2}_{3.7}$ \\
$[2500,3000]$ & $38.2$ & $^+_-$ & $^{1.1}_{1.1}$&$^+_-$ & $^{7.8}_{6.5}$ & $44.1$ & $^+_-$ & $^{0.6}_{0.6}$&$^+_-$ & $^{4.9}_{5.5}$ & $38.3$ & $^+_-$ & $^{0.5}_{0.5}$&$^+_-$ & $^{4.2}_{3.3}$ & $33.9$ & $^+_-$ & $^{0.7}_{0.7}$&$^+_-$ & $^{3.5}_{3.2}$ & $23.7$ & $^+_-$ & $^{1.1}_{1.1}$&$^+_-$ & $^{3.2}_{3.1}$ \\
$[3000,3500]$ & $27.1$ & $^+_-$ & $^{0.7}_{0.7}$&$^+_-$ & $^{4.9}_{4.2}$ & $30.1$ & $^+_-$ & $^{0.4}_{0.4}$&$^+_-$ & $^{3.2}_{3.4}$ & $27.0$ & $^+_-$ & $^{0.4}_{0.4}$&$^+_-$ & $^{2.9}_{2.3}$ & $22.4$ & $^+_-$ & $^{0.4}_{0.4}$&$^+_-$ & $^{2.3}_{2.1}$ & $13.9$ & $^+_-$ & $^{0.7}_{0.7}$&$^+_-$ & $^{1.7}_{1.6}$ \\
$[3500,4000]$ & $21.2$ & $^+_-$ & $^{0.5}_{0.5}$&$^+_-$ & $^{3.6}_{3.0}$ & $20.9$ & $^+_-$ & $^{0.3}_{0.3}$&$^+_-$ & $^{2.2}_{2.2}$ & $19.0$ & $^+_-$ & $^{0.3}_{0.3}$&$^+_-$ & $^{2.0}_{1.6}$ & $16.6$ & $^+_-$ & $^{0.3}_{0.3}$&$^+_-$ & $^{1.7}_{1.5}$ & $10.7$ & $^+_-$ & $^{0.6}_{0.6}$&$^+_-$ & $^{1.5}_{1.3}$ \\
$[4000,5000]$ & $12.4$ & $^+_-$ & $^{0.2}_{0.2}$&$^+_-$ & $^{1.8}_{1.6}$ & $13.4$ & $^+_-$ & $^{0.2}_{0.2}$&$^+_-$ & $^{1.4}_{1.3}$ & $10.7$ & $^+_-$ & $^{0.1}_{0.1}$&$^+_-$ & $^{1.1}_{0.9}$ & $9.60$ & $^+_-$ & $^{0.17}_{0.17}$&$^+_-$ & $^{0.93}_{0.87}$ & $6.36$ & $^+_-$ & $^{0.28}_{0.28}$&$^+_-$ & $^{0.78}_{0.61}$ \\
$[5000,6000]$ & $7.7$ & $^+_-$ & $^{0.2}_{0.2}$&$^+_-$ & $^{0.9}_{1.0}$ & $7.11$ & $^+_-$ & $^{0.11}_{0.11}$&$^+_-$ & $^{0.79}_{0.59}$ & $5.89$ & $^+_-$ & $^{0.10}_{0.10}$&$^+_-$ & $^{0.58}_{0.49}$ & $4.62$ & $^+_-$ & $^{0.10}_{0.10}$&$^+_-$ & $^{0.54}_{0.41}$ & $3.10$ & $^+_-$ & $^{0.20}_{0.20}$&$^+_-$ & $^{0.45}_{0.40}$ \\
$[6000,7000]$ & $3.83$ & $^+_-$ & $^{0.10}_{0.10}$&$^+_-$ & $^{0.44}_{0.46}$ & $3.85$ & $^+_-$ & $^{0.08}_{0.08}$&$^+_-$ & $^{0.41}_{0.34}$ & $3.03$ & $^+_-$ & $^{0.07}_{0.07}$&$^+_-$ & $^{0.31}_{0.27}$ & $2.10$ & $^+_-$ & $^{0.07}_{0.07}$&$^+_-$ & $^{0.25}_{0.20}$ & $1.39$ & $^+_-$ & $^{0.16}_{0.16}$&$^+_-$ & $^{0.25}_{0.21}$ \\
$[7000,8000]$ & $2.43$ & $^+_-$ & $^{0.08}_{0.08}$&$^+_-$ & $^{0.32}_{0.26}$ & $1.79$ & $^+_-$ & $^{0.05}_{0.05}$&$^+_-$ & $^{0.18}_{0.17}$ & $1.60$ & $^+_-$ & $^{0.05}_{0.05}$&$^+_-$ & $^{0.18}_{0.14}$ & $1.33$ & $^+_-$ & $^{0.06}_{0.06}$&$^+_-$ & $^{0.18}_{0.14}$ & $0.68$ & $^+_-$ & $^{0.18}_{0.18}$&$^+_-$ & $^{0.18}_{0.18}$ \\
$[8000,9000]$ & $1.33$ & $^+_-$ & $^{0.05}_{0.05}$&$^+_-$ & $^{0.17}_{0.14}$ & $1.19$ & $^+_-$ & $^{0.04}_{0.04}$&$^+_-$ & $^{0.13}_{0.12}$ & $1.22$ & $^+_-$ & $^{0.04}_{0.04}$&$^+_-$ & $^{0.16}_{0.12}$ & $0.72$ & $^+_-$ & $^{0.05}_{0.05}$&$^+_-$ & $^{0.10}_{0.09}$ & \multicolumn{5}{c}{ } \\
$[9000,10000]$ & $0.754$ & $^+_-$ & $^{0.036}_{0.036}$&$^+_-$ & $^{0.097}_{0.067}$ & $0.639$ & $^+_-$ & $^{0.026}_{0.026}$&$^+_-$ & $^{0.069}_{0.054}$ & $0.588$ & $^+_-$ & $^{0.027}_{0.027}$&$^+_-$ & $^{0.066}_{0.050}$ & $0.433$ & $^+_-$ & $^{0.041}_{0.041}$&$^+_-$ & $^{0.053}_{0.044}$ & \multicolumn{5}{c}{ } \\
$[10000,11000]$ & $0.553$ & $^+_-$ & $^{0.032}_{0.032}$&$^+_-$ & $^{0.063}_{0.037}$ & $0.450$ & $^+_-$ & $^{0.023}_{0.023}$&$^+_-$ & $^{0.045}_{0.031}$ & $0.346$ & $^+_-$ & $^{0.022}_{0.022}$&$^+_-$ & $^{0.043}_{0.023}$ & $0.192$ & $^+_-$ & $^{0.033}_{0.033}$&$^+_-$ & $^{0.025}_{0.010}$ & \multicolumn{5}{c}{ } \\
$[11000,12000]$ & $0.331$ & $^+_-$ & $^{0.025}_{0.025}$&$^+_-$ & $^{0.038}_{0.023}$ & $0.289$ & $^+_-$ & $^{0.018}_{0.018}$&$^+_-$ & $^{0.031}_{0.017}$ & $0.240$ & $^+_-$ & $^{0.018}_{0.018}$&$^+_-$ & $^{0.028}_{0.012}$ & \multicolumn{5}{c}{ } & \multicolumn{5}{c}{ } \\
$[12000,13000]$ & $0.247$ & $^+_-$ & $^{0.020}_{0.020}$&$^+_-$ & $^{0.030}_{0.016}$ & $0.164$ & $^+_-$ & $^{0.014}_{0.014}$&$^+_-$ & $^{0.020}_{0.008}$ & $0.169$ & $^+_-$ & $^{0.016}_{0.016}$&$^+_-$ & $^{0.022}_{0.006}$ & \multicolumn{5}{c}{ } & \multicolumn{5}{c}{ } \\
$[13000,14000]$ & $0.137$ & $^+_-$ & $^{0.016}_{0.016}$&$^+_-$ & $^{0.019}_{0.009}$ & $0.159$ & $^+_-$ & $^{0.014}_{0.014}$&$^+_-$ & $^{0.022}_{0.006}$ & $0.1090$ & $^+_-$ & $^{0.0146}_{0.0146}$&$^+_-$ & $^{0.0212}_{0.0010}$ & \multicolumn{5}{c}{ } & \multicolumn{5}{c}{ } \\
$[14000,15000]$ & \multicolumn{5}{c}{ } & \multicolumn{5}{c}{ } & \multicolumn{5}{c}{ } & \multicolumn{5}{c}{ } & \multicolumn{5}{c}{ } \\
\bottomrule\end{tabular}

%% file: tables/differential/Dstp.tex
\hvFloat[
  nonFloat=true,
  objectAngle=90,
  capPos=l,
  capAngle=90,
  capWidth=h]{table}{
    \centering
    \input{./tables/differential/DstToD0pi_D0ToKpi.tex}
}{Differential production cross-sections, $\text{d}^2\sigma/(\text{d}\pT\,\text{d}y)$, in 
$\si{\micro\barn}/(\si{\gevc})$ for prompt $\Dstp + \Dstm$ mesons in bins 
of \pTy.
The first uncertainty is statistical, and the second is the total 
systematic.}{table:differential:Dstp}

%% file: tables/differential/DstToD0pi_D0ToKpi.tex
\renewcommand{\arraystretch}{1.3}
\begin{tabular}{l|r@{\hskip+0.2em}c@{\hskip+0.2em}r@{\hskip+0.2em}c@{\hskip+0.2em}rr@{\hskip+0.2em}c@{\hskip+0.2em}r@{\hskip+0.2em}c@{\hskip+0.2em}rr@{\hskip+0.2em}c@{\hskip+0.2em}r@{\hskip+0.2em}c@{\hskip+0.2em}rr@{\hskip+0.2em}c@{\hskip+0.2em}r@{\hskip+0.2em}c@{\hskip+0.2em}rr@{\hskip+0.2em}c@{\hskip+0.2em}r@{\hskip+0.2em}c@{\hskip+0.2em}r}
\toprule&\multicolumn{25}{c}{$\text{$y$}$}\\
$\text{$p_{\text{T}}$ [\text{MeV}/c]}$ & \multicolumn{5}{c}{$[2,2.5]$} & \multicolumn{5}{c}{$[2.5,3]$} & \multicolumn{5}{c}{$[3,3.5]$} & \multicolumn{5}{c}{$[3.5,4]$} & \multicolumn{5}{c}{$[4,4.5]$} \\
\midrule$[0,1000]$ & \multicolumn{5}{c}{ } & \multicolumn{5}{c}{ } & \multicolumn{5}{c}{ } & \multicolumn{5}{c}{ } & $67$ & $^+_-$ & $^{9}_{9}$&$^+_-$ & $^{20}_{19}$ \\
$[1000,1500]$ & \multicolumn{5}{c}{ } & $124$ & $^+_-$ & $^{5}_{5}$&$^+_-$ & $^{16}_{17}$ & $152$ & $^+_-$ & $^{2}_{2}$&$^+_-$ & $^{13}_{15}$ & $123$ & $^+_-$ & $^{2}_{2}$&$^+_-$ & $^{12}_{12}$ & $127$ & $^+_-$ & $^{4}_{4}$&$^+_-$ & $^{20}_{16}$ \\
$[1500,2000]$ & \multicolumn{5}{c}{ } & $159$ & $^+_-$ & $^{3}_{3}$&$^+_-$ & $^{18}_{16}$ & $148$ & $^+_-$ & $^{1}_{1}$&$^+_-$ & $^{12}_{14}$ & $125$ & $^+_-$ & $^{1}_{1}$&$^+_-$ & $^{11}_{12}$ & $100$ & $^+_-$ & $^{2}_{2}$&$^+_-$ & $^{11}_{\phantom{0}9}$ \\
$[2000,2500]$ & $117$ & $^+_-$ & $^{11}_{11}$&$^+_-$ & $^{27}_{29}$ & $123$ & $^+_-$ & $^{1}_{1}$&$^+_-$ & $^{15}_{10}$ & $118.2$ & $^+_-$ & $^{0.8}_{0.8}$&$^+_-$ & $^{\phantom{0}9.4}_{11.1}$ & $98.4$ & $^+_-$ & $^{0.8}_{0.8}$&$^+_-$ & $^{9.2}_{8.2}$ & $71.3$ & $^+_-$ & $^{1.3}_{1.3}$&$^+_-$ & $^{8.1}_{7.1}$ \\
$[2500,3000]$ & $90$ & $^+_-$ & $^{3}_{3}$&$^+_-$ & $^{13}_{14}$ & $91.7$ & $^+_-$ & $^{0.8}_{0.8}$&$^+_-$ & $^{8.3}_{9.3}$ & $83.2$ & $^+_-$ & $^{0.5}_{0.5}$&$^+_-$ & $^{7.6}_{6.7}$ & $67.3$ & $^+_-$ & $^{0.6}_{0.6}$&$^+_-$ & $^{6.3}_{5.6}$ & $50.9$ & $^+_-$ & $^{1.0}_{1.0}$&$^+_-$ & $^{4.4}_{4.8}$ \\
$[3000,3500]$ & $64.8$ & $^+_-$ & $^{1.6}_{1.6}$&$^+_-$ & $^{7.7}_{9.0}$ & $65.2$ & $^+_-$ & $^{0.5}_{0.5}$&$^+_-$ & $^{5.6}_{6.6}$ & $57.9$ & $^+_-$ & $^{0.4}_{0.4}$&$^+_-$ & $^{5.4}_{4.3}$ & $45.7$ & $^+_-$ & $^{0.4}_{0.4}$&$^+_-$ & $^{4.3}_{3.7}$ & $32.8$ & $^+_-$ & $^{0.7}_{0.7}$&$^+_-$ & $^{4.2}_{3.7}$ \\
$[3500,4000]$ & $46.8$ & $^+_-$ & $^{1.0}_{1.0}$&$^+_-$ & $^{5.3}_{6.1}$ & $48.9$ & $^+_-$ & $^{0.4}_{0.4}$&$^+_-$ & $^{4.2}_{4.9}$ & $39.9$ & $^+_-$ & $^{0.3}_{0.3}$&$^+_-$ & $^{3.7}_{2.9}$ & $33.2$ & $^+_-$ & $^{0.3}_{0.3}$&$^+_-$ & $^{3.1}_{2.7}$ & $23.9$ & $^+_-$ & $^{0.6}_{0.6}$&$^+_-$ & $^{2.1}_{2.1}$ \\
$[4000,5000]$ & $28.2$ & $^+_-$ & $^{0.4}_{0.4}$&$^+_-$ & $^{2.8}_{3.3}$ & $27.8$ & $^+_-$ & $^{0.2}_{0.2}$&$^+_-$ & $^{2.3}_{2.7}$ & $22.9$ & $^+_-$ & $^{0.1}_{0.1}$&$^+_-$ & $^{2.0}_{1.6}$ & $19.1$ & $^+_-$ & $^{0.2}_{0.2}$&$^+_-$ & $^{1.8}_{1.6}$ & $11.7$ & $^+_-$ & $^{0.3}_{0.3}$&$^+_-$ & $^{1.5}_{1.3}$ \\
$[5000,6000]$ & $14.3$ & $^+_-$ & $^{0.2}_{0.2}$&$^+_-$ & $^{1.4}_{1.6}$ & $13.7$ & $^+_-$ & $^{0.1}_{0.1}$&$^+_-$ & $^{1.1}_{1.3}$ & $11.96$ & $^+_-$ & $^{0.09}_{0.09}$&$^+_-$ & $^{1.02}_{0.88}$ & $9.91$ & $^+_-$ & $^{0.12}_{0.12}$&$^+_-$ & $^{0.95}_{0.67}$ & $5.1$ & $^+_-$ & $^{0.3}_{0.3}$&$^+_-$ & $^{1.0}_{0.8}$ \\
$[6000,7000]$ & $8.45$ & $^+_-$ & $^{0.14}_{0.14}$&$^+_-$ & $^{0.79}_{0.94}$ & $7.83$ & $^+_-$ & $^{0.07}_{0.07}$&$^+_-$ & $^{0.64}_{0.73}$ & $6.33$ & $^+_-$ & $^{0.06}_{0.06}$&$^+_-$ & $^{0.50}_{0.49}$ & $4.44$ & $^+_-$ & $^{0.08}_{0.08}$&$^+_-$ & $^{0.35}_{0.35}$ & $1.93$ & $^+_-$ & $^{0.32}_{0.32}$&$^+_-$ & $^{0.40}_{0.36}$ \\
$[7000,8000]$ & $4.96$ & $^+_-$ & $^{0.09}_{0.09}$&$^+_-$ & $^{0.46}_{0.54}$ & $4.25$ & $^+_-$ & $^{0.05}_{0.05}$&$^+_-$ & $^{0.37}_{0.41}$ & $3.70$ & $^+_-$ & $^{0.05}_{0.05}$&$^+_-$ & $^{0.32}_{0.27}$ & $2.68$ & $^+_-$ & $^{0.07}_{0.07}$&$^+_-$ & $^{0.30}_{0.27}$ & \multicolumn{5}{c}{ } \\
$[8000,9000]$ & $2.91$ & $^+_-$ & $^{0.06}_{0.06}$&$^+_-$ & $^{0.28}_{0.32}$ & $2.53$ & $^+_-$ & $^{0.04}_{0.04}$&$^+_-$ & $^{0.22}_{0.23}$ & $2.08$ & $^+_-$ & $^{0.04}_{0.04}$&$^+_-$ & $^{0.21}_{0.15}$ & $1.19$ & $^+_-$ & $^{0.06}_{0.06}$&$^+_-$ & $^{0.21}_{0.15}$ & \multicolumn{5}{c}{ } \\
$[9000,10000]$ & $1.26$ & $^+_-$ & $^{0.03}_{0.03}$&$^+_-$ & $^{0.12}_{0.13}$ & $1.59$ & $^+_-$ & $^{0.03}_{0.03}$&$^+_-$ & $^{0.14}_{0.14}$ & $1.23$ & $^+_-$ & $^{0.03}_{0.03}$&$^+_-$ & $^{0.10}_{0.09}$ & $0.92$ & $^+_-$ & $^{0.08}_{0.08}$&$^+_-$ & $^{0.15}_{0.12}$ & \multicolumn{5}{c}{ } \\
$[10000,11000]$ & $1.17$ & $^+_-$ & $^{0.04}_{0.04}$&$^+_-$ & $^{0.10}_{0.11}$ & $1.034$ & $^+_-$ & $^{0.025}_{0.025}$&$^+_-$ & $^{0.089}_{0.083}$ & $0.830$ & $^+_-$ & $^{0.025}_{0.025}$&$^+_-$ & $^{0.066}_{0.048}$ & $0.382$ & $^+_-$ & $^{0.067}_{0.067}$&$^+_-$ & $^{0.079}_{0.067}$ & \multicolumn{5}{c}{ } \\
$[11000,12000]$ & $0.721$ & $^+_-$ & $^{0.027}_{0.027}$&$^+_-$ & $^{0.065}_{0.069}$ & $0.661$ & $^+_-$ & $^{0.020}_{0.020}$&$^+_-$ & $^{0.061}_{0.052}$ & $0.472$ & $^+_-$ & $^{0.021}_{0.021}$&$^+_-$ & $^{0.046}_{0.033}$ & \multicolumn{5}{c}{ } & \multicolumn{5}{c}{ } \\
$[12000,13000]$ & $0.494$ & $^+_-$ & $^{0.023}_{0.023}$&$^+_-$ & $^{0.048}_{0.047}$ & $0.491$ & $^+_-$ & $^{0.017}_{0.017}$&$^+_-$ & $^{0.047}_{0.036}$ & $0.347$ & $^+_-$ & $^{0.021}_{0.021}$&$^+_-$ & $^{0.030}_{0.022}$ & \multicolumn{5}{c}{ } & \multicolumn{5}{c}{ } \\
$[13000,14000]$ & $0.374$ & $^+_-$ & $^{0.020}_{0.020}$&$^+_-$ & $^{0.043}_{0.037}$ & $0.292$ & $^+_-$ & $^{0.014}_{0.014}$&$^+_-$ & $^{0.031}_{0.021}$ & $0.272$ & $^+_-$ & $^{0.022}_{0.022}$&$^+_-$ & $^{0.033}_{0.020}$ & \multicolumn{5}{c}{ } & \multicolumn{5}{c}{ } \\
$[14000,15000]$ & $0.218$ & $^+_-$ & $^{0.015}_{0.015}$&$^+_-$ & $^{0.034}_{0.024}$ & $0.248$ & $^+_-$ & $^{0.014}_{0.014}$&$^+_-$ & $^{0.026}_{0.016}$ & $0.110$ & $^+_-$ & $^{0.016}_{0.016}$&$^+_-$ & $^{0.023}_{0.017}$ & \multicolumn{5}{c}{ } & \multicolumn{5}{c}{ } \\
\bottomrule\end{tabular}

%% file: appendix_ratios.tex
\section{Cross-section ratios at different energies}
\label{app:ratios}

Tables~\ref{table:differential_ratio:D0}--\ref{table:differential_ratio:Dstp} give the numerical results of the cross-section ratios between $\sqrt{s}=13$ and \SI{7}{\TeV}.
\vspace{1cm}

\input{tables/differential_ratios/D0.tex}

\input{tables/differential_ratios/Dp.tex}

\input{tables/differential_ratios/Dsp.tex}

\input{tables/differential_ratios/Dstp.tex}

%% file: tables/differential_ratios/D0.tex
\hvFloat[
  nonFloat=true,
  objectAngle=90,
  capPos=l,
  capAngle=90,
  capWidth=h]{table}{
    \centering
    \input{./tables/differential_ratios/D0ToKpi.tex}
}{The ratios of differential production cross-sections, $R_{13/7}$, for 
prompt $\Dz + \Dzbar$ mesons in bins of \pTy.
The first uncertainty is statistical, and the second is the total 
systematic.}{table:differential_ratio:D0}

%% file: tables/differential_ratios/D0ToKpi.tex
\renewcommand{\arraystretch}{1.3}
\begin{tabular}{l|r@{\hskip+0.2em}c@{\hskip+0.2em}r@{\hskip+0.2em}c@{\hskip+0.2em}rr@{\hskip+0.2em}c@{\hskip+0.2em}r@{\hskip+0.2em}c@{\hskip+0.2em}rr@{\hskip+0.2em}c@{\hskip+0.2em}r@{\hskip+0.2em}c@{\hskip+0.2em}rr@{\hskip+0.2em}c@{\hskip+0.2em}r@{\hskip+0.2em}c@{\hskip+0.2em}rr@{\hskip+0.2em}c@{\hskip+0.2em}r@{\hskip+0.2em}c@{\hskip+0.2em}r}
\toprule&\multicolumn{25}{c}{$\text{$y$}$}\\
$\text{$p_{\text{T}}$ [\text{MeV}/c]}$ & \multicolumn{5}{c}{$[2,2.5]$} & \multicolumn{5}{c}{$[2.5,3]$} & \multicolumn{5}{c}{$[3,3.5]$} & \multicolumn{5}{c}{$[3.5,4]$} & \multicolumn{5}{c}{$[4,4.5]$} \\
\midrule$[0,1000]$ & $1.12$ & $^+_-$ & $^{0.06}_{0.05}$&$^+_-$ & $^{0.19}_{0.15}$ & $1.44$ & $^+_-$ & $^{0.05}_{0.05}$&$^+_-$ & $^{0.18}_{0.17}$ & $1.54$ & $^+_-$ & $^{0.07}_{0.06}$&$^+_-$ & $^{0.18}_{0.14}$ & $1.59$ & $^+_-$ & $^{0.09}_{0.08}$&$^+_-$ & $^{0.18}_{0.14}$ & $1.80$ & $^+_-$ & $^{0.19}_{0.15}$&$^+_-$ & $^{0.24}_{0.18}$ \\
$[1000,2000]$ & $1.40$ & $^+_-$ & $^{0.06}_{0.05}$&$^+_-$ & $^{0.21}_{0.17}$ & $1.48$ & $^+_-$ & $^{0.04}_{0.04}$&$^+_-$ & $^{0.17}_{0.15}$ & $1.61$ & $^+_-$ & $^{0.05}_{0.05}$&$^+_-$ & $^{0.17}_{0.15}$ & $1.67$ & $^+_-$ & $^{0.07}_{0.06}$&$^+_-$ & $^{0.18}_{0.15}$ & $1.85$ & $^+_-$ & $^{0.13}_{0.11}$&$^+_-$ & $^{0.22}_{0.17}$ \\
$[2000,3000]$ & $1.53$ & $^+_-$ & $^{0.06}_{0.05}$&$^+_-$ & $^{0.18}_{0.18}$ & $1.65$ & $^+_-$ & $^{0.04}_{0.04}$&$^+_-$ & $^{0.19}_{0.15}$ & $1.83$ & $^+_-$ & $^{0.05}_{0.05}$&$^+_-$ & $^{0.20}_{0.16}$ & $1.86$ & $^+_-$ & $^{0.07}_{0.06}$&$^+_-$ & $^{0.21}_{0.16}$ & $2.00$ & $^+_-$ & $^{0.14}_{0.12}$&$^+_-$ & $^{0.25}_{0.20}$ \\
$[3000,4000]$ & $1.62$ & $^+_-$ & $^{0.07}_{0.07}$&$^+_-$ & $^{0.20}_{0.18}$ & $1.92$ & $^+_-$ & $^{0.06}_{0.06}$&$^+_-$ & $^{0.21}_{0.17}$ & $1.84$ & $^+_-$ & $^{0.06}_{0.06}$&$^+_-$ & $^{0.20}_{0.16}$ & $2.25$ & $^+_-$ & $^{0.10}_{0.10}$&$^+_-$ & $^{0.29}_{0.22}$ & $2.32$ & $^+_-$ & $^{0.21}_{0.18}$&$^+_-$ & $^{0.35}_{0.26}$ \\
$[4000,5000]$ & $1.79$ & $^+_-$ & $^{0.10}_{0.09}$&$^+_-$ & $^{0.24}_{0.18}$ & $2.00$ & $^+_-$ & $^{0.09}_{0.08}$&$^+_-$ & $^{0.25}_{0.19}$ & $1.90$ & $^+_-$ & $^{0.09}_{0.08}$&$^+_-$ & $^{0.21}_{0.16}$ & $2.32$ & $^+_-$ & $^{0.16}_{0.14}$&$^+_-$ & $^{0.28}_{0.23}$ & $2.30$ & $^+_-$ & $^{0.41}_{0.31}$&$^+_-$ & $^{0.51}_{0.31}$ \\
$[5000,6000]$ & $1.90$ & $^+_-$ & $^{0.14}_{0.13}$&$^+_-$ & $^{0.26}_{0.19}$ & $1.78$ & $^+_-$ & $^{0.10}_{0.09}$&$^+_-$ & $^{0.23}_{0.17}$ & $2.18$ & $^+_-$ & $^{0.15}_{0.14}$&$^+_-$ & $^{0.28}_{0.22}$ & $2.14$ & $^+_-$ & $^{0.21}_{0.17}$&$^+_-$ & $^{0.29}_{0.20}$ & $1.69$ & $^+_-$ & $^{0.73}_{0.40}$&$^+_-$ & $^{0.72}_{0.29}$ \\
$[6000,7000]$ & $1.88$ & $^+_-$ & $^{0.22}_{0.18}$&$^+_-$ & $^{0.27}_{0.18}$ & $2.00$ & $^+_-$ & $^{0.19}_{0.16}$&$^+_-$ & $^{0.27}_{0.20}$ & $2.41$ & $^+_-$ & $^{0.30}_{0.24}$&$^+_-$ & $^{0.32}_{0.25}$ & $2.19$ & $^+_-$ & $^{0.38}_{0.28}$&$^+_-$ & $^{0.37}_{0.25}$ & \multicolumn{5}{c}{ } \\
$[7000,8000]$ & $2.17$ & $^+_-$ & $^{0.48}_{0.33}$&$^+_-$ & $^{0.39}_{0.22}$ & $2.01$ & $^+_-$ & $^{0.36}_{0.26}$&$^+_-$ & $^{0.32}_{0.19}$ & $3.15$ & $^+_-$ & $^{0.87}_{0.56}$&$^+_-$ & $^{0.51}_{0.33}$ & $2.16$ & $^+_-$ & $^{0.95}_{0.51}$&$^+_-$ & $^{0.67}_{0.30}$ & \multicolumn{5}{c}{ } \\
\bottomrule\end{tabular}

%% file: tables/differential_ratios/Dp.tex
\hvFloat[
  floatPos=p,
  objectAngle=90,
  capPos=l,
  capAngle=90,
  capWidth=h]{table}{
    \centering
    \input{./tables/differential_ratios/DpToKpipi.tex}
}{The ratios of differential production cross-sections, $R_{13/7}$, for 
prompt $\Dp + \Dm$ mesons in bins of \pTy.
The first uncertainty is statistical, and the second is the total 
systematic.}{table:differential_ratio:Dp}

%% file: tables/differential_ratios/DpToKpipi.tex
\renewcommand{\arraystretch}{1.3}
\begin{tabular}{l|r@{\hskip+0.2em}c@{\hskip+0.2em}r@{\hskip+0.2em}c@{\hskip+0.2em}rr@{\hskip+0.2em}c@{\hskip+0.2em}r@{\hskip+0.2em}c@{\hskip+0.2em}rr@{\hskip+0.2em}c@{\hskip+0.2em}r@{\hskip+0.2em}c@{\hskip+0.2em}rr@{\hskip+0.2em}c@{\hskip+0.2em}r@{\hskip+0.2em}c@{\hskip+0.2em}rr@{\hskip+0.2em}c@{\hskip+0.2em}r@{\hskip+0.2em}c@{\hskip+0.2em}r}
\toprule&\multicolumn{25}{c}{$\text{$y$}$}\\
$\text{$p_{\text{T}}$ [\text{MeV}/c]}$ & \multicolumn{5}{c}{$[2,2.5]$} & \multicolumn{5}{c}{$[2.5,3]$} & \multicolumn{5}{c}{$[3,3.5]$} & \multicolumn{5}{c}{$[3.5,4]$} & \multicolumn{5}{c}{$[4,4.5]$} \\
\midrule$[0,1000]$ & \multicolumn{5}{c}{ } & $1.85$ & $^+_-$ & $^{0.16}_{0.14}$&$^+_-$ & $^{0.66}_{0.43}$ & $1.57$ & $^+_-$ & $^{0.10}_{0.09}$&$^+_-$ & $^{0.46}_{0.31}$ & $1.66$ & $^+_-$ & $^{0.13}_{0.11}$&$^+_-$ & $^{0.48}_{0.33}$ & $1.50$ & $^+_-$ & $^{0.23}_{0.19}$&$^+_-$ & $^{0.62}_{0.38}$ \\
$[1000,2000]$ & $1.48$ & $^+_-$ & $^{0.21}_{0.16}$&$^+_-$ & $^{0.52}_{0.32}$ & $1.51$ & $^+_-$ & $^{0.07}_{0.06}$&$^+_-$ & $^{0.43}_{0.26}$ & $1.66$ & $^+_-$ & $^{0.06}_{0.05}$&$^+_-$ & $^{0.36}_{0.27}$ & $1.61$ & $^+_-$ & $^{0.06}_{0.06}$&$^+_-$ & $^{0.38}_{0.26}$ & $1.86$ & $^+_-$ & $^{0.15}_{0.13}$&$^+_-$ & $^{0.53}_{0.33}$ \\
$[2000,3000]$ & $1.65$ & $^+_-$ & $^{0.14}_{0.12}$&$^+_-$ & $^{0.42}_{0.30}$ & $1.73$ & $^+_-$ & $^{0.06}_{0.06}$&$^+_-$ & $^{0.38}_{0.27}$ & $1.71$ & $^+_-$ & $^{0.05}_{0.05}$&$^+_-$ & $^{0.32}_{0.25}$ & $1.83$ & $^+_-$ & $^{0.07}_{0.06}$&$^+_-$ & $^{0.37}_{0.26}$ & $2.09$ & $^+_-$ & $^{0.17}_{0.14}$&$^+_-$ & $^{0.64}_{0.39}$ \\
$[3000,4000]$ & $1.75$ & $^+_-$ & $^{0.13}_{0.11}$&$^+_-$ & $^{0.40}_{0.31}$ & $1.80$ & $^+_-$ & $^{0.07}_{0.06}$&$^+_-$ & $^{0.33}_{0.27}$ & $1.87$ & $^+_-$ & $^{0.07}_{0.06}$&$^+_-$ & $^{0.32}_{0.24}$ & $2.04$ & $^+_-$ & $^{0.09}_{0.09}$&$^+_-$ & $^{0.40}_{0.28}$ & $2.47$ & $^+_-$ & $^{0.26}_{0.21}$&$^+_-$ & $^{0.74}_{0.44}$ \\
$[4000,5000]$ & $2.01$ & $^+_-$ & $^{0.16}_{0.14}$&$^+_-$ & $^{0.42}_{0.34}$ & $1.90$ & $^+_-$ & $^{0.09}_{0.08}$&$^+_-$ & $^{0.32}_{0.26}$ & $1.98$ & $^+_-$ & $^{0.10}_{0.09}$&$^+_-$ & $^{0.35}_{0.24}$ & $2.34$ & $^+_-$ & $^{0.16}_{0.14}$&$^+_-$ & $^{0.52}_{0.33}$ & $3.00$ & $^+_-$ & $^{0.48}_{0.37}$&$^+_-$ & $^{0.93}_{0.54}$ \\
$[5000,6000]$ & $2.24$ & $^+_-$ & $^{0.23}_{0.19}$&$^+_-$ & $^{0.52}_{0.40}$ & $2.07$ & $^+_-$ & $^{0.13}_{0.12}$&$^+_-$ & $^{0.38}_{0.28}$ & $2.17$ & $^+_-$ & $^{0.15}_{0.13}$&$^+_-$ & $^{0.38}_{0.26}$ & $2.48$ & $^+_-$ & $^{0.23}_{0.19}$&$^+_-$ & $^{0.57}_{0.36}$ & $5.4$ & $^+_-$ & $^{2.1}_{1.2}$&$^+_-$ & $^{2.2}_{0.9}$ \\
$[6000,7000]$ & $2.16$ & $^+_-$ & $^{0.27}_{0.21}$&$^+_-$ & $^{0.49}_{0.35}$ & $2.14$ & $^+_-$ & $^{0.19}_{0.16}$&$^+_-$ & $^{0.38}_{0.27}$ & $1.74$ & $^+_-$ & $^{0.15}_{0.13}$&$^+_-$ & $^{0.48}_{0.29}$ & $3.06$ & $^+_-$ & $^{0.43}_{0.34}$&$^+_-$ & $^{0.88}_{0.53}$ & \multicolumn{5}{c}{ } \\
$[7000,8000]$ & $2.04$ & $^+_-$ & $^{0.33}_{0.25}$&$^+_-$ & $^{0.53}_{0.33}$ & $2.26$ & $^+_-$ & $^{0.30}_{0.24}$&$^+_-$ & $^{0.45}_{0.28}$ & $2.51$ & $^+_-$ & $^{0.37}_{0.28}$&$^+_-$ & $^{0.76}_{0.40}$ & $4.0$ & $^+_-$ & $^{1.1}_{0.7}$&$^+_-$ & $^{1.4}_{0.7}$ & \multicolumn{5}{c}{ } \\
\bottomrule\end{tabular}

%% file: tables/differential_ratios/Dsp.tex
\hvFloat[
  floatPos=p,
  objectAngle=90,
  capPos=l,
  capAngle=90,
  capWidth=h]{table}{
    \centering
    \input{./tables/differential_ratios/DsTophipi.tex}
}{The ratios of differential production cross-sections, $R_{13/7}$, for 
prompt $\Dsp + \Dsm$ mesons in bins of \pTy.
The first uncertainty is statistical, and the second is the total 
systematic.}{table:differential_ratio:Dsp}

%% file: tables/differential_ratios/DsTophipi.tex
\renewcommand{\arraystretch}{1.3}
\begin{tabular}{l|r@{\hskip+0.2em}c@{\hskip+0.2em}r@{\hskip+0.2em}c@{\hskip+0.2em}rr@{\hskip+0.2em}c@{\hskip+0.2em}r@{\hskip+0.2em}c@{\hskip+0.2em}rr@{\hskip+0.2em}c@{\hskip+0.2em}r@{\hskip+0.2em}c@{\hskip+0.2em}rr@{\hskip+0.2em}c@{\hskip+0.2em}r@{\hskip+0.2em}c@{\hskip+0.2em}rr@{\hskip+0.2em}c@{\hskip+0.2em}r@{\hskip+0.2em}c@{\hskip+0.2em}r}
\toprule&\multicolumn{25}{c}{$\text{$y$}$}\\
$\text{$p_{\text{T}}$ [\text{MeV}/c]}$ & \multicolumn{5}{c}{$[2,2.5]$} & \multicolumn{5}{c}{$[2.5,3]$} & \multicolumn{5}{c}{$[3,3.5]$} & \multicolumn{5}{c}{$[3.5,4]$} & \multicolumn{5}{c}{$[4,4.5]$} \\
\midrule
$[1000,2000]$ & $0.86$ & $^+_-$ & $^{0.47}_{0.23}$&$^+_-$ & $^{0.61}_{0.24}$ & $1.51$ & $^+_-$ & $^{0.28}_{0.21}$&$^+_-$ & $^{0.63}_{0.31}$ & $2.74$ & $^+_-$ & $^{0.60}_{0.41}$&$^+_-$ & $^{0.97}_{0.53}$ & $2.6$ & $^+_-$ & $^{1.0}_{0.6}$&$^+_-$ & $^{1.3}_{0.5}$ & \multicolumn{5}{c}{ } \\
$[2000,3000]$ & $3.6$ & $^+_-$ & $^{1.6}_{0.8}$&$^+_-$ & $^{1.9}_{0.8}$ & $2.89$ & $^+_-$ & $^{0.47}_{0.35}$&$^+_-$ & $^{0.76}_{0.46}$ & $2.28$ & $^+_-$ & $^{0.29}_{0.24}$&$^+_-$ & $^{0.56}_{0.35}$ & $2.76$ & $^+_-$ & $^{0.60}_{0.42}$&$^+_-$ & $^{0.79}_{0.43}$ & $2.1$ & $^+_-$ & $^{1.3}_{0.6}$&$^+_-$ & $^{1.0}_{0.4}$ \\
$[3000,4000]$ & $3.6$ & $^+_-$ & $^{1.4}_{0.8}$&$^+_-$ & $^{1.5}_{0.7}$ & $2.29$ & $^+_-$ & $^{0.34}_{0.27}$&$^+_-$ & $^{0.55}_{0.38}$ & $2.41$ & $^+_-$ & $^{0.40}_{0.30}$&$^+_-$ & $^{0.69}_{0.37}$ & $2.55$ & $^+_-$ & $^{0.55}_{0.39}$&$^+_-$ & $^{0.73}_{0.38}$ & $2.9$ & $^+_-$ & $^{2.2}_{0.9}$&$^+_-$ & $^{1.3}_{0.5}$ \\
$[4000,5000]$ & $3.5$ & $^+_-$ & $^{1.3}_{0.8}$&$^+_-$ & $^{1.3}_{0.6}$ & $3.08$ & $^+_-$ & $^{0.63}_{0.45}$&$^+_-$ & $^{0.77}_{0.44}$ & $3.61$ & $^+_-$ & $^{0.87}_{0.59}$&$^+_-$ & $^{0.94}_{0.48}$ & $2.96$ & $^+_-$ & $^{0.93}_{0.57}$&$^+_-$ & $^{0.84}_{0.43}$ & \multicolumn{5}{c}{ } \\
$[5000,6000]$ & $4.2$ & $^+_-$ & $^{2.5}_{1.1}$&$^+_-$ & $^{1.8}_{0.8}$ & $3.21$ & $^+_-$ & $^{0.88}_{0.58}$&$^+_-$ & $^{0.98}_{0.44}$ & $3.3$ & $^+_-$ & $^{1.0}_{0.7}$&$^+_-$ & $^{0.9}_{0.4}$ & $3.4$ & $^+_-$ & $^{1.6}_{0.8}$&$^+_-$ & $^{1.5}_{0.5}$ & \multicolumn{5}{c}{ } \\
$[6000,7000]$ & $2.8$ & $^+_-$ & $^{1.4}_{0.7}$&$^+_-$ & $^{1.2}_{0.5}$ & $3.8$ & $^+_-$ & $^{1.7}_{0.9}$&$^+_-$ & $^{1.2}_{0.5}$ & $2.44$ & $^+_-$ & $^{1.00}_{0.55}$&$^+_-$ & $^{0.78}_{0.33}$ & \multicolumn{5}{c}{ } & \multicolumn{5}{c}{ } \\
$[7000,8000]$ & $2.0$ & $^+_-$ & $^{1.1}_{0.5}$&$^+_-$ & $^{1.1}_{0.4}$ & \multicolumn{5}{c}{ } & $3.4$ & $^+_-$ & $^{2.5}_{1.0}$&$^+_-$ & $^{1.4}_{0.5}$ & \multicolumn{5}{c}{ } & \multicolumn{5}{c}{ } \\
\bottomrule\end{tabular}

%% file: tables/differential_ratios/Dstp.tex
\hvFloat[
  floatPos=p,
  objectAngle=90,
  capPos=l,
  capAngle=90,
  capWidth=h]{table}{
    \centering
    \input{./tables/differential_ratios/DstToD0pi_D0ToKpi.tex}
}{The ratios of differential production cross-sections, $R_{13/7}$, for 
prompt $\Dstp + \Dstm$ mesons in bins of \pTy.
The first uncertainty is statistical, and the second is the total 
systematic.}{table:differential_ratio:Dstp}

%% file: tables/differential_ratios/DstToD0pi_D0ToKpi.tex
\renewcommand{\arraystretch}{1.3}
\begin{tabular}{l|r@{\hskip+0.2em}c@{\hskip+0.2em}r@{\hskip+0.2em}c@{\hskip+0.2em}rr@{\hskip+0.2em}c@{\hskip+0.2em}r@{\hskip+0.2em}c@{\hskip+0.2em}rr@{\hskip+0.2em}c@{\hskip+0.2em}r@{\hskip+0.2em}c@{\hskip+0.2em}rr@{\hskip+0.2em}c@{\hskip+0.2em}r@{\hskip+0.2em}c@{\hskip+0.2em}rr@{\hskip+0.2em}c@{\hskip+0.2em}r@{\hskip+0.2em}c@{\hskip+0.2em}r}
\toprule&\multicolumn{25}{c}{$\text{$y$}$}\\
$\text{$p_{\text{T}}$ [\text{MeV}/c]}$ & \multicolumn{5}{c}{$[2,2.5]$} & \multicolumn{5}{c}{$[2.5,3]$} & \multicolumn{5}{c}{$[3,3.5]$} & \multicolumn{5}{c}{$[3.5,4]$} & \multicolumn{5}{c}{$[4,4.5]$} \\
\midrule$[0,1000]$ & \multicolumn{5}{c}{ } & \multicolumn{5}{c}{ } & \multicolumn{5}{c}{ } & \multicolumn{5}{c}{ } & $0.71$ & $^+_-$ & $^{0.30}_{0.18}$&$^+_-$ & $^{0.33}_{0.20}$ \\
$[1000,2000]$ & \multicolumn{5}{c}{ } & $1.13$ & $^+_-$ & $^{0.18}_{0.13}$&$^+_-$ & $^{0.24}_{0.16}$ & $1.53$ & $^+_-$ & $^{0.10}_{0.09}$&$^+_-$ & $^{0.25}_{0.21}$ & $1.58$ & $^+_-$ & $^{0.14}_{0.12}$&$^+_-$ & $^{0.27}_{0.22}$ & $1.75$ & $^+_-$ & $^{0.28}_{0.21}$&$^+_-$ & $^{0.39}_{0.22}$ \\
$[2000,3000]$ & \multicolumn{5}{c}{ } & $1.75$ & $^+_-$ & $^{0.18}_{0.15}$&$^+_-$ & $^{0.34}_{0.22}$ & $2.02$ & $^+_-$ & $^{0.13}_{0.12}$&$^+_-$ & $^{0.33}_{0.26}$ & $1.72$ & $^+_-$ & $^{0.14}_{0.12}$&$^+_-$ & $^{0.30}_{0.21}$ & $1.65$ & $^+_-$ & $^{0.24}_{0.19}$&$^+_-$ & $^{0.33}_{0.22}$ \\
$[3000,4000]$ & $1.82$ & $^+_-$ & $^{0.46}_{0.31}$&$^+_-$ & $^{0.40}_{0.28}$ & $1.67$ & $^+_-$ & $^{0.14}_{0.12}$&$^+_-$ & $^{0.28}_{0.23}$ & $1.76$ & $^+_-$ & $^{0.13}_{0.11}$&$^+_-$ & $^{0.32}_{0.20}$ & $1.89$ & $^+_-$ & $^{0.18}_{0.15}$&$^+_-$ & $^{0.37}_{0.24}$ & $2.77$ & $^+_-$ & $^{0.72}_{0.48}$&$^+_-$ & $^{0.65}_{0.35}$ \\
$[4000,5000]$ & $1.43$ & $^+_-$ & $^{0.28}_{0.20}$&$^+_-$ & $^{0.29}_{0.22}$ & $2.21$ & $^+_-$ & $^{0.26}_{0.21}$&$^+_-$ & $^{0.39}_{0.31}$ & $1.85$ & $^+_-$ & $^{0.19}_{0.16}$&$^+_-$ & $^{0.34}_{0.22}$ & $2.11$ & $^+_-$ & $^{0.28}_{0.22}$&$^+_-$ & $^{0.42}_{0.27}$ & $1.71$ & $^+_-$ & $^{0.74}_{0.39}$&$^+_-$ & $^{0.53}_{0.23}$ \\
$[5000,6000]$ & $1.83$ & $^+_-$ & $^{0.50}_{0.32}$&$^+_-$ & $^{0.37}_{0.26}$ & $1.80$ & $^+_-$ & $^{0.25}_{0.20}$&$^+_-$ & $^{0.32}_{0.25}$ & $1.74$ & $^+_-$ & $^{0.24}_{0.19}$&$^+_-$ & $^{0.35}_{0.22}$ & $2.53$ & $^+_-$ & $^{0.55}_{0.39}$&$^+_-$ & $^{0.59}_{0.27}$ & \multicolumn{5}{c}{ } \\
$[6000,7000]$ & $1.76$ & $^+_-$ & $^{0.57}_{0.35}$&$^+_-$ & $^{0.40}_{0.26}$ & $2.20$ & $^+_-$ & $^{0.48}_{0.34}$&$^+_-$ & $^{0.46}_{0.30}$ & $3.02$ & $^+_-$ & $^{0.94}_{0.58}$&$^+_-$ & $^{0.77}_{0.36}$ & $3.3$ & $^+_-$ & $^{1.8}_{0.8}$&$^+_-$ & $^{0.9}_{0.4}$ & \multicolumn{5}{c}{ } \\
$[7000,8000]$ & $1.42$ & $^+_-$ & $^{0.71}_{0.36}$&$^+_-$ & $^{0.42}_{0.22}$ & $1.93$ & $^+_-$ & $^{0.76}_{0.42}$&$^+_-$ & $^{0.50}_{0.26}$ & \multicolumn{5}{c}{ } & \multicolumn{5}{c}{ } & \multicolumn{5}{c}{ } \\
\bottomrule\end{tabular}

%% file: appendix_mesonRatios.tex
\section{Cross-section ratios for different mesons}
\label{app:MesonRatio}

\begin{figure}[h!]
  \centering
  \includegraphics[width=0.55\textwidth]{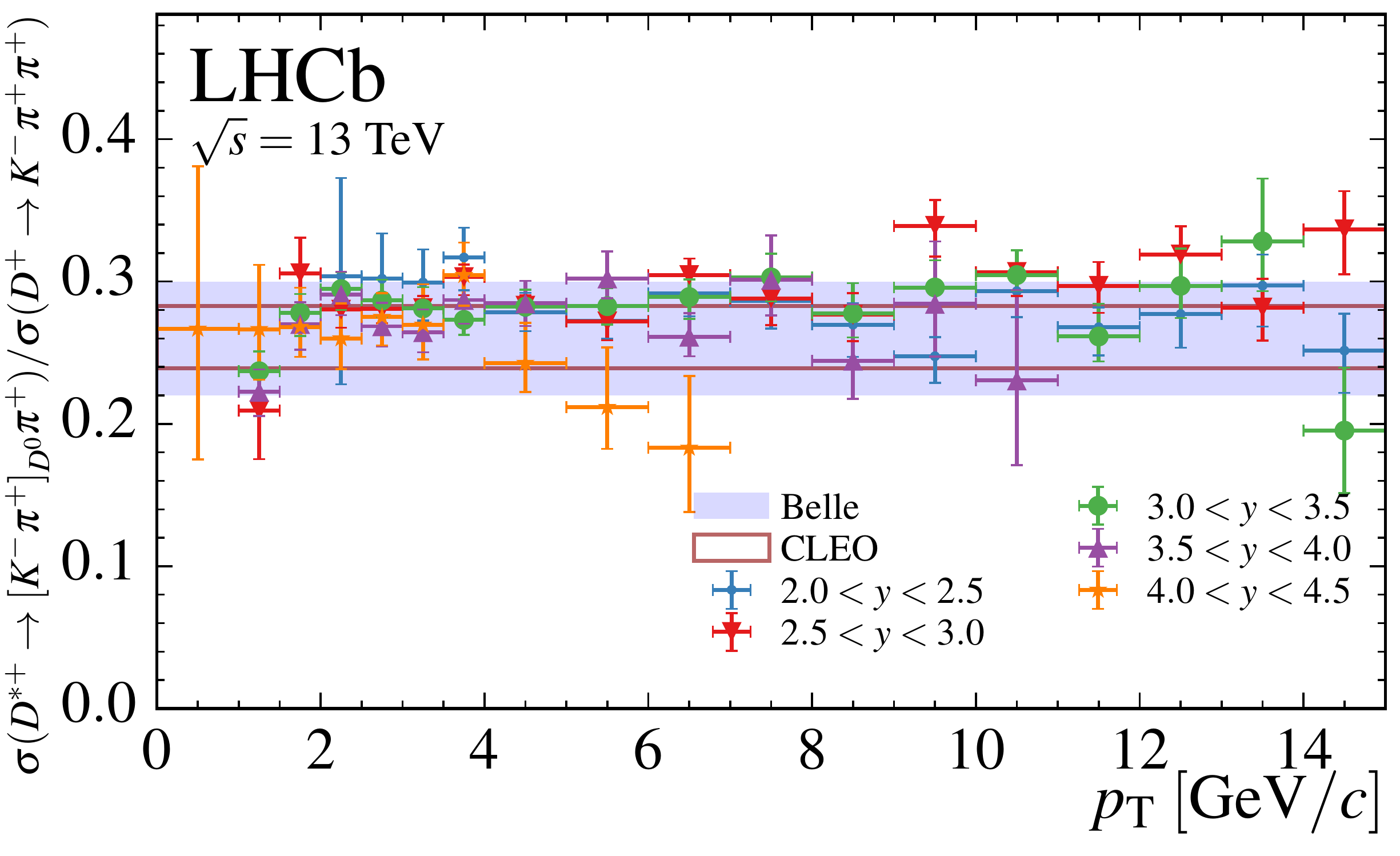}
  \includegraphics[width=0.55\textwidth]{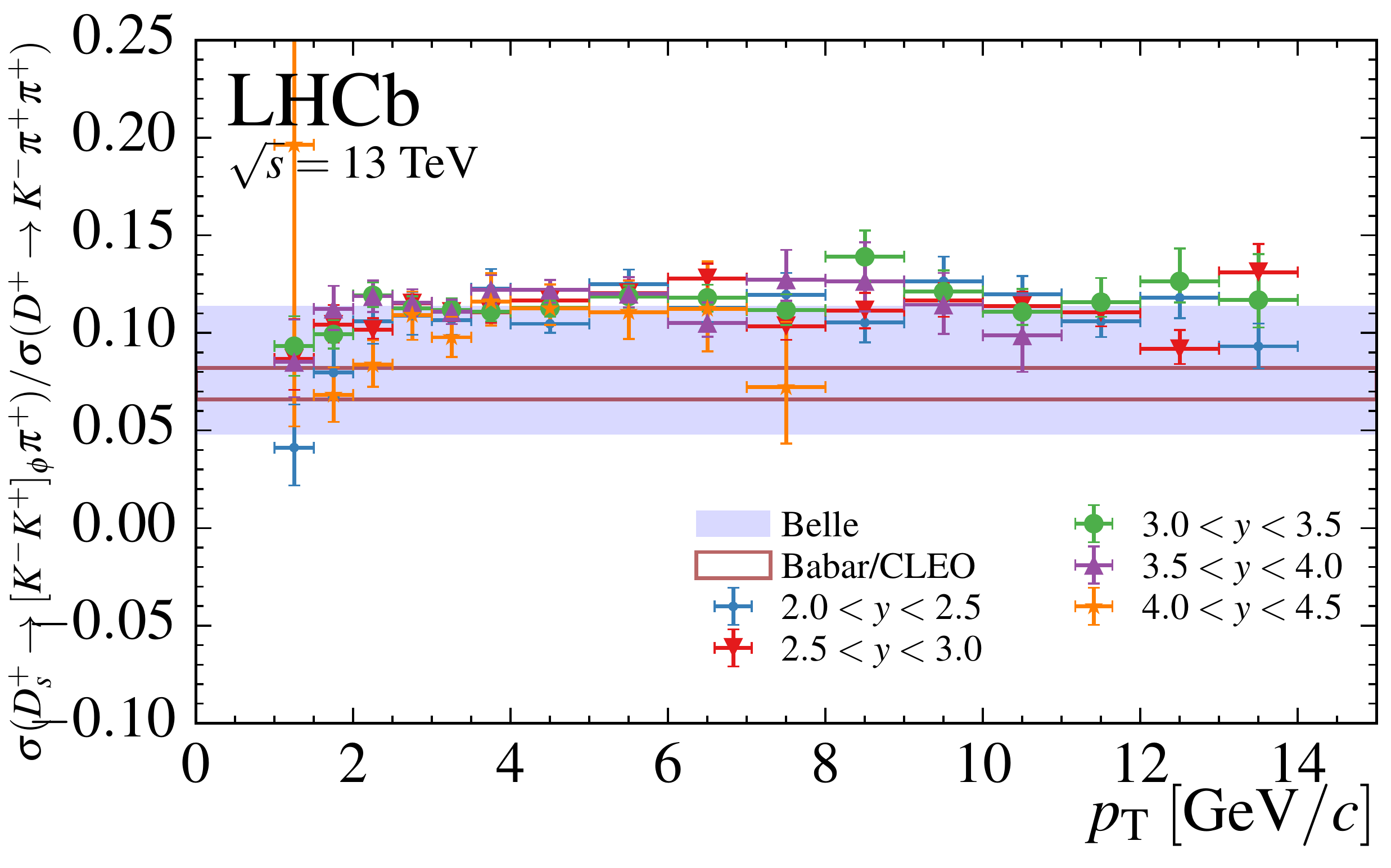}
  \includegraphics[width=0.55\textwidth]{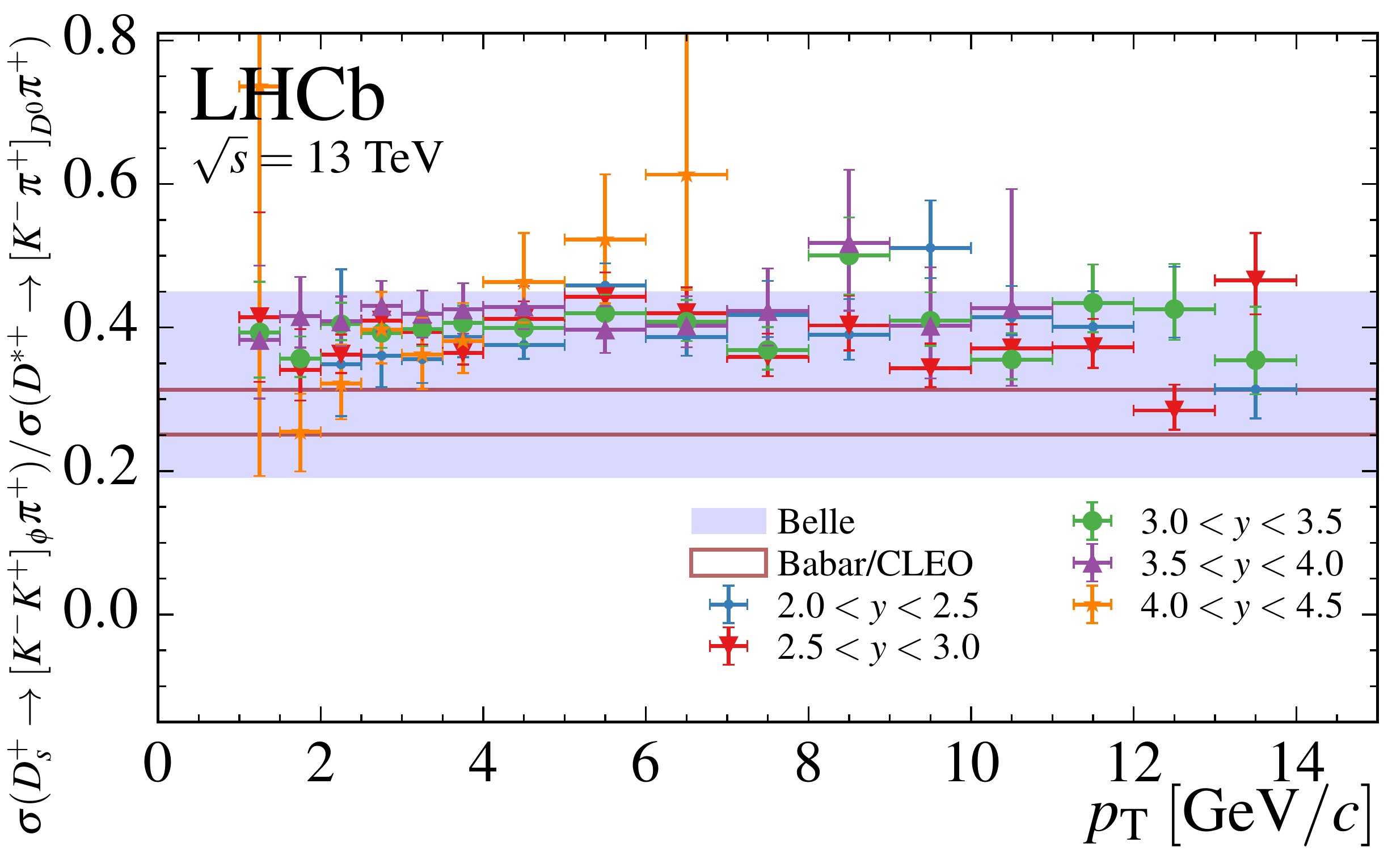}
    \caption{%
      Ratios of cross-section-times-branching-fraction measurements of (top) 
      \Dstarp, and (middle) \Dsp mesons with respect to \Dp cross-sections, and 
      (bottom) \Dsp over \Dstarp mesons.
      The bands indicate the corresponding ratios computed using measurements 
      from $\ep\en$ collider 
      experiments~\cite{Artuso:2004pj,Seuster:2005tr,Aubert:2002ue}.
      The ratios are given as a function of \pT and different symbols indicate 
      different ranges in \rapidity.
      The notation $\sigma(\decay{D}{f})$ is shorthand for \xsectimesbfrac.
      \label{fig:MesonRatio2}}
\end{figure}

Figure~\ref{fig:MesonRatio2} shows the remaining three ratios of cross-section-times-branching-fraction measurements for different mesons, completing those shown and discussed in Sec.~\ref{sec:ratios}.
The numerical values of these ratios are given in Tables~\ref{table:meson_ratio:DpD0}--\ref{table:meson_ratio:DsDst}.

\clearpage

\hvFloat[
  floatPos=p,
  objectAngle=90,
  capPos=l,
  capAngle=90,
  capWidth=h]{table}{
    \centering
    \input{./tables/mode_ratios/Dp_by_D0_nbc.tex}
}{%
  The ratios of differential production cross-section-times-branching fraction 
  measurements for prompt \Dp and \Dz mesons in bins of \pTy.
  The first uncertainty is statistical, and the second is the total 
  systematic. All values are given in percent.
}{table:meson_ratio:DpD0}

\hvFloat[
  floatPos=p,
  objectAngle=90,
  capPos=l,
  capAngle=90,
  capWidth=h]{table}{
    \centering
    \input{./tables/mode_ratios/Ds_by_D0_nbc.tex}
}{%
  The ratios of differential production cross-section-times-branching-fraction 
  measurements for prompt \Dsp and \Dz mesons in bins of \pTy.
  The first uncertainty is statistical, and the second is the total 
  systematic. All values are given in percent.
}{table:meson_ratio:DsD0}

\hvFloat[
  floatPos=p,
  objectAngle=90,
  capPos=l,
  capAngle=90,
  capWidth=h]{table}{
    \centering
    \input{./tables/mode_ratios/Dst_by_D0_nbc.tex}
}{%
  The ratios of differential production cross-section-times-branching-fraction 
  measurements for prompt \Dstarp and \Dz mesons in bins of \pTy.
  The first uncertainty is statistical, and the second is the total 
  systematic. All values are given in percent.
}{table:meson_ratio:DstD0}

\hvFloat[
  floatPos=p,
  objectAngle=90,
  capPos=l,
  capAngle=90,
  capWidth=h]{table}{
    \centering
    \input{./tables/mode_ratios/Ds_by_Dp_nbc.tex}
}{%
  The ratios of differential production cross-section-times-branching-fraction 
  measurements for prompt \Dsp and \Dp mesons in bins of \pTy.
  The first uncertainty is statistical, and the second is the total 
  systematic. All values are given in percent.
}{table:meson_ratio:DsDp}

\hvFloat[
  floatPos=p,
  objectAngle=90,
  capPos=l,
  capAngle=90,
  capWidth=h]{table}{
    \centering
    \input{./tables/mode_ratios/Dst_by_Dp_nbc.tex}
}{%
  The ratios of differential production cross-section-times-branching-fraction 
  for prompt \Dstarp and \Dp mesons in bins of \pTy.
  The first uncertainty is statistical, and the second is the total 
  systematic. All values are given in percent.
}{table:meson_ratio:DstDp}

\hvFloat[
  floatPos=p,
  objectAngle=90,
  capPos=l,
  capAngle=90,
  capWidth=h]{table}{
    \centering
    \input{./tables/mode_ratios/Ds_by_Dst_nbc.tex}
}{%
  The ratios of differential production cross-section-times-branching-fraction 
  for prompt \Dsp and \Dstarp mesons in bins of \pTy.
  The first uncertainty is statistical, and the second is the total 
  systematic. All values are given in percent.
}{table:meson_ratio:DsDst}

%% file: tables/mode_ratios/Dp_by_D0_nbc.tex
\renewcommand{\arraystretch}{1.3}
\begin{tabular}{l|r@{\hskip+0.2em}c@{\hskip+0.2em}r@{\hskip+0.2em}c@{\hskip+0.2em}rr@{\hskip+0.2em}c@{\hskip+0.2em}r@{\hskip+0.2em}c@{\hskip+0.2em}rr@{\hskip+0.2em}c@{\hskip+0.2em}r@{\hskip+0.2em}c@{\hskip+0.2em}rr@{\hskip+0.2em}c@{\hskip+0.2em}r@{\hskip+0.2em}c@{\hskip+0.2em}rr@{\hskip+0.2em}c@{\hskip+0.2em}r@{\hskip+0.2em}c@{\hskip+0.2em}r}
\toprule&\multicolumn{25}{c}{$\text{$y$}$}\\
$\text{$p_{\text{T}}$ [\text{MeV}/c]}$ & \multicolumn{5}{c}{$[2,2.5]$} & \multicolumn{5}{c}{$[2.5,3]$} & \multicolumn{5}{c}{$[3,3.5]$} & \multicolumn{5}{c}{$[3.5,4]$} & \multicolumn{5}{c}{$[4,4.5]$} \\
\midrule$[0,1000]$ & \multicolumn{5}{c}{ } & $132$ & $^+_-$ & $^{5}_{5}$&$^+_-$ & $^{29}_{23}$ & $89$ & $^+_-$ & $^{2}_{2}$&$^+_-$ & $^{10}_{10}$ & $89$ & $^+_-$ & $^{3}_{3}$&$^+_-$ & $^{10}_{10}$ & $82$ & $^+_-$ & $^{5}_{5}$&$^+_-$ & $^{17}_{16}$ \\
$[1000,1500]$ & $99$ & $^+_-$ & $^{4}_{4}$&$^+_-$ & $^{20}_{19}$ & $87.3$ & $^+_-$ & $^{0.9}_{1.0}$&$^+_-$ & $^{11.8}_{\phantom{0}7.0}$ & $95.5$ & $^+_-$ & $^{0.8}_{0.8}$&$^+_-$ & $^{6.3}_{6.1}$ & $91.1$ & $^+_-$ & $^{1.0}_{1.0}$&$^+_-$ & $^{5.9}_{5.6}$ & $95.8$ & $^+_-$ & $^{2.1}_{2.1}$&$^+_-$ & $^{9.2}_{8.8}$ \\
$[1500,2000]$ & $88.3$ & $^+_-$ & $^{1.5}_{1.5}$&$^+_-$ & $^{9.9}_{9.1}$ & $86.1$ & $^+_-$ & $^{0.5}_{0.5}$&$^+_-$ & $^{8.5}_{6.0}$ & $95.1$ & $^+_-$ & $^{0.5}_{0.5}$&$^+_-$ & $^{4.8}_{5.5}$ & $91.3$ & $^+_-$ & $^{0.6}_{0.6}$&$^+_-$ & $^{4.6}_{4.5}$ & $95.2$ & $^+_-$ & $^{1.4}_{1.4}$&$^+_-$ & $^{6.8}_{6.9}$ \\
$[2000,2500]$ & $88.0$ & $^+_-$ & $^{1.0}_{1.0}$&$^+_-$ & $^{8.8}_{7.0}$ & $94.4$ & $^+_-$ & $^{0.5}_{0.5}$&$^+_-$ & $^{7.4}_{5.7}$ & $92.9$ & $^+_-$ & $^{0.4}_{0.4}$&$^+_-$ & $^{3.7}_{4.6}$ & $88.4$ & $^+_-$ & $^{0.5}_{0.5}$&$^+_-$ & $^{3.7}_{3.9}$ & $98.0$ & $^+_-$ & $^{1.2}_{1.2}$&$^+_-$ & $^{6.0}_{5.1}$ \\
$[2500,3000]$ & $88.8$ & $^+_-$ & $^{0.9}_{0.9}$&$^+_-$ & $^{8.2}_{7.0}$ & $96.6$ & $^+_-$ & $^{0.5}_{0.5}$&$^+_-$ & $^{4.5}_{6.5}$ & $94.1$ & $^+_-$ & $^{0.4}_{0.4}$&$^+_-$ & $^{3.5}_{4.3}$ & $96.9$ & $^+_-$ & $^{0.6}_{0.6}$&$^+_-$ & $^{4.0}_{4.7}$ & $96.6$ & $^+_-$ & $^{1.2}_{1.2}$&$^+_-$ & $^{5.8}_{4.6}$ \\
$[3000,3500]$ & $90.9$ & $^+_-$ & $^{0.9}_{0.9}$&$^+_-$ & $^{8.1}_{7.4}$ & $100.5$ & $^+_-$ & $^{0.5}_{0.5}$&$^+_-$ & $^{4.5}_{6.1}$ & $101.6$ & $^+_-$ & $^{0.5}_{0.5}$&$^+_-$ & $^{3.7}_{4.2}$ & $96.3$ & $^+_-$ & $^{0.6}_{0.6}$&$^+_-$ & $^{4.2}_{4.8}$ & $96.1$ & $^+_-$ & $^{1.4}_{1.3}$&$^+_-$ & $^{6.0}_{5.2}$ \\
$[3500,4000]$ & $91.0$ & $^+_-$ & $^{0.9}_{0.9}$&$^+_-$ & $^{7.2}_{7.5}$ & $98.4$ & $^+_-$ & $^{0.5}_{0.5}$&$^+_-$ & $^{4.6}_{5.3}$ & $105.7$ & $^+_-$ & $^{0.6}_{0.6}$&$^+_-$ & $^{4.2}_{4.1}$ & $94.8$ & $^+_-$ & $^{0.7}_{0.7}$&$^+_-$ & $^{4.3}_{4.0}$ & $91.8$ & $^+_-$ & $^{1.5}_{1.5}$&$^+_-$ & $^{7.2}_{7.2}$ \\
$[4000,5000]$ & $103.2$ & $^+_-$ & $^{0.8}_{0.8}$&$^+_-$ & $^{5.9}_{8.7}$ & $103.0$ & $^+_-$ & $^{0.5}_{0.5}$&$^+_-$ & $^{4.3}_{5.0}$ & $100.9$ & $^+_-$ & $^{0.5}_{0.5}$&$^+_-$ & $^{4.0}_{3.3}$ & $98.9$ & $^+_-$ & $^{0.7}_{0.6}$&$^+_-$ & $^{4.6}_{2.7}$ & $125$ & $^+_-$ & $^{2}_{2}$&$^+_-$ & $^{13}_{14}$ \\
$[5000,6000]$ & $100.2$ & $^+_-$ & $^{0.9}_{0.9}$&$^+_-$ & $^{4.9}_{8.1}$ & $108.2$ & $^+_-$ & $^{0.6}_{0.6}$&$^+_-$ & $^{4.6}_{4.8}$ & $105.6$ & $^+_-$ & $^{0.7}_{0.7}$&$^+_-$ & $^{4.7}_{3.3}$ & $106.6$ & $^+_-$ & $^{1.0}_{1.0}$&$^+_-$ & $^{4.6}_{4.2}$ & $149$ & $^+_-$ & $^{6}_{6}$&$^+_-$ & $^{26}_{27}$ \\
$[6000,7000]$ & $106.0$ & $^+_-$ & $^{1.1}_{1.1}$&$^+_-$ & $^{5.5}_{7.6}$ & $101.1$ & $^+_-$ & $^{0.8}_{0.8}$&$^+_-$ & $^{4.5}_{4.3}$ & $103.0$ & $^+_-$ & $^{0.9}_{0.9}$&$^+_-$ & $^{5.1}_{3.3}$ & $112.8$ & $^+_-$ & $^{1.6}_{1.5}$&$^+_-$ & $^{7.7}_{8.3}$ & $210$ & $^+_-$ & $^{40}_{30}$&$^+_-$ & $^{130}_{\phantom{0}90}$ \\
$[7000,8000]$ & $115.5$ & $^+_-$ & $^{1.6}_{1.5}$&$^+_-$ & $^{6.9}_{7.8}$ & $102.8$ & $^+_-$ & $^{1.0}_{1.0}$&$^+_-$ & $^{4.7}_{4.7}$ & $104.8$ & $^+_-$ & $^{1.2}_{1.2}$&$^+_-$ & $^{6.1}_{3.6}$ & $122$ & $^+_-$ & $^{3}_{3}$&$^+_-$ & $^{12}_{14}$ & \multicolumn{5}{c}{ } \\
$[8000,9000]$ & $119.5$ & $^+_-$ & $^{2.0}_{1.9}$&$^+_-$ & $^{8.5}_{7.6}$ & $108.2$ & $^+_-$ & $^{1.4}_{1.4}$&$^+_-$ & $^{5.6}_{5.7}$ & $113.0$ & $^+_-$ & $^{1.7}_{1.7}$&$^+_-$ & $^{7.2}_{5.0}$ & $114$ & $^+_-$ & $^{5}_{5}$&$^+_-$ & $^{17}_{17}$ & \multicolumn{5}{c}{ } \\
$[9000,10000]$ & $87.8$ & $^+_-$ & $^{1.8}_{1.7}$&$^+_-$ & $^{4.8}_{4.2}$ & $91.0$ & $^+_-$ & $^{1.5}_{1.5}$&$^+_-$ & $^{3.5}_{3.8}$ & $102.0$ & $^+_-$ & $^{2.1}_{2.0}$&$^+_-$ & $^{3.2}_{4.4}$ & $143$ & $^+_-$ & $^{12}_{11}$&$^+_-$ & $^{24}_{28}$ & \multicolumn{5}{c}{ } \\
$[10000,11000]$ & $108.4$ & $^+_-$ & $^{2.7}_{2.7}$&$^+_-$ & $^{5.3}_{4.5}$ & $104.7$ & $^+_-$ & $^{2.2}_{2.2}$&$^+_-$ & $^{2.7}_{4.1}$ & $113.7$ & $^+_-$ & $^{3.2}_{3.1}$&$^+_-$ & $^{4.2}_{6.8}$ & $146$ & $^+_-$ & $^{32}_{23}$&$^+_-$ & $^{32}_{32}$ & \multicolumn{5}{c}{ } \\
$[11000,12000]$ & $114.9$ & $^+_-$ & $^{3.6}_{3.4}$&$^+_-$ & $^{5.5}_{5.2}$ & $101.4$ & $^+_-$ & $^{2.7}_{2.6}$&$^+_-$ & $^{2.6}_{3.8}$ & $119.3$ & $^+_-$ & $^{4.9}_{4.4}$&$^+_-$ & $^{4.8}_{7.3}$ & \multicolumn{5}{c}{ } & \multicolumn{5}{c}{ } \\
$[12000,13000]$ & $119.2$ & $^+_-$ & $^{4.6}_{4.5}$&$^+_-$ & $^{7.2}_{6.9}$ & $101.0$ & $^+_-$ & $^{3.4}_{3.2}$&$^+_-$ & $^{2.8}_{4.2}$ & $110.4$ & $^+_-$ & $^{6.3}_{6.0}$&$^+_-$ & $^{6.8}_{6.6}$ & \multicolumn{5}{c}{ } & \multicolumn{5}{c}{ } \\
$[13000,14000]$ & $114.2$ & $^+_-$ & $^{5.3}_{5.0}$&$^+_-$ & $^{6.6}_{6.5}$ & $97.8$ & $^+_-$ & $^{4.0}_{3.9}$&$^+_-$ & $^{4.2}_{4.0}$ & $125$ & $^+_-$ & $^{10}_{\phantom{0}9}$&$^+_-$ & $^{15}_{16}$ & \multicolumn{5}{c}{ } & \multicolumn{5}{c}{ } \\
$[14000,15000]$ & $118.1$ & $^+_-$ & $^{6.9}_{6.3}$&$^+_-$ & $^{7.2}_{5.7}$ & $100.9$ & $^+_-$ & $^{5.2}_{4.9}$&$^+_-$ & $^{5.9}_{3.8}$ & $142$ & $^+_-$ & $^{20}_{16}$&$^+_-$ & $^{25}_{18}$ & \multicolumn{5}{c}{ } & \multicolumn{5}{c}{ } \\
\bottomrule\end{tabular}

%% file: tables/mode_ratios/Ds_by_D0_nbc.tex
\renewcommand{\arraystretch}{1.3}
\begin{tabular}{l|r@{\hskip+0.2em}c@{\hskip+0.2em}r@{\hskip+0.2em}c@{\hskip+0.2em}rr@{\hskip+0.2em}c@{\hskip+0.2em}r@{\hskip+0.2em}c@{\hskip+0.2em}rr@{\hskip+0.2em}c@{\hskip+0.2em}r@{\hskip+0.2em}c@{\hskip+0.2em}rr@{\hskip+0.2em}c@{\hskip+0.2em}r@{\hskip+0.2em}c@{\hskip+0.2em}rr@{\hskip+0.2em}c@{\hskip+0.2em}r@{\hskip+0.2em}c@{\hskip+0.2em}r}
\toprule&\multicolumn{25}{c}{$\text{$y$}$}\\
$\text{$p_{\text{T}}$ [\text{MeV}/c]}$ & \multicolumn{5}{c}{$[2,2.5]$} & \multicolumn{5}{c}{$[2.5,3]$} & \multicolumn{5}{c}{$[3,3.5]$} & \multicolumn{5}{c}{$[3.5,4]$} & \multicolumn{5}{c}{$[4,4.5]$} \\
\midrule$[1000,1500]$ & $4.1$ & $^+_-$ & $^{1.0}_{1.0}$&$^+_-$ & $^{1.8}_{1.6}$ & $7.6$ & $^+_-$ & $^{0.5}_{0.5}$&$^+_-$ & $^{2.3}_{1.4}$ & $8.9$ & $^+_-$ & $^{0.5}_{0.5}$&$^+_-$ & $^{1.4}_{1.4}$ & $7.7$ & $^+_-$ & $^{0.8}_{0.8}$&$^+_-$ & $^{1.9}_{1.5}$ & $19$ & $^+_-$ & $^{6}_{6}$&$^+_-$ & $^{\phantom{0}5}_{12}$ \\
$[1500,2000]$ & $7.0$ & $^+_-$ & $^{0.4}_{0.4}$&$^+_-$ & $^{1.5}_{1.2}$ & $9.0$ & $^+_-$ & $^{0.2}_{0.2}$&$^+_-$ & $^{1.5}_{1.1}$ & $9.43$ & $^+_-$ & $^{0.24}_{0.24}$&$^+_-$ & $^{0.72}_{0.78}$ & $10.3$ & $^+_-$ & $^{0.4}_{0.4}$&$^+_-$ & $^{1.0}_{1.0}$ & $6.5$ & $^+_-$ & $^{0.7}_{0.7}$&$^+_-$ & $^{1.1}_{1.1}$ \\
$[2000,2500]$ & $9.3$ & $^+_-$ & $^{0.3}_{0.3}$&$^+_-$ & $^{1.6}_{1.2}$ & $9.59$ & $^+_-$ & $^{0.17}_{0.17}$&$^+_-$ & $^{0.99}_{0.65}$ & $11.07$ & $^+_-$ & $^{0.19}_{0.19}$&$^+_-$ & $^{0.66}_{0.70}$ & $10.50$ & $^+_-$ & $^{0.25}_{0.25}$&$^+_-$ & $^{0.71}_{0.74}$ & $8.2$ & $^+_-$ & $^{0.5}_{0.5}$&$^+_-$ & $^{1.1}_{1.0}$ \\
$[2500,3000]$ & $9.7$ & $^+_-$ & $^{0.3}_{0.3}$&$^+_-$ & $^{1.4}_{1.1}$ & $11.12$ & $^+_-$ & $^{0.16}_{0.16}$&$^+_-$ & $^{0.58}_{0.90}$ & $10.59$ & $^+_-$ & $^{0.15}_{0.15}$&$^+_-$ & $^{0.61}_{0.50}$ & $11.19$ & $^+_-$ & $^{0.22}_{0.23}$&$^+_-$ & $^{0.65}_{0.73}$ & $10.6$ & $^+_-$ & $^{0.5}_{0.5}$&$^+_-$ & $^{1.1}_{1.1}$ \\
$[3000,3500]$ & $9.7$ & $^+_-$ & $^{0.3}_{0.3}$&$^+_-$ & $^{1.2}_{1.0}$ & $11.13$ & $^+_-$ & $^{0.16}_{0.17}$&$^+_-$ & $^{0.57}_{0.78}$ & $11.36$ & $^+_-$ & $^{0.17}_{0.17}$&$^+_-$ & $^{0.62}_{0.51}$ & $10.66$ & $^+_-$ & $^{0.22}_{0.21}$&$^+_-$ & $^{0.62}_{0.70}$ & $9.40$ & $^+_-$ & $^{0.46}_{0.46}$&$^+_-$ & $^{0.96}_{0.87}$ \\
$[3500,4000]$ & $11.2$ & $^+_-$ & $^{0.3}_{0.3}$&$^+_-$ & $^{1.2}_{1.0}$ & $10.88$ & $^+_-$ & $^{0.17}_{0.17}$&$^+_-$ & $^{0.55}_{0.72}$ & $11.72$ & $^+_-$ & $^{0.19}_{0.19}$&$^+_-$ & $^{0.60}_{0.54}$ & $11.58$ & $^+_-$ & $^{0.25}_{0.25}$&$^+_-$ & $^{0.73}_{0.71}$ & $10.6$ & $^+_-$ & $^{0.6}_{0.6}$&$^+_-$ & $^{1.3}_{1.1}$ \\
$[4000,5000]$ & $10.80$ & $^+_-$ & $^{0.22}_{0.21}$&$^+_-$ & $^{0.90}_{0.97}$ & $12.01$ & $^+_-$ & $^{0.15}_{0.15}$&$^+_-$ & $^{0.60}_{0.61}$ & $11.37$ & $^+_-$ & $^{0.15}_{0.15}$&$^+_-$ & $^{0.48}_{0.42}$ & $12.08$ & $^+_-$ & $^{0.22}_{0.22}$&$^+_-$ & $^{0.57}_{0.54}$ & $14.0$ & $^+_-$ & $^{0.7}_{0.6}$&$^+_-$ & $^{2.1}_{1.7}$ \\
$[5000,6000]$ & $12.5$ & $^+_-$ & $^{0.3}_{0.3}$&$^+_-$ & $^{0.8}_{1.2}$ & $13.02$ & $^+_-$ & $^{0.21}_{0.20}$&$^+_-$ & $^{0.77}_{0.58}$ & $12.53$ & $^+_-$ & $^{0.22}_{0.22}$&$^+_-$ & $^{0.65}_{0.51}$ & $12.79$ & $^+_-$ & $^{0.30}_{0.29}$&$^+_-$ & $^{0.88}_{0.67}$ & $16.5$ & $^+_-$ & $^{1.2}_{1.2}$&$^+_-$ & $^{2.8}_{2.7}$ \\
$[6000,7000]$ & $11.95$ & $^+_-$ & $^{0.33}_{0.33}$&$^+_-$ & $^{0.81}_{1.00}$ & $12.94$ & $^+_-$ & $^{0.27}_{0.26}$&$^+_-$ & $^{0.77}_{0.62}$ & $12.15$ & $^+_-$ & $^{0.28}_{0.27}$&$^+_-$ & $^{0.73}_{0.59}$ & $11.86$ & $^+_-$ & $^{0.40}_{0.39}$&$^+_-$ & $^{0.96}_{0.66}$ & $23$ & $^+_-$ & $^{5}_{4}$&$^+_-$ & $^{15}_{10}$ \\
$[7000,8000]$ & $13.8$ & $^+_-$ & $^{0.5}_{0.5}$&$^+_-$ & $^{1.2}_{1.1}$ & $10.64$ & $^+_-$ & $^{0.29}_{0.29}$&$^+_-$ & $^{0.58}_{0.68}$ & $11.71$ & $^+_-$ & $^{0.35}_{0.34}$&$^+_-$ & $^{0.94}_{0.62}$ & $15.5$ & $^+_-$ & $^{0.8}_{0.8}$&$^+_-$ & $^{2.2}_{1.8}$ & \multicolumn{5}{c}{ } \\
$[8000,9000]$ & $12.6$ & $^+_-$ & $^{0.5}_{0.5}$&$^+_-$ & $^{1.3}_{1.1}$ & $12.06$ & $^+_-$ & $^{0.41}_{0.41}$&$^+_-$ & $^{0.84}_{0.93}$ & $15.7$ & $^+_-$ & $^{0.6}_{0.6}$&$^+_-$ & $^{1.6}_{1.2}$ & $14.5$ & $^+_-$ & $^{1.1}_{1.1}$&$^+_-$ & $^{3.0}_{2.6}$ & \multicolumn{5}{c}{ } \\
$[9000,10000]$ & $11.11$ & $^+_-$ & $^{0.56}_{0.55}$&$^+_-$ & $^{0.96}_{0.74}$ & $10.60$ & $^+_-$ & $^{0.46}_{0.45}$&$^+_-$ & $^{0.57}_{0.57}$ & $12.35$ & $^+_-$ & $^{0.61}_{0.59}$&$^+_-$ & $^{0.79}_{0.65}$ & $16.3$ & $^+_-$ & $^{2.1}_{1.9}$&$^+_-$ & $^{3.7}_{3.7}$ & \multicolumn{5}{c}{ } \\
$[10000,11000]$ & $12.84$ & $^+_-$ & $^{0.80}_{0.78}$&$^+_-$ & $^{0.82}_{0.23}$ & $11.92$ & $^+_-$ & $^{0.63}_{0.63}$&$^+_-$ & $^{0.24}_{0.36}$ & $12.30$ & $^+_-$ & $^{0.82}_{0.82}$&$^+_-$ & $^{0.77}_{0.14}$ & $14.5$ & $^+_-$ & $^{4.1}_{3.2}$&$^+_-$ & $^{4.8}_{3.9}$ & \multicolumn{5}{c}{ } \\
$[11000,12000]$ & $12.05$ & $^+_-$ & $^{0.94}_{0.94}$&$^+_-$ & $^{0.69}_{0.33}$ & $11.21$ & $^+_-$ & $^{0.74}_{0.74}$&$^+_-$ & $^{0.42}_{0.27}$ & $13.5$ & $^+_-$ & $^{1.1}_{1.1}$&$^+_-$ & $^{1.0}_{0.4}$ & \multicolumn{5}{c}{ } & \multicolumn{5}{c}{ } \\
$[12000,13000]$ & $14.0$ & $^+_-$ & $^{1.2}_{1.2}$&$^+_-$ & $^{0.9}_{0.6}$ & $9.17$ & $^+_-$ & $^{0.81}_{0.80}$&$^+_-$ & $^{0.43}_{0.25}$ & $14.0$ & $^+_-$ & $^{1.5}_{1.4}$&$^+_-$ & $^{1.0}_{1.0}$ & \multicolumn{5}{c}{ } & \multicolumn{5}{c}{ } \\
$[13000,14000]$ & $10.6$ & $^+_-$ & $^{1.3}_{1.2}$&$^+_-$ & $^{0.6}_{0.5}$ & $12.8$ & $^+_-$ & $^{1.2}_{1.1}$&$^+_-$ & $^{0.8}_{0.3}$ & $14.1$ & $^+_-$ & $^{2.2}_{2.0}$&$^+_-$ & $^{1.9}_{0.2}$ & \multicolumn{5}{c}{ } & \multicolumn{5}{c}{ } \\
$[14000,15000]$ & \multicolumn{5}{c}{ } & \multicolumn{5}{c}{ } & \multicolumn{5}{c}{ } & \multicolumn{5}{c}{ } & \multicolumn{5}{c}{ } \\
\bottomrule\end{tabular}

%% file: tables/mode_ratios/Dst_by_D0_nbc.tex
\renewcommand{\arraystretch}{1.3}
\begin{tabular}{l|r@{\hskip+0.2em}c@{\hskip+0.2em}r@{\hskip+0.2em}c@{\hskip+0.2em}rr@{\hskip+0.2em}c@{\hskip+0.2em}r@{\hskip+0.2em}c@{\hskip+0.2em}rr@{\hskip+0.2em}c@{\hskip+0.2em}r@{\hskip+0.2em}c@{\hskip+0.2em}rr@{\hskip+0.2em}c@{\hskip+0.2em}r@{\hskip+0.2em}c@{\hskip+0.2em}rr@{\hskip+0.2em}c@{\hskip+0.2em}r@{\hskip+0.2em}c@{\hskip+0.2em}r}
\toprule&\multicolumn{25}{c}{$\text{$y$}$}\\
$\text{$p_{\text{T}}$ [\text{MeV}/c]}$ & \multicolumn{5}{c}{$[2,2.5]$} & \multicolumn{5}{c}{$[2.5,3]$} & \multicolumn{5}{c}{$[3,3.5]$} & \multicolumn{5}{c}{$[3.5,4]$} & \multicolumn{5}{c}{$[4,4.5]$} \\
\midrule$[0,1000]$ & \multicolumn{5}{c}{ } & \multicolumn{5}{c}{ } & \multicolumn{5}{c}{ } & \multicolumn{5}{c}{ } & $21.9$ & $^+_-$ & $^{3.0}_{3.0}$&$^+_-$ & $^{6.7}_{6.3}$ \\
$[1000,1500]$ & \multicolumn{5}{c}{ } & $18.3$ & $^+_-$ & $^{0.8}_{0.8}$&$^+_-$ & $^{2.0}_{2.0}$ & $22.6$ & $^+_-$ & $^{0.3}_{0.3}$&$^+_-$ & $^{1.3}_{1.6}$ & $20.3$ & $^+_-$ & $^{0.3}_{0.3}$&$^+_-$ & $^{1.4}_{1.5}$ & $25.5$ & $^+_-$ & $^{0.8}_{0.8}$&$^+_-$ & $^{3.7}_{3.1}$ \\
$[1500,2000]$ & \multicolumn{5}{c}{ } & $26.3$ & $^+_-$ & $^{0.5}_{0.4}$&$^+_-$ & $^{1.8}_{1.6}$ & $26.4$ & $^+_-$ & $^{0.2}_{0.2}$&$^+_-$ & $^{1.3}_{1.9}$ & $24.7$ & $^+_-$ & $^{0.3}_{0.3}$&$^+_-$ & $^{1.5}_{1.9}$ & $25.5$ & $^+_-$ & $^{0.6}_{0.6}$&$^+_-$ & $^{2.5}_{2.0}$ \\
$[2000,2500]$ & $26.8$ & $^+_-$ & $^{2.4}_{2.4}$&$^+_-$ & $^{5.7}_{6.0}$ & $26.5$ & $^+_-$ & $^{0.3}_{0.3}$&$^+_-$ & $^{1.9}_{1.1}$ & $27.4$ & $^+_-$ & $^{0.2}_{0.2}$&$^+_-$ & $^{1.3}_{1.8}$ & $25.7$ & $^+_-$ & $^{0.2}_{0.2}$&$^+_-$ & $^{1.6}_{1.6}$ & $25.5$ & $^+_-$ & $^{0.5}_{0.5}$&$^+_-$ & $^{2.6}_{2.1}$ \\
$[2500,3000]$ & $26.8$ & $^+_-$ & $^{0.9}_{0.9}$&$^+_-$ & $^{2.9}_{3.1}$ & $27.1$ & $^+_-$ & $^{0.3}_{0.3}$&$^+_-$ & $^{1.2}_{1.7}$ & $27.0$ & $^+_-$ & $^{0.2}_{0.2}$&$^+_-$ & $^{1.6}_{1.5}$ & $26.0$ & $^+_-$ & $^{0.2}_{0.3}$&$^+_-$ & $^{1.7}_{1.7}$ & $26.6$ & $^+_-$ & $^{0.6}_{0.6}$&$^+_-$ & $^{1.8}_{2.0}$ \\
$[3000,3500]$ & $27.2$ & $^+_-$ & $^{0.7}_{0.7}$&$^+_-$ & $^{2.2}_{2.5}$ & $28.3$ & $^+_-$ & $^{0.2}_{0.2}$&$^+_-$ & $^{1.2}_{1.9}$ & $28.6$ & $^+_-$ & $^{0.2}_{0.2}$&$^+_-$ & $^{1.7}_{1.4}$ & $25.5$ & $^+_-$ & $^{0.3}_{0.3}$&$^+_-$ & $^{1.8}_{1.8}$ & $25.9$ & $^+_-$ & $^{0.6}_{0.6}$&$^+_-$ & $^{2.7}_{2.2}$ \\
$[3500,4000]$ & $28.9$ & $^+_-$ & $^{0.6}_{0.6}$&$^+_-$ & $^{2.1}_{2.5}$ & $29.8$ & $^+_-$ & $^{0.3}_{0.3}$&$^+_-$ & $^{1.4}_{2.1}$ & $28.9$ & $^+_-$ & $^{0.2}_{0.2}$&$^+_-$ & $^{1.8}_{1.4}$ & $27.2$ & $^+_-$ & $^{0.3}_{0.3}$&$^+_-$ & $^{1.9}_{1.7}$ & $27.9$ & $^+_-$ & $^{0.8}_{0.8}$&$^+_-$ & $^{2.7}_{2.7}$ \\
$[4000,5000]$ & $28.8$ & $^+_-$ & $^{0.4}_{0.4}$&$^+_-$ & $^{1.4}_{2.3}$ & $29.2$ & $^+_-$ & $^{0.2}_{0.2}$&$^+_-$ & $^{1.3}_{2.0}$ & $28.5$ & $^+_-$ & $^{0.2}_{0.2}$&$^+_-$ & $^{1.6}_{1.4}$ & $28.2$ & $^+_-$ & $^{0.3}_{0.3}$&$^+_-$ & $^{2.0}_{1.5}$ & $30.3$ & $^+_-$ & $^{0.9}_{0.9}$&$^+_-$ & $^{2.3}_{2.2}$ \\
$[5000,6000]$ & $27.3$ & $^+_-$ & $^{0.4}_{0.4}$&$^+_-$ & $^{1.3}_{2.3}$ & $29.4$ & $^+_-$ & $^{0.2}_{0.2}$&$^+_-$ & $^{1.4}_{2.0}$ & $29.9$ & $^+_-$ & $^{0.3}_{0.3}$&$^+_-$ & $^{1.7}_{1.3}$ & $32.2$ & $^+_-$ & $^{0.4}_{0.5}$&$^+_-$ & $^{2.3}_{1.6}$ & $31.5$ & $^+_-$ & $^{2.1}_{2.0}$&$^+_-$ & $^{4.1}_{3.7}$ \\
$[6000,7000]$ & $30.9$ & $^+_-$ & $^{0.5}_{0.5}$&$^+_-$ & $^{1.8}_{2.6}$ & $30.8$ & $^+_-$ & $^{0.3}_{0.3}$&$^+_-$ & $^{1.6}_{2.1}$ & $29.8$ & $^+_-$ & $^{0.4}_{0.4}$&$^+_-$ & $^{1.7}_{1.5}$ & $29.5$ & $^+_-$ & $^{0.6}_{0.6}$&$^+_-$ & $^{2.2}_{2.2}$ & $38$ & $^+_-$ & $^{10}_{\phantom{0}8}$&$^+_-$ & $^{24}_{17}$ \\
$[7000,8000]$ & $33.1$ & $^+_-$ & $^{0.7}_{0.7}$&$^+_-$ & $^{2.0}_{2.9}$ & $29.6$ & $^+_-$ & $^{0.4}_{0.4}$&$^+_-$ & $^{1.7}_{2.3}$ & $31.7$ & $^+_-$ & $^{0.5}_{0.5}$&$^+_-$ & $^{2.2}_{1.6}$ & $36.8$ & $^+_-$ & $^{1.3}_{1.2}$&$^+_-$ & $^{4.4}_{4.3}$ & \multicolumn{5}{c}{ } \\
$[8000,9000]$ & $32.2$ & $^+_-$ & $^{0.8}_{0.8}$&$^+_-$ & $^{2.2}_{2.9}$ & $29.9$ & $^+_-$ & $^{0.5}_{0.5}$&$^+_-$ & $^{1.9}_{2.3}$ & $31.4$ & $^+_-$ & $^{0.7}_{0.7}$&$^+_-$ & $^{2.8}_{1.7}$ & $28.0$ & $^+_-$ & $^{1.8}_{1.7}$&$^+_-$ & $^{4.1}_{3.4}$ & \multicolumn{5}{c}{ } \\
$[9000,10000]$ & $21.8$ & $^+_-$ & $^{0.7}_{0.7}$&$^+_-$ & $^{1.2}_{1.7}$ & $30.9$ & $^+_-$ & $^{0.7}_{0.7}$&$^+_-$ & $^{1.6}_{2.1}$ & $30.2$ & $^+_-$ & $^{0.8}_{0.8}$&$^+_-$ & $^{1.6}_{1.5}$ & $40.6$ & $^+_-$ & $^{4.8}_{4.3}$&$^+_-$ & $^{6.1}_{7.0}$ & \multicolumn{5}{c}{ } \\
$[10000,11000]$ & $31.8$ & $^+_-$ & $^{1.2}_{1.1}$&$^+_-$ & $^{1.6}_{2.2}$ & $32.1$ & $^+_-$ & $^{0.9}_{0.9}$&$^+_-$ & $^{1.6}_{2.1}$ & $34.6$ & $^+_-$ & $^{1.4}_{1.3}$&$^+_-$ & $^{1.8}_{1.8}$ & $34$ & $^+_-$ & $^{10}_{\phantom{0}8}$&$^+_-$ & $^{13}_{11}$ & \multicolumn{5}{c}{ } \\
$[11000,12000]$ & $30.8$ & $^+_-$ & $^{1.3}_{1.4}$&$^+_-$ & $^{1.6}_{2.5}$ & $30.1$ & $^+_-$ & $^{1.1}_{1.1}$&$^+_-$ & $^{1.4}_{2.0}$ & $31.2$ & $^+_-$ & $^{1.8}_{1.7}$&$^+_-$ & $^{1.6}_{1.7}$ & \multicolumn{5}{c}{ } & \multicolumn{5}{c}{ } \\
$[12000,13000]$ & $33.0$ & $^+_-$ & $^{1.8}_{1.8}$&$^+_-$ & $^{2.1}_{3.1}$ & $32.2$ & $^+_-$ & $^{1.4}_{1.4}$&$^+_-$ & $^{1.6}_{2.2}$ & $32.8$ & $^+_-$ & $^{2.5}_{2.4}$&$^+_-$ & $^{2.2}_{1.8}$ & \multicolumn{5}{c}{ } & \multicolumn{5}{c}{ } \\
$[13000,14000]$ & $34.0$ & $^+_-$ & $^{2.1}_{2.1}$&$^+_-$ & $^{1.8}_{3.0}$ & $27.6$ & $^+_-$ & $^{1.6}_{1.6}$&$^+_-$ & $^{1.5}_{2.0}$ & $41.2$ & $^+_-$ & $^{4.7}_{4.3}$&$^+_-$ & $^{3.1}_{3.4}$ & \multicolumn{5}{c}{ } & \multicolumn{5}{c}{ } \\
$[14000,15000]$ & $29.7$ & $^+_-$ & $^{2.5}_{2.4}$&$^+_-$ & $^{1.9}_{2.4}$ & $34.0$ & $^+_-$ & $^{2.5}_{2.3}$&$^+_-$ & $^{2.3}_{2.5}$ & $27.8$ & $^+_-$ & $^{5.6}_{4.8}$&$^+_-$ & $^{6.7}_{5.7}$ & \multicolumn{5}{c}{ } & \multicolumn{5}{c}{ } \\
\bottomrule\end{tabular}

%% file: tables/mode_ratios/Ds_by_Dp_nbc.tex
\renewcommand{\arraystretch}{1.3}
\begin{tabular}{l|r@{\hskip+0.2em}c@{\hskip+0.2em}r@{\hskip+0.2em}c@{\hskip+0.2em}rr@{\hskip+0.2em}c@{\hskip+0.2em}r@{\hskip+0.2em}c@{\hskip+0.2em}rr@{\hskip+0.2em}c@{\hskip+0.2em}r@{\hskip+0.2em}c@{\hskip+0.2em}rr@{\hskip+0.2em}c@{\hskip+0.2em}r@{\hskip+0.2em}c@{\hskip+0.2em}rr@{\hskip+0.2em}c@{\hskip+0.2em}r@{\hskip+0.2em}c@{\hskip+0.2em}r}
\toprule&\multicolumn{25}{c}{$\text{$y$}$}\\
$\text{$p_{\text{T}}$ [\text{MeV}/c]}$ & \multicolumn{5}{c}{$[2,2.5]$} & \multicolumn{5}{c}{$[2.5,3]$} & \multicolumn{5}{c}{$[3,3.5]$} & \multicolumn{5}{c}{$[3.5,4]$} & \multicolumn{5}{c}{$[4,4.5]$} \\
\midrule$[1000,1500]$ & $4.1$ & $^+_-$ & $^{1.0}_{1.0}$&$^+_-$ & $^{2.0}_{1.7}$ & $8.7$ & $^+_-$ & $^{0.6}_{0.6}$&$^+_-$ & $^{1.9}_{1.5}$ & $9.3$ & $^+_-$ & $^{0.6}_{0.6}$&$^+_-$ & $^{1.4}_{1.4}$ & $8.5$ & $^+_-$ & $^{0.8}_{0.8}$&$^+_-$ & $^{2.0}_{1.6}$ & $20$ & $^+_-$ & $^{6}_{6}$&$^+_-$ & $^{\phantom{0}5}_{13}$ \\
$[1500,2000]$ & $8.0$ & $^+_-$ & $^{0.5}_{0.5}$&$^+_-$ & $^{1.5}_{1.2}$ & $10.42$ & $^+_-$ & $^{0.28}_{0.28}$&$^+_-$ & $^{0.98}_{0.86}$ & $9.92$ & $^+_-$ & $^{0.25}_{0.25}$&$^+_-$ & $^{0.68}_{0.67}$ & $11.2$ & $^+_-$ & $^{0.4}_{0.4}$&$^+_-$ & $^{1.1}_{1.0}$ & $6.8$ & $^+_-$ & $^{0.7}_{0.7}$&$^+_-$ & $^{1.2}_{1.2}$ \\
$[2000,2500]$ & $10.6$ & $^+_-$ & $^{0.4}_{0.4}$&$^+_-$ & $^{1.4}_{1.1}$ & $10.15$ & $^+_-$ & $^{0.18}_{0.18}$&$^+_-$ & $^{0.58}_{0.42}$ & $11.93$ & $^+_-$ & $^{0.20}_{0.21}$&$^+_-$ & $^{0.64}_{0.56}$ & $11.88$ & $^+_-$ & $^{0.28}_{0.28}$&$^+_-$ & $^{0.76}_{0.76}$ & $8.4$ & $^+_-$ & $^{0.5}_{0.6}$&$^+_-$ & $^{1.1}_{1.0}$ \\
$[2500,3000]$ & $10.9$ & $^+_-$ & $^{0.3}_{0.3}$&$^+_-$ & $^{1.2}_{0.9}$ & $11.51$ & $^+_-$ & $^{0.17}_{0.17}$&$^+_-$ & $^{0.44}_{0.52}$ & $11.25$ & $^+_-$ & $^{0.17}_{0.16}$&$^+_-$ & $^{0.60}_{0.36}$ & $11.55$ & $^+_-$ & $^{0.23}_{0.23}$&$^+_-$ & $^{0.65}_{0.64}$ & $10.9$ & $^+_-$ & $^{0.5}_{0.5}$&$^+_-$ & $^{1.1}_{1.2}$ \\
$[3000,3500]$ & $10.66$ & $^+_-$ & $^{0.30}_{0.29}$&$^+_-$ & $^{0.93}_{0.70}$ & $11.07$ & $^+_-$ & $^{0.16}_{0.17}$&$^+_-$ & $^{0.42}_{0.45}$ & $11.18$ & $^+_-$ & $^{0.17}_{0.17}$&$^+_-$ & $^{0.56}_{0.39}$ & $11.07$ & $^+_-$ & $^{0.22}_{0.22}$&$^+_-$ & $^{0.59}_{0.58}$ & $9.77$ & $^+_-$ & $^{0.48}_{0.48}$&$^+_-$ & $^{0.93}_{0.88}$ \\
$[3500,4000]$ & $12.27$ & $^+_-$ & $^{0.33}_{0.33}$&$^+_-$ & $^{0.95}_{0.67}$ & $11.05$ & $^+_-$ & $^{0.17}_{0.17}$&$^+_-$ & $^{0.43}_{0.49}$ & $11.09$ & $^+_-$ & $^{0.18}_{0.18}$&$^+_-$ & $^{0.54}_{0.46}$ & $12.21$ & $^+_-$ & $^{0.27}_{0.26}$&$^+_-$ & $^{0.69}_{0.69}$ & $11.6$ & $^+_-$ & $^{0.6}_{0.6}$&$^+_-$ & $^{1.3}_{1.0}$ \\
$[4000,5000]$ & $10.47$ & $^+_-$ & $^{0.21}_{0.21}$&$^+_-$ & $^{0.66}_{0.41}$ & $11.66$ & $^+_-$ & $^{0.15}_{0.15}$&$^+_-$ & $^{0.46}_{0.36}$ & $11.27$ & $^+_-$ & $^{0.15}_{0.15}$&$^+_-$ & $^{0.42}_{0.42}$ & $12.21$ & $^+_-$ & $^{0.22}_{0.22}$&$^+_-$ & $^{0.45}_{0.63}$ & $11.2$ & $^+_-$ & $^{0.5}_{0.5}$&$^+_-$ & $^{1.1}_{0.7}$ \\
$[5000,6000]$ & $12.49$ & $^+_-$ & $^{0.29}_{0.29}$&$^+_-$ & $^{0.70}_{0.65}$ & $12.03$ & $^+_-$ & $^{0.19}_{0.19}$&$^+_-$ & $^{0.62}_{0.41}$ & $11.87$ & $^+_-$ & $^{0.20}_{0.20}$&$^+_-$ & $^{0.51}_{0.54}$ & $12.00$ & $^+_-$ & $^{0.28}_{0.27}$&$^+_-$ & $^{0.81}_{0.63}$ & $11.1$ & $^+_-$ & $^{0.7}_{0.7}$&$^+_-$ & $^{1.4}_{1.2}$ \\
$[6000,7000]$ & $11.28$ & $^+_-$ & $^{0.31}_{0.31}$&$^+_-$ & $^{0.71}_{0.65}$ & $12.80$ & $^+_-$ & $^{0.26}_{0.26}$&$^+_-$ & $^{0.70}_{0.57}$ & $11.80$ & $^+_-$ & $^{0.27}_{0.27}$&$^+_-$ & $^{0.62}_{0.66}$ & $10.51$ & $^+_-$ & $^{0.34}_{0.33}$&$^+_-$ & $^{0.99}_{0.62}$ & $11.2$ & $^+_-$ & $^{1.3}_{1.3}$&$^+_-$ & $^{2.0}_{1.7}$ \\
$[7000,8000]$ & $12.0$ & $^+_-$ & $^{0.4}_{0.4}$&$^+_-$ & $^{1.0}_{0.8}$ & $10.35$ & $^+_-$ & $^{0.28}_{0.27}$&$^+_-$ & $^{0.58}_{0.65}$ & $11.17$ & $^+_-$ & $^{0.33}_{0.33}$&$^+_-$ & $^{0.74}_{0.69}$ & $12.7$ & $^+_-$ & $^{0.6}_{0.6}$&$^+_-$ & $^{1.4}_{0.9}$ & $7.2$ & $^+_-$ & $^{2.0}_{1.9}$&$^+_-$ & $^{2.7}_{2.2}$ \\
$[8000,9000]$ & $10.53$ & $^+_-$ & $^{0.43}_{0.44}$&$^+_-$ & $^{0.98}_{0.91}$ & $11.15$ & $^+_-$ & $^{0.38}_{0.38}$&$^+_-$ & $^{0.81}_{0.83}$ & $13.9$ & $^+_-$ & $^{0.5}_{0.5}$&$^+_-$ & $^{1.2}_{1.2}$ & $12.7$ & $^+_-$ & $^{0.9}_{0.8}$&$^+_-$ & $^{1.8}_{1.3}$ & \multicolumn{5}{c}{ } \\
$[9000,10000]$ & $12.6$ & $^+_-$ & $^{0.6}_{0.6}$&$^+_-$ & $^{1.1}_{0.8}$ & $11.65$ & $^+_-$ & $^{0.49}_{0.50}$&$^+_-$ & $^{0.76}_{0.66}$ & $12.11$ & $^+_-$ & $^{0.58}_{0.58}$&$^+_-$ & $^{0.92}_{0.62}$ & $11.4$ & $^+_-$ & $^{1.1}_{1.1}$&$^+_-$ & $^{1.2}_{1.0}$ & \multicolumn{5}{c}{ } \\
$[10000,11000]$ & $11.98$ & $^+_-$ & $^{0.72}_{0.72}$&$^+_-$ & $^{0.37}_{0.36}$ & $11.38$ & $^+_-$ & $^{0.59}_{0.61}$&$^+_-$ & $^{0.43}_{0.17}$ & $11.07$ & $^+_-$ & $^{0.70}_{0.71}$&$^+_-$ & $^{0.60}_{0.59}$ & $9.9$ & $^+_-$ & $^{1.8}_{1.7}$&$^+_-$ & $^{1.1}_{0.7}$ & \multicolumn{5}{c}{ } \\
$[11000,12000]$ & $10.58$ & $^+_-$ & $^{0.80}_{0.81}$&$^+_-$ & $^{0.36}_{0.35}$ & $11.05$ & $^+_-$ & $^{0.73}_{0.71}$&$^+_-$ & $^{0.52}_{0.14}$ & $11.57$ & $^+_-$ & $^{0.89}_{0.89}$&$^+_-$ & $^{0.45}_{0.44}$ & \multicolumn{5}{c}{ } & \multicolumn{5}{c}{ } \\
$[12000,13000]$ & $11.8$ & $^+_-$ & $^{1.0}_{1.0}$&$^+_-$ & $^{0.6}_{0.2}$ & $9.17$ & $^+_-$ & $^{0.79}_{0.78}$&$^+_-$ & $^{0.38}_{0.38}$ & $12.6$ & $^+_-$ & $^{1.2}_{1.2}$&$^+_-$ & $^{0.6}_{0.6}$ & \multicolumn{5}{c}{ } & \multicolumn{5}{c}{ } \\
$[13000,14000]$ & $9.3$ & $^+_-$ & $^{1.1}_{1.1}$&$^+_-$ & $^{0.4}_{0.4}$ & $13.1$ & $^+_-$ & $^{1.2}_{1.2}$&$^+_-$ & $^{0.8}_{0.2}$ & $11.7$ & $^+_-$ & $^{1.6}_{1.6}$&$^+_-$ & $^{1.1}_{1.0}$ & \multicolumn{5}{c}{ } & \multicolumn{5}{c}{ } \\
$[14000,15000]$ & \multicolumn{5}{c}{ } & \multicolumn{5}{c}{ } & \multicolumn{5}{c}{ } & \multicolumn{5}{c}{ } & \multicolumn{5}{c}{ } \\
\bottomrule\end{tabular}

%% file: tables/mode_ratios/Dst_by_Dp_nbc.tex
\renewcommand{\arraystretch}{1.3}
\begin{tabular}{l|r@{\hskip+0.2em}c@{\hskip+0.2em}r@{\hskip+0.2em}c@{\hskip+0.2em}rr@{\hskip+0.2em}c@{\hskip+0.2em}r@{\hskip+0.2em}c@{\hskip+0.2em}rr@{\hskip+0.2em}c@{\hskip+0.2em}r@{\hskip+0.2em}c@{\hskip+0.2em}rr@{\hskip+0.2em}c@{\hskip+0.2em}r@{\hskip+0.2em}c@{\hskip+0.2em}rr@{\hskip+0.2em}c@{\hskip+0.2em}r@{\hskip+0.2em}c@{\hskip+0.2em}r}
\toprule&\multicolumn{25}{c}{$\text{$y$}$}\\
$\text{$p_{\text{T}}$ [\text{MeV}/c]}$ & \multicolumn{5}{c}{$[2,2.5]$} & \multicolumn{5}{c}{$[2.5,3]$} & \multicolumn{5}{c}{$[3,3.5]$} & \multicolumn{5}{c}{$[3.5,4]$} & \multicolumn{5}{c}{$[4,4.5]$} \\
\midrule$[0,1000]$ & \multicolumn{5}{c}{ } & \multicolumn{5}{c}{ } & \multicolumn{5}{c}{ } & \multicolumn{5}{c}{ } & $27$ & $^+_-$ & $^{4}_{4}$&$^+_-$ & $^{11}_{\phantom{0}8}$ \\
$[1000,1500]$ & \multicolumn{5}{c}{ } & $20.9$ & $^+_-$ & $^{0.9}_{0.9}$&$^+_-$ & $^{2.5}_{3.3}$ & $23.7$ & $^+_-$ & $^{0.3}_{0.4}$&$^+_-$ & $^{1.4}_{1.6}$ & $22.3$ & $^+_-$ & $^{0.4}_{0.4}$&$^+_-$ & $^{1.5}_{1.6}$ & $26.7$ & $^+_-$ & $^{1.0}_{1.0}$&$^+_-$ & $^{4.4}_{3.4}$ \\
$[1500,2000]$ & \multicolumn{5}{c}{ } & $30.6$ & $^+_-$ & $^{0.5}_{0.5}$&$^+_-$ & $^{2.4}_{2.8}$ & $27.8$ & $^+_-$ & $^{0.2}_{0.2}$&$^+_-$ & $^{1.3}_{1.6}$ & $27.0$ & $^+_-$ & $^{0.3}_{0.3}$&$^+_-$ & $^{1.5}_{1.8}$ & $26.8$ & $^+_-$ & $^{0.7}_{0.6}$&$^+_-$ & $^{2.7}_{2.0}$ \\
$[2000,2500]$ & $30.4$ & $^+_-$ & $^{2.8}_{2.7}$&$^+_-$ & $^{6.3}_{7.1}$ & $28.0$ & $^+_-$ & $^{0.3}_{0.3}$&$^+_-$ & $^{1.7}_{1.2}$ & $29.5$ & $^+_-$ & $^{0.2}_{0.2}$&$^+_-$ & $^{1.2}_{1.4}$ & $29.1$ & $^+_-$ & $^{0.3}_{0.3}$&$^+_-$ & $^{1.6}_{1.4}$ & $26.0$ & $^+_-$ & $^{0.5}_{0.5}$&$^+_-$ & $^{2.4}_{2.1}$ \\
$[2500,3000]$ & $30.2$ & $^+_-$ & $^{1.1}_{1.1}$&$^+_-$ & $^{3.0}_{3.5}$ & $28.1$ & $^+_-$ & $^{0.3}_{0.3}$&$^+_-$ & $^{1.0}_{1.0}$ & $28.7$ & $^+_-$ & $^{0.2}_{0.2}$&$^+_-$ & $^{1.5}_{1.1}$ & $26.9$ & $^+_-$ & $^{0.3}_{0.3}$&$^+_-$ & $^{1.6}_{1.4}$ & $27.5$ & $^+_-$ & $^{0.6}_{0.6}$&$^+_-$ & $^{1.6}_{1.9}$ \\
$[3000,3500]$ & $29.9$ & $^+_-$ & $^{0.8}_{0.8}$&$^+_-$ & $^{2.2}_{2.6}$ & $28.1$ & $^+_-$ & $^{0.2}_{0.3}$&$^+_-$ & $^{0.8}_{1.0}$ & $28.1$ & $^+_-$ & $^{0.2}_{0.2}$&$^+_-$ & $^{1.5}_{1.0}$ & $26.4$ & $^+_-$ & $^{0.3}_{0.3}$&$^+_-$ & $^{1.6}_{1.4}$ & $27.0$ & $^+_-$ & $^{0.6}_{0.6}$&$^+_-$ & $^{2.8}_{2.4}$ \\
$[3500,4000]$ & $31.7$ & $^+_-$ & $^{0.7}_{0.7}$&$^+_-$ & $^{2.0}_{2.2}$ & $30.3$ & $^+_-$ & $^{0.3}_{0.3}$&$^+_-$ & $^{0.8}_{1.2}$ & $27.3$ & $^+_-$ & $^{0.2}_{0.2}$&$^+_-$ & $^{1.4}_{1.0}$ & $28.7$ & $^+_-$ & $^{0.3}_{0.3}$&$^+_-$ & $^{1.7}_{1.6}$ & $30.4$ & $^+_-$ & $^{0.8}_{0.8}$&$^+_-$ & $^{2.1}_{2.0}$ \\
$[4000,5000]$ & $27.9$ & $^+_-$ & $^{0.4}_{0.4}$&$^+_-$ & $^{1.3}_{1.3}$ & $28.3$ & $^+_-$ & $^{0.2}_{0.2}$&$^+_-$ & $^{0.7}_{1.2}$ & $28.2$ & $^+_-$ & $^{0.2}_{0.2}$&$^+_-$ & $^{1.2}_{1.1}$ & $28.5$ & $^+_-$ & $^{0.3}_{0.3}$&$^+_-$ & $^{1.6}_{1.6}$ & $24.3$ & $^+_-$ & $^{0.7}_{0.7}$&$^+_-$ & $^{2.7}_{1.9}$ \\
$[5000,6000]$ & $27.2$ & $^+_-$ & $^{0.4}_{0.4}$&$^+_-$ & $^{1.2}_{1.2}$ & $27.2$ & $^+_-$ & $^{0.2}_{0.2}$&$^+_-$ & $^{0.8}_{1.3}$ & $28.3$ & $^+_-$ & $^{0.2}_{0.2}$&$^+_-$ & $^{1.3}_{1.3}$ & $30.2$ & $^+_-$ & $^{0.4}_{0.4}$&$^+_-$ & $^{1.9}_{1.3}$ & $21.2$ & $^+_-$ & $^{1.2}_{1.2}$&$^+_-$ & $^{4.0}_{2.7}$ \\
$[6000,7000]$ & $29.2$ & $^+_-$ & $^{0.5}_{0.5}$&$^+_-$ & $^{1.4}_{1.5}$ & $30.5$ & $^+_-$ & $^{0.3}_{0.3}$&$^+_-$ & $^{1.1}_{1.6}$ & $28.9$ & $^+_-$ & $^{0.3}_{0.3}$&$^+_-$ & $^{1.2}_{1.5}$ & $26.1$ & $^+_-$ & $^{0.5}_{0.5}$&$^+_-$ & $^{1.6}_{1.3}$ & $18.3$ & $^+_-$ & $^{3.0}_{3.0}$&$^+_-$ & $^{4.0}_{3.4}$ \\
$[7000,8000]$ & $28.7$ & $^+_-$ & $^{0.6}_{0.6}$&$^+_-$ & $^{1.4}_{1.9}$ & $28.8$ & $^+_-$ & $^{0.4}_{0.4}$&$^+_-$ & $^{1.4}_{1.8}$ & $30.3$ & $^+_-$ & $^{0.5}_{0.5}$&$^+_-$ & $^{1.6}_{1.6}$ & $30.1$ & $^+_-$ & $^{0.9}_{0.9}$&$^+_-$ & $^{3.0}_{2.3}$ & \multicolumn{5}{c}{ } \\
$[8000,9000]$ & $27.0$ & $^+_-$ & $^{0.7}_{0.7}$&$^+_-$ & $^{1.4}_{2.2}$ & $27.7$ & $^+_-$ & $^{0.5}_{0.5}$&$^+_-$ & $^{1.4}_{1.8}$ & $27.8$ & $^+_-$ & $^{0.6}_{0.5}$&$^+_-$ & $^{2.1}_{1.6}$ & $24.4$ & $^+_-$ & $^{1.2}_{1.2}$&$^+_-$ & $^{3.8}_{2.4}$ & \multicolumn{5}{c}{ } \\
$[9000,10000]$ & $24.7$ & $^+_-$ & $^{0.8}_{0.8}$&$^+_-$ & $^{1.1}_{1.7}$ & $33.9$ & $^+_-$ & $^{0.7}_{0.7}$&$^+_-$ & $^{1.7}_{2.0}$ & $29.6$ & $^+_-$ & $^{0.8}_{0.8}$&$^+_-$ & $^{1.7}_{1.3}$ & $28.4$ & $^+_-$ & $^{2.6}_{2.5}$&$^+_-$ & $^{3.6}_{2.8}$ & \multicolumn{5}{c}{ } \\
$[10000,11000]$ & $29.3$ & $^+_-$ & $^{1.0}_{1.0}$&$^+_-$ & $^{0.6}_{1.5}$ & $30.7$ & $^+_-$ & $^{0.8}_{0.8}$&$^+_-$ & $^{1.3}_{1.5}$ & $30.5$ & $^+_-$ & $^{1.1}_{1.0}$&$^+_-$ & $^{1.4}_{0.3}$ & $23.1$ & $^+_-$ & $^{4.2}_{4.1}$&$^+_-$ & $^{4.8}_{4.3}$ & \multicolumn{5}{c}{ } \\
$[11000,12000]$ & $26.8$ & $^+_-$ & $^{1.2}_{1.1}$&$^+_-$ & $^{0.8}_{1.6}$ & $29.7$ & $^+_-$ & $^{1.0}_{1.0}$&$^+_-$ & $^{1.3}_{1.6}$ & $26.1$ & $^+_-$ & $^{1.3}_{1.3}$&$^+_-$ & $^{1.9}_{1.2}$ & \multicolumn{5}{c}{ } & \multicolumn{5}{c}{ } \\
$[12000,13000]$ & $27.7$ & $^+_-$ & $^{1.4}_{1.4}$&$^+_-$ & $^{1.0}_{1.9}$ & $31.9$ & $^+_-$ & $^{1.3}_{1.3}$&$^+_-$ & $^{1.5}_{1.7}$ & $29.7$ & $^+_-$ & $^{2.0}_{2.0}$&$^+_-$ & $^{1.7}_{1.1}$ & \multicolumn{5}{c}{ } & \multicolumn{5}{c}{ } \\
$[13000,14000]$ & $29.7$ & $^+_-$ & $^{1.8}_{1.7}$&$^+_-$ & $^{1.2}_{2.3}$ & $28.2$ & $^+_-$ & $^{1.6}_{1.5}$&$^+_-$ & $^{1.3}_{1.8}$ & $32.8$ & $^+_-$ & $^{3.0}_{2.8}$&$^+_-$ & $^{3.2}_{2.0}$ & \multicolumn{5}{c}{ } & \multicolumn{5}{c}{ } \\
$[14000,15000]$ & $25.1$ & $^+_-$ & $^{2.0}_{1.9}$&$^+_-$ & $^{1.7}_{2.2}$ & $33.7$ & $^+_-$ & $^{2.2}_{2.1}$&$^+_-$ & $^{1.6}_{2.4}$ & $19.5$ & $^+_-$ & $^{3.0}_{2.9}$&$^+_-$ & $^{3.2}_{3.3}$ & \multicolumn{5}{c}{ } & \multicolumn{5}{c}{ } \\
\bottomrule\end{tabular}

%% file: tables/mode_ratios/Ds_by_Dst_nbc.tex
\renewcommand{\arraystretch}{1.3}
\begin{tabular}{l|r@{\hskip+0.2em}c@{\hskip+0.2em}r@{\hskip+0.2em}c@{\hskip+0.2em}rr@{\hskip+0.2em}c@{\hskip+0.2em}r@{\hskip+0.2em}c@{\hskip+0.2em}rr@{\hskip+0.2em}c@{\hskip+0.2em}r@{\hskip+0.2em}c@{\hskip+0.2em}rr@{\hskip+0.2em}c@{\hskip+0.2em}r@{\hskip+0.2em}c@{\hskip+0.2em}rr@{\hskip+0.2em}c@{\hskip+0.2em}r@{\hskip+0.2em}c@{\hskip+0.2em}r}
\toprule&\multicolumn{25}{c}{$\text{$y$}$}\\
$\text{$p_{\text{T}}$ [\text{MeV}/c]}$ & \multicolumn{5}{c}{$[2,2.5]$} & \multicolumn{5}{c}{$[2.5,3]$} & \multicolumn{5}{c}{$[3,3.5]$} & \multicolumn{5}{c}{$[3.5,4]$} & \multicolumn{5}{c}{$[4,4.5]$} \\
\midrule$[1000,1500]$ & \multicolumn{5}{c}{ } & $41$ & $^+_-$ & $^{3}_{3}$&$^+_-$ & $^{14}_{\phantom{0}8}$ & $39.3$ & $^+_-$ & $^{2.4}_{2.4}$&$^+_-$ & $^{6.6}_{5.8}$ & $38.3$ & $^+_-$ & $^{3.9}_{3.8}$&$^+_-$ & $^{9.6}_{7.2}$ & $74$ & $^+_-$ & $^{24}_{24}$&$^+_-$ & $^{22}_{49}$ \\
$[1500,2000]$ & \multicolumn{5}{c}{ } & $34.1$ & $^+_-$ & $^{1.1}_{1.0}$&$^+_-$ & $^{5.6}_{4.1}$ & $35.7$ & $^+_-$ & $^{0.9}_{0.9}$&$^+_-$ & $^{2.9}_{2.4}$ & $41.6$ & $^+_-$ & $^{1.5}_{1.5}$&$^+_-$ & $^{5.2}_{4.1}$ & $25.4$ & $^+_-$ & $^{2.7}_{2.6}$&$^+_-$ & $^{4.6}_{4.8}$ \\
$[2000,2500]$ & $35$ & $^+_-$ & $^{4}_{3}$&$^+_-$ & $^{13}_{\phantom{0}7}$ & $36.2$ & $^+_-$ & $^{0.7}_{0.7}$&$^+_-$ & $^{2.7}_{2.5}$ & $40.5$ & $^+_-$ & $^{0.7}_{0.7}$&$^+_-$ & $^{2.9}_{2.1}$ & $40.8$ & $^+_-$ & $^{1.0}_{1.0}$&$^+_-$ & $^{3.3}_{3.3}$ & $32.2$ & $^+_-$ & $^{2.2}_{2.2}$&$^+_-$ & $^{5.0}_{4.5}$ \\
$[2500,3000]$ & $36.1$ & $^+_-$ & $^{1.7}_{1.6}$&$^+_-$ & $^{6.7}_{4.1}$ & $41.0$ & $^+_-$ & $^{0.7}_{0.7}$&$^+_-$ & $^{1.8}_{2.1}$ & $39.3$ & $^+_-$ & $^{0.6}_{0.6}$&$^+_-$ & $^{2.6}_{2.0}$ & $43.0$ & $^+_-$ & $^{0.9}_{0.9}$&$^+_-$ & $^{3.4}_{3.4}$ & $39.7$ & $^+_-$ & $^{2.1}_{2.1}$&$^+_-$ & $^{4.8}_{4.2}$ \\
$[3000,3500]$ & $35.6$ & $^+_-$ & $^{1.3}_{1.3}$&$^+_-$ & $^{4.9}_{3.1}$ & $39.4$ & $^+_-$ & $^{0.7}_{0.6}$&$^+_-$ & $^{1.8}_{1.6}$ & $39.8$ & $^+_-$ & $^{0.6}_{0.6}$&$^+_-$ & $^{2.3}_{2.3}$ & $41.9$ & $^+_-$ & $^{0.9}_{0.9}$&$^+_-$ & $^{3.1}_{3.2}$ & $36.2$ & $^+_-$ & $^{1.9}_{1.9}$&$^+_-$ & $^{4.7}_{4.5}$ \\
$[3500,4000]$ & $38.7$ & $^+_-$ & $^{1.3}_{1.3}$&$^+_-$ & $^{4.1}_{2.6}$ & $36.4$ & $^+_-$ & $^{0.6}_{0.6}$&$^+_-$ & $^{1.9}_{1.5}$ & $40.6$ & $^+_-$ & $^{0.7}_{0.7}$&$^+_-$ & $^{2.4}_{2.5}$ & $42.5$ & $^+_-$ & $^{1.0}_{1.0}$&$^+_-$ & $^{3.5}_{3.3}$ & $38.1$ & $^+_-$ & $^{2.3}_{2.3}$&$^+_-$ & $^{4.8}_{3.8}$ \\
$[4000,5000]$ & $37.6$ & $^+_-$ & $^{0.9}_{0.9}$&$^+_-$ & $^{2.8}_{1.8}$ & $41.2$ & $^+_-$ & $^{0.6}_{0.6}$&$^+_-$ & $^{2.4}_{1.2}$ & $39.9$ & $^+_-$ & $^{0.6}_{0.6}$&$^+_-$ & $^{2.1}_{2.2}$ & $42.9$ & $^+_-$ & $^{0.8}_{0.8}$&$^+_-$ & $^{2.7}_{3.0}$ & $46.3$ & $^+_-$ & $^{2.4}_{2.3}$&$^+_-$ & $^{6.4}_{5.2}$ \\
$[5000,6000]$ & $45.8$ & $^+_-$ & $^{1.2}_{1.2}$&$^+_-$ & $^{2.9}_{2.7}$ & $44.2$ & $^+_-$ & $^{0.8}_{0.8}$&$^+_-$ & $^{3.4}_{1.5}$ & $42.0$ & $^+_-$ & $^{0.8}_{0.8}$&$^+_-$ & $^{2.5}_{2.4}$ & $39.7$ & $^+_-$ & $^{1.0}_{1.0}$&$^+_-$ & $^{3.1}_{3.1}$ & $52.2$ & $^+_-$ & $^{4.6}_{4.2}$&$^+_-$ & $^{7.9}_{7.8}$ \\
$[6000,7000]$ & $38.7$ & $^+_-$ & $^{1.2}_{1.2}$&$^+_-$ & $^{2.9}_{2.3}$ & $42.0$ & $^+_-$ & $^{0.9}_{0.9}$&$^+_-$ & $^{3.4}_{2.0}$ & $40.8$ & $^+_-$ & $^{1.0}_{1.0}$&$^+_-$ & $^{2.9}_{2.5}$ & $40.2$ & $^+_-$ & $^{1.5}_{1.4}$&$^+_-$ & $^{3.8}_{2.6}$ & $61$ & $^+_-$ & $^{14}_{11}$&$^+_-$ & $^{19}_{12}$ \\
$[7000,8000]$ & $41.7$ & $^+_-$ & $^{1.6}_{1.6}$&$^+_-$ & $^{4.5}_{2.6}$ & $35.9$ & $^+_-$ & $^{1.0}_{1.0}$&$^+_-$ & $^{3.1}_{2.5}$ & $36.9$ & $^+_-$ & $^{1.2}_{1.1}$&$^+_-$ & $^{3.0}_{2.5}$ & $42.3$ & $^+_-$ & $^{2.3}_{2.2}$&$^+_-$ & $^{5.5}_{4.2}$ & \multicolumn{5}{c}{ } \\
$[8000,9000]$ & $39.0$ & $^+_-$ & $^{1.8}_{1.7}$&$^+_-$ & $^{4.6}_{3.1}$ & $40.3$ & $^+_-$ & $^{1.5}_{1.4}$&$^+_-$ & $^{3.9}_{3.2}$ & $50.0$ & $^+_-$ & $^{2.0}_{2.0}$&$^+_-$ & $^{4.9}_{5.1}$ & $51.8$ & $^+_-$ & $^{4.2}_{4.0}$&$^+_-$ & $^{9.3}_{8.5}$ & \multicolumn{5}{c}{ } \\
$[9000,10000]$ & $51.1$ & $^+_-$ & $^{2.8}_{2.8}$&$^+_-$ & $^{6.0}_{3.1}$ & $34.3$ & $^+_-$ & $^{1.5}_{1.5}$&$^+_-$ & $^{3.1}_{2.2}$ & $40.9$ & $^+_-$ & $^{2.1}_{2.1}$&$^+_-$ & $^{3.4}_{2.8}$ & $40.2$ & $^+_-$ & $^{5.5}_{4.8}$&$^+_-$ & $^{6.1}_{5.5}$ & \multicolumn{5}{c}{ } \\
$[10000,11000]$ & $41.4$ & $^+_-$ & $^{2.7}_{2.7}$&$^+_-$ & $^{2.1}_{2.1}$ & $37.1$ & $^+_-$ & $^{2.1}_{2.1}$&$^+_-$ & $^{2.6}_{1.6}$ & $35.5$ & $^+_-$ & $^{2.5}_{2.5}$&$^+_-$ & $^{2.8}_{1.2}$ & $43$ & $^+_-$ & $^{12}_{\phantom{0}9}$&$^+_-$ & $^{11}_{\phantom{0}5}$ & \multicolumn{5}{c}{ } \\
$[11000,12000]$ & $40.1$ & $^+_-$ & $^{3.3}_{3.2}$&$^+_-$ & $^{2.4}_{2.4}$ & $37.3$ & $^+_-$ & $^{2.6}_{2.6}$&$^+_-$ & $^{3.0}_{1.3}$ & $43.4$ & $^+_-$ & $^{3.9}_{3.7}$&$^+_-$ & $^{3.7}_{1.7}$ & \multicolumn{5}{c}{ } & \multicolumn{5}{c}{ } \\
$[12000,13000]$ & $42.5$ & $^+_-$ & $^{4.1}_{3.8}$&$^+_-$ & $^{4.3}_{1.2}$ & $28.4$ & $^+_-$ & $^{2.6}_{2.5}$&$^+_-$ & $^{2.5}_{0.9}$ & $42.6$ & $^+_-$ & $^{4.7}_{4.6}$&$^+_-$ & $^{2.5}_{2.5}$ & \multicolumn{5}{c}{ } & \multicolumn{5}{c}{ } \\
$[13000,14000]$ & $31.4$ & $^+_-$ & $^{4.0}_{3.9}$&$^+_-$ & $^{3.3}_{1.0}$ & $46.5$ & $^+_-$ & $^{4.7}_{4.5}$&$^+_-$ & $^{4.7}_{1.5}$ & $35.5$ & $^+_-$ & $^{5.5}_{5.4}$&$^+_-$ & $^{2.8}_{2.7}$ & \multicolumn{5}{c}{ } & \multicolumn{5}{c}{ } \\
$[14000,15000]$ & \multicolumn{5}{c}{ } & \multicolumn{5}{c}{ } & \multicolumn{5}{c}{ } & \multicolumn{5}{c}{ } & \multicolumn{5}{c}{ } \\
\bottomrule\end{tabular}

%% file: LHCb_HD_authorlist_2015-07-21.tex
\centerline{\large\bf LHCb collaboration}
\begin{flushleft}
\small
R.~Aaij$^{38}$, 
C.~Abell\'{a}n~Beteta$^{40}$, 
B.~Adeva$^{37}$, 
M.~Adinolfi$^{46}$, 
A.~Affolder$^{52}$, 
Z.~Ajaltouni$^{5}$, 
S.~Akar$^{6}$, 
J.~Albrecht$^{9}$, 
F.~Alessio$^{38}$, 
M.~Alexander$^{51}$
% ...

{\footnotesize \it
% ...
$ ^{5}$Clermont Universit\'{e}, Universit\'{e} Blaise Pascal, CNRS/IN2P3, LPC, Clermont-Ferrand, France\\
$ ^{6}$CPPM, Aix-Marseille Universit\'{e}, CNRS/IN2P3, Marseille, France\\
% ...
$ ^{9}$Fakult\"{a}t Physik, Technische Universit\"{a}t Dortmund, Dortmund, Germany\\
% ...
$ ^{37}$Universidad de Santiago de Compostela, Santiago de Compostela, Spain\\
$ ^{38}$European Organization for Nuclear Research (CERN), Geneva, Switzerland\\
% ...
$ ^{40}$Physik-Institut, Universit\"{a}t Z\"{u}rich, Z\"{u}rich, Switzerland\\
% ...
$ ^{46}$H.H. Wills Physics Laboratory, University of Bristol, Bristol, United Kingdom\\
% ...
$ ^{51}$School of Physics and Astronomy, University of Glasgow, Glasgow, United Kingdom\\
$ ^{52}$Oliver Lodge Laboratory, University of Liverpool, Liverpool, United Kingdom\\
% ...
}
\end{flushleft}